\documentclass[twoside]{elsart}
\journal{Physics Reports}
\usepackage[dvips]{epsfig}
\usepackage{hangcaption}
\usepackage[dvips]{graphicx}
\usepackage[dvips]{color}
\usepackage{amssymb,amsmath}
\usepackage{calc}
\usepackage{natbib}
\usepackage{pifont}
\def\deg{^{\circ}}
\def\be{\begin{equation}}
\def\ee{\end{equation}}
\def\dl{\Phi^{\mbox{DL}}}
\def\stl{\Phi^{\mbox{SL}}}
\def\sdl{\Phi^{\mbox{SDL}}}
\def\ds{\Phi^{\mbox{DS}}}
\def\ps{\Phi^{\mbox{PS}}}
\def\diffflux{\mbox{GeV}^{-1}\,\mbox{cm}^{-2}\,\mbox{s}^{-1}\,\mbox{sr}^{-1}}
\def\pointflux{\mbox{GeV}^{-1}\,\mbox{cm}^{-2}\,\mbox{s}^{-1}}
\def\diffunits{\mbox{GeV cm}^{-2}\,\mbox{s}^{-1}\,\mbox{sr}^{-1}}
\def\pointunits{\mbox{GeV cm}^{-2}\,\mbox{s}^{-1}}
\def\en{E_{\nu}}
\def\eg{E_{\gamma}}
\def\ep{E_{p}}
\def\enb{\epsilon_{\nu}^{b}}
\def\enbG{\epsilon_{\nu,GeV}^{b}}

\def\ens{\epsilon_{\nu}^{s}}
\def\ensG{\epsilon_{\nu,GeV}^{s}}
\def\egb{\epsilon_{\gamma}^{b}}
\def\egbM{\epsilon_{\gamma,MeV}^{b}}
\def\exb{\epsilon_{X}^{b}}
\def\g25{\Gamma_{2.5}}
\def\lumi{L_{\gamma}^{52}}
\usepackage{amssymb}  % Mathesymbole
\begin{document}% HEADER
%%%%%%%%%%%%%%%%%%%%%%%%%%%%%%%%%%%%%%%%%%%%%%%%
\pagenumbering{arabic}
\begin{frontmatter}
\title{High-energy neutrinos in the context of multimessenger astrophysics}
\author[dort]{Julia K.~Becker\corauthref{cor}}
\address[dort]{Institut f\"ur Physik, Universit\"at Dortmund, 44221 Dortmund, Germany}
\corauth[cor]{{\scriptsize Contact: julia.becker@tu-dortmund.de, phone: +49-231-7553667}}
\date{\today}
%\linenumbers
\begin{abstract}
\noindent
The field of astroparticle physics is currently developing rapidly, since new
experiments challenge our understanding of the investigated processes. Three
messengers can be used to extract information on the properties of astrophysical
sources: photons, charged Cosmic Rays and neutrinos. This review focuses on
high-energy neutrinos ($E_{\nu}>100$~GeV) with the main topics as follows.
\begin{itemize}
\item The production mechanism of high-energy neutrinos in astrophysical
  shocks. The connection between the observed photon
  spectra and charged Cosmic Rays is described and the source properties
  as they are known from photon observations and from charged Cosmic Rays are
  presented.
\item High-energy neutrino detection. Current detection methods are described and the status of the
  next generation neutrino telescopes are reviewed. In particular, water
  and ice Cherenkov detectors as well as radio measurements in ice and with
  balloon experiments are presented. In addition, future perspectives for
  optical, radio and acoustic detection of neutrinos
  are reviewed.
\item Sources of neutrino emission. The main source classes are
  reviewed, i.e.~galactic sources, Active Galactic Nuclei, starburst galaxies and Gamma Ray
  Bursts. The interaction of high energy protons with the cosmic microwave
  background implies the production of neutrinos, referred to as GZK
  neutrinos. 
\item Implications of neutrino flux limits. Recent limits given by the AMANDA
  experiment and their implications regarding the physics of the sources are presented.
\end{itemize}
\end{abstract}
\end{frontmatter}
\clearpage
\tableofcontents
\parindent=0cm
\parskip=0.2cm
%\linenumbers
%%%%%%%%%%%%%%%%%%%%%%%%%%%%%%%%%%%%%%%%%%%%
% introductory sections,  experiments and galactic
\section{Introduction}
The multimessenger connection between Cosmic Rays, photons and neutrinos of
different wavelengths is crucial to comprehend in the pursuit of a deeper understanding of the fundamental processes driving
non-thermal astrophysical sources. None of these messengers alone is able to
give a complete picture. While photons reveal the surface of those
objects due to the optical thickness of the sources, charged Cosmic Rays
give a direct sight into the inner acceleration processes. However, they do
not point back to their origin, since scrambled by intergalactic magnetic
fields. Neutrinos are produced in interactions of protons with a photon field
or other protons. The detection of an extraterrestrial neutrino signal at the
highest energies ($\en>100$~GeV) renders possible the exploration of the
particle acceleration region itself. Neutrinos escape the acceleration region
and propagate through space basically untouched and this advantage is directly
connected to the drawback of the very low detection probability of neutrinos. In this review, the connection of the
three messengers will be reviewed with the focus on the resulting neutrino
fluxes and their detection probabilities. 

\begin{figure}[h!]
\centering{
\epsfig{file=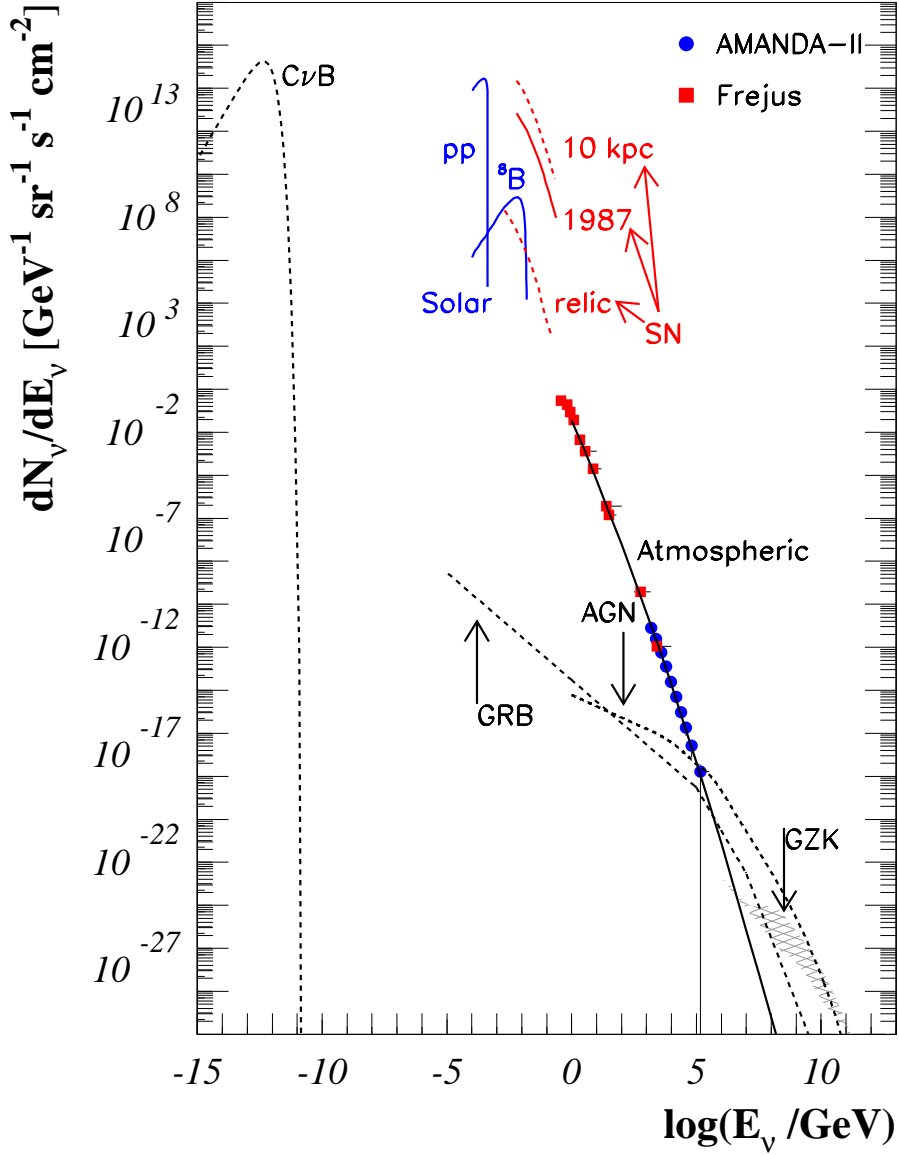,width=\linewidth}
\caption[Astrophysical neutrino spectra]{The astrophysical neutrino spectrum including different source
  predictions ranging from meV up to EeV energies. Point source fluxes have
  been scaled by $1/(4\,\pi)$ in order to be comparable to diffuse
  spectra. Figure after~\cite{koshiba1992,roulet00}. The individual spectra
  are explained and referenced in the text. The atmospheric prediction, averaged over the
  solid angle, is taken from~\cite{volkova80}, the atmospheric data are from
  the {\sc Fr{\'e}jus} experiment~\cite{frejus_spectrum} (red squares) and from the
  {\sc AMANDA} experiment (blue
  circles)~\cite{kirsten_icrc07,kirsten_phd}. The fluxes based on mere
  predictions are shown as dashed lines. The solid lines represent those
  fluxes already measured.}
\label{diffuse_all}
}
\end{figure}

Astrophysical neutrinos are produced over the broad energy range, covering 15
orders of magnitude. An overview
of the neutrino spectrum  is
shown in Fig.~\ref{diffuse_all}. From the lowest energies of $\en\sim$~meV to
the highest energies of $\en\sim$~EeV, the intensity of the signal decreases
by 42 orders of magnitude, making it necessary to introduce new methods of
neutrino detection and analysis in order to increase the sensitivity to the
neutrino fluxes especially at the highest energies.

The Cosmic Neutrino
Background (C$\nu$B) is an isotropic neutrino flux having decoupled in the
early Universe, only $1$~s after the Big Bang, see~\cite{weinberg1972} and
references therein. It is the neutrino equivalent
of the cosmic microwave background (CMB). The
temperature of the blackbody spectrum has dropped to $\sim 1.9$~K today due to
the expansion of the Universe, and the flux peaks at milli-eV energies. While this background is essentially
predicted in the standard model of cosmology, it was not possible to test it
experimentally yet. Recently, the measurement of the C$\nu$B was
proposed by using the interaction of cosmic neutrinos with nuclei
undergoing beta decay~\cite{cocco2007}, with $^3$H as an optimal candidate. For the case of a $100$~g $^3$H-detector, an event rate of a few tens
to a few hundreds is expected. The exact number depends on the distribution of cosmic
neutrinos and the neutrino mass. This is currently not within experimental
reach, but can probably explored in the future with improved experimental techniques.

The sun emits neutrinos
in different fusion processes in the MeV range. In the figure, neutrinos from
$p\,p$ interactions and the $^7$B spectrum is shown. 
Massive stars ($M\sim 10^{20}\cdot M_{\odot}$) are expected to emit neutrinos
at $\sim 1$~MeV due to Silicon burning~\cite{odrzywolek04,odrzywolek06}.

At slightly higher
energies lies the neutrino spectrum from SN 1987A. The expected flux from a
SN at a distance of $10$~kpc is indicated. Such a close supernova will be
observed by todays neutrino telescopes. The total diffuse flux from SNRs
(``relic''), however, is about four orders of magnitude lower and could not be
tested yet. 

At energies of $E_{\nu}>0.1$~GeV, the measured spectrum of atmospheric
neutrinos is indicated, squares are measurements from {\sc Fr{\'e}jus}~\cite{frejus,frejus_spectrum}
and dots are {\sc AMANDA}
measurements from the years
2000-2003~\cite{kirsten_icrc07,kirsten_phd}. At the highest energies, a generic spectrum from
GRBs is indicated~\cite{wb97,wb99}, as well as the maximum contribution from
AGN~\cite{mpr} and the expected neutrino flux from the absorption of protons
by the GZK effect~\cite{yoshida_teshima}. These sources have not been observed
yet due to the high atmospheric background and limited sensitivities of the
detectors. During the past ten years, high-energy neutrino detection still succeeded
in improving the instruments in a way that made it possible to cover five
decades of energy and to improve the sensitivity by 17 (seventeen!) orders of
magnitude. We have today reached the point where the most optimistic models
could already be used to constrain the physics of different sources classes and where the more realistic models
are being challenged by the upcoming generation of neutrino detectors. This
review discusses their options and possibilities.

In particular, the different source types being able to accelerate neutrinos
to the highest energies are examined in
more detail. The hypothesis of neutrino emission from these objects is
reviewed quantitatively in the context of multimessenger physics. Section~\ref{multimessenger}
focuses on what is known about the non-thermal Universe from photons and
charged Cosmic Rays. The connection from these two messengers to the third,
complementary particle, the neutrino, is drawn. In Section~\ref{nu_detection_methods}, the
focus lies on the methods used to detect neutrinos at the highest
energies. What neutrino signal to expect from extragalactic sources is
considered in the following sections. Possible neutrino emission from galactic
sources is presented in Section~\ref{galactic}. Active
Galactic Nuclei (AGN) are reviewed in
Section~\ref{nus_agn},
the neutrino flux models from Gamma Ray Bursts (GRBs) are discussed in
Section~\ref{nus_grbs} and starburst galaxies are presented in Section~\ref{starbursts}. Finally, production scenarios of neutrinos generated by the interaction of
high-energy protons with the CMB are presented in
Section~\ref{nus_gzk} before closing with a summary in Section~\ref{summary}.
\clearpage
%%%%%%%%%%%%%%%%%%%%%%%%%%%%%%%%%%%
\section{Introductory notes}
%%%%%%%%%%%%%%%%%%%%%%%%%%%%%%%%%%%
%==============================================
\subsection{On the representation of flux models in figures}\vspace{-0.5cm}
%==============================================
Figure~\ref{diffuse_all} shows the neutrino flux models in a double-logarithmic
representation, implying that the dependence of the differential flux $dN/d\en$
with the energy is shown as $\log\left[dN_{\nu}/d\en\right](\log\en)$. In general,
non-thermal particle spectra can usually be approximated by powerlaws, $dN/dE=A\cdot
E^{-\alpha}$. In a double-logarithmic representation, this leads to a
straight line,
\be
\log\left[dN/dE\right]\left(\log E \right)=\log\left[A_{\nu}\cdot
  {E}^{-\alpha} \right]=\log\left[ A\right]-\alpha\cdot \log\left[E\right]\,.\nonumber
\ee
The slope of the line is given by the spectral index $\alpha$ and the y-axis
intercept represents the normalization $A$. Such powerlaw spectra are observed in
the case of photons and charged Cosmic Rays, and are expected in the case of
neutrinos. In the following, all spectra will be shown in the double-logarithmic
representation. Since many spectra are very steep ($dN/dE\sim E^{-2}-E^{-4}$)
for all three messengers, it is useful to weight the y-axis with a power of
the energy, $E^{w}$,
straight line,
\begin{eqnarray}
\log\left[E^{w}\cdot dN/dE\right]\left(\log E \right)&=&\log\left[A_{\nu}\cdot
  {E}^{-\alpha+w} \right]\\
&=&\log\left[ A\right]-(\alpha-w)\cdot \log\left[E\right]\,.\nonumber
\end{eqnarray}
For $w=2$, which is often used in the case of neutrino spectra, an
$E^{-2}$-spectrum is then represented by a flat line, which is convenient,
since neutrino spectra at $\en>100$~GeV are often expected to be close to
${\en}^{-1.5}-{\en}^{-2.5}$. The double-logarithmic representation with a
weight is chosen to simplify to read off the spectral index of the models and
measured fluxes.
%==============================================
\subsection{Flux and limit notations}\vspace{-0.5cm}
%==============================================
Throughout this review, different fluxes and flux limits
  will appear. 
Usually, the flux at Earth is given per energy interval,
$dN/dE$, in units of $\left[dN/dE\right]=\diffflux$. Point source fluxes have
the same notation, but the unit $\left[dN/dE\right]=\pointflux$. 
The convention for the different particle and electromagnetic fluxes is as follows:
\begin{itemize}
\item {\it Charged Cosmic Rays}\\
For the total spectrum of charged particles, the spectrum is written as
\be
\frac{dN_{CR}}{dE_{CR}}=A_{CR}\cdot E_{CR}^{-\alpha_{CR}}\,.
\ee
If only protons or electrons are considered, the following notation is used:
\be
\frac{dN_{p}}{d\ep}=A_{p}\cdot {\ep}^{-\alpha_{p}}\,.
\ee
\item {\it Photons}\\
Photon spectra are usually given as:
\be
\frac{dN_{\gamma}}{d\eg}=A_{\gamma}\cdot {\eg}^{\alpha_{\gamma}}\,.
\ee
The total power $P_{tot}(\omega)$, depending on the angular frequency
$\omega$, is written as
\be
P_{tot}\propto \omega^{s}
\ee
with $s=-\alpha_{\gamma}-1$.

For the special case of TeV photons, the notation
\be
\frac{dN_{TeV}}{dE_{TeV}}=F_{TeV}\cdot {\left(\frac{E_{TeV}}{1 \mbox{TeV}}\right)}^{\alpha_{TeV}}
\ee
is used. The spectrum is normalized at 1 TeV.
\item {\it Neutrinos}\\
The neutrino spectrum is usually multiplied by the energy squared, ${\en}^{2}$,
\be
{\en}^{2}\,\frac{dN_{\nu}}{d\en}=A_{\nu}\cdot {\en}^{-\alpha_{\nu}}\,.
\ee
Alternatively, the spectrum itself is given as
\be
\frac{dN_{\nu}}{d\en}=A_{\nu}'\cdot {\en}^{-\alpha_{\nu}'}
\ee
with $\alpha_{\nu}=\alpha_{\nu}'-2$.
\end{itemize}
Neutrino flux limits are usually given in the form of $\Phi={\en}^2\cdot dN_{\nu}/dE_{\nu}$
and are denoted as follows:
\begin{itemize}
\item $\dl$: Diffuse Limit (DL) given in units of $\diffunits$. 
\item $\stl$: Stacking Limit (SL) in units of $\pointunits$, obtained for the point source flux from a
  certain class of AGN. The stacking method is explained in Section~\ref{nu_detection_methods}.
\item $\sdl$: Stacking Diffuse Limit (SDL), derived from the stacking limit in
  the same units as the diffuse limit, $\diffunits$. It is determined by
  taking into account the contribution from weaker sources as well as yet
  unidentified sources, present in a diffuse background.
\end{itemize}
A similar convention is used to denote the corresponding sensitivities:
\begin{itemize}
\item $\ds$: Diffuse Sensitivity (DS) in units of $\diffunits$. 
\item $\ps$: the sensitivity to a single point source in units of $\pointunits$.
\end{itemize}
\clearpage
\section{The multimessenger connection \label{multimessenger}}
Neutrino flux predictions are built on the direct connection between the observed
Cosmic Ray spectrum and the non-thermal emission from astronomical
sources. Here, it is reviewed how the different parts of the Cosmic Ray
spectrum can be connected to the various objects and source classes and how this in
turn leads to the production of neutrinos.
%==============================================
\subsection{Cosmic Rays}
%==============================================
Charged Cosmic Rays (CRs) have been observed from energies of $E_{CR}\sim 10^9$~eV up to $E_{CR}\sim
10^{20}$~eV. The spectrum is altered by the Solar wind for energies
below about $E_{CR}\sim 20\cdot Z$~GeV, where $Z$ is the charge of the nucleus. At
higher energies, the energy spectrum follows a powerlaw with two, possibly three
breaks. The powerlaw structure can in general be explained by 
shock acceleration in astrophysical sources. In this section, the observed
spectrum with its features is discussed as well as the acceleration mechanism
built on stochastic acceleration of test particles at magnetic field
inhomogeneities in astrophysical shocks. 
%----------------------------------------------
\subsubsection{Observation of charged Cosmic Rays}
%----------------------------------------------
Already in the early 20th century, it was discovered that the Earth is exposed to a
continuous flux of charged particles from outer space, see e.g.~\cite{v_hess12,kohlhoerster1913}.
Viktor Hess and others performed balloon flights proving that
the ionization of the atmosphere increases with height. This contradicted the
hypothesis that the flux of ionizing particles arises from radioactive matter in the Earth's rocks
exclusively. 

\begin{figure}[h!]
\centering{
\epsfig{file=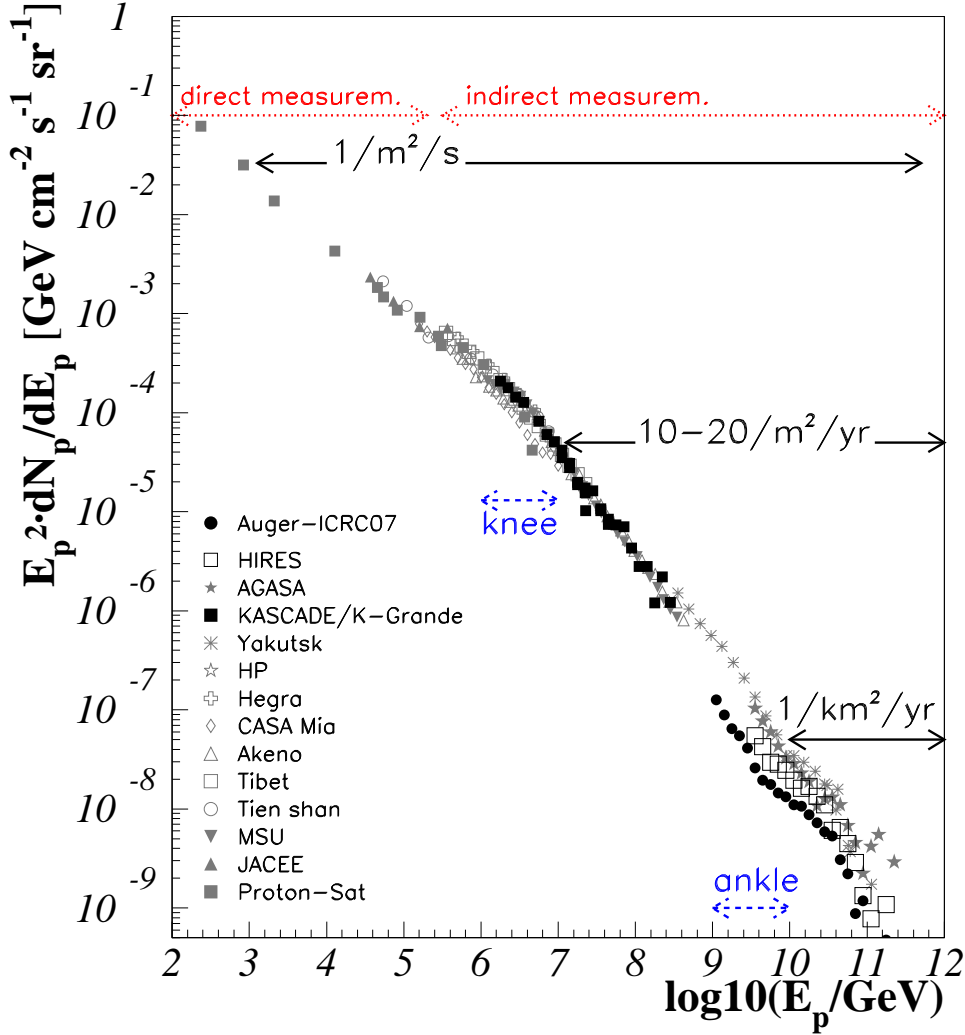,width=\linewidth}
\caption[All particle Cosmic Ray spectrum]{All particle Cosmic Ray spectrum. Data points come
  from the experiments as listed in the bottom left corner:
  {\sc Auger}~\cite{yamamoto_icrc2007}, {\sc HiRes}~\cite{hires_spect2002},
  {\sc AGASA}~\cite{agasa_spect1995}, {\sc
  KASCADE}~\cite{kascade_spect2005,kascade_grande}. {\sc
  Yakutsk}~\cite{yakutsk_spect1985}, {\sc Haverah Park}
  \cite{haverah_park_spect2001}, {\sc HEGRA} \cite{hegra_spect1999}, {\sc CASA-MIA}
\cite{casa_mia_spect1999}, {\sc Akeno}~\cite{yakutsk_spect1985}, {\sc Tibet}
  \cite{tibet_spect2003}, {\sc Tien Shan} 
\cite{tien_shan_spect1995}, {\sc MSU} \cite{msu_spect1994}, {\sc JACEE}
  \cite{jacee_spect1995}, {\sc Proton-Sat}
\cite{proton_sat_spect1975}.}
\label{cr_spect}
}
\end{figure}

\begin{figure}
\centering{
\epsfig{file=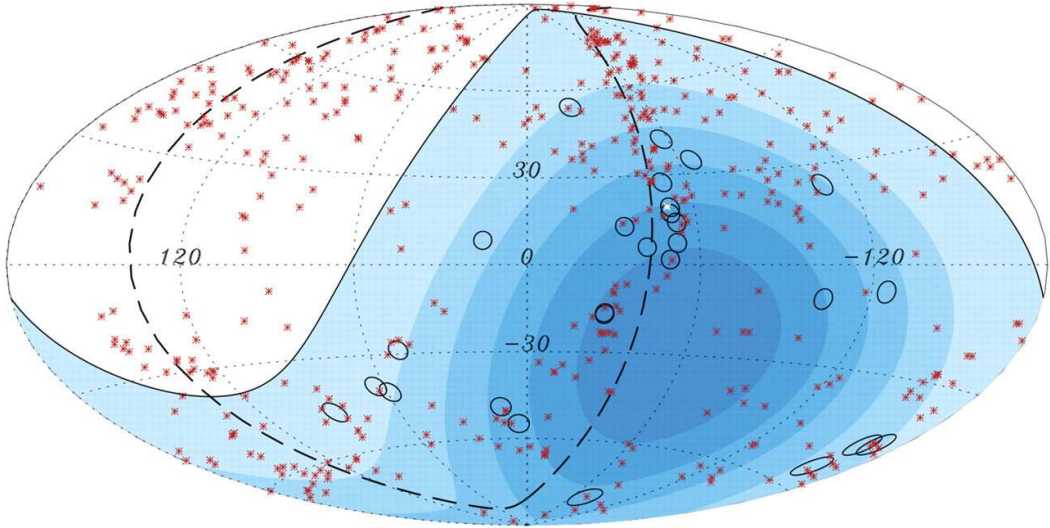,width=\linewidth}
\caption{Skyplot of 27 {\sc Auger} events at the highest energies ($E>5.7\cdot
  10^{19}$~eV,
  circles with radius $\psi=3.1\deg$)
  and the sources in the V-C catalog up to a redshift
  $z_{\max}=0.018$ (red stars). Within this distance, 318 of 472 AGN in the
  V-C catalog are in the field of view of {\sc Auger}. The blue outlines show the
  sensitivity of {\sc Auger}. Darker color indicates larger relative exposure.
  The black, dashed line represents the supergalactic plane. The closest AGN,
  Cen~A, is marked in white. From~\cite{auger_science2007}. Reprinted with permission from AAAS.\label{auger_skyplot}}
}
\end{figure}

Today, the spectrum  of charged CRs, $dN_{CR}/dE_{CR}$, has been examined over a wide range of
energies $E_{CR}$, using balloon experiments and satellites for low energies
($E_{CR}<10^{14}$~eV) and Earth-bound experiments for high energies
($E_{CR}>10^{14}$~eV). The all particle energy spectrum of CRs is shown in
Fig.~\ref{cr_spect}. The spectrum is weighted by $E_{CR}^{2}$ in order to have a
flatter representation of the very steep spectrum. The powerlaw behavior of
the spectrum is clearly visible, 
\begin{equation}
\frac{dN_{CR}}{dE_{CR}}\propto E_{CR}^{-\alpha_{CR}}\,.
\end{equation}
Two kinks can also be seen, referred to as 
{\em knee} and {\em ankle}.
The spectral indices for the different parts of the spectrum are~\cite{wiebel_crs,wiebel_data}
\begin{equation}
\alpha_{CR} \approx \left\{
\begin{array}{ll}
  2.67& \mbox{ for } \quad \qquad \log(E_{CR}/\mbox{eV})<15.4\\
  3.10& \mbox{ for }15.4<\log(E_{CR}/\mbox{eV})<18.5\\
 2.75 & \mbox{ for }18.5<\log(E_{CR}/\mbox{eV}) \,. 
\end{array}
\right.
\end{equation}
A second knee around $E_{CR} \sim 4\cdot 10^{17}$~eV is discussed today, with an
even steeper behavior up to the ankle, see e.g.~\cite{hoerandel03,hoerandel07} and
references therein.
The general spectral powerlaw-like behavior can be explained by stochastic
particle acceleration in collision-less plasmas. 
%----------------------------------------------
\paragraph{Cosmic Rays \& directional information}
%----------------------------------------------
Charged Cosmic Rays below $E_{CR}<Z\cdot 10^{17}$~eV do not point back to their origin, since they are
scrambled by interstellar magnetic fields. The strong connection between
non-thermal emission from astrophysical sources and particle acceleration can,
however, be used to establish a model for different sources and source types to
explain the Cosmic Ray spectrum. Sources within our Galaxy can produce the
Cosmic Ray spectrum up to the ankle. Events at higher energies have to be
extragalactic:
\begin{itemize}
\item[(a)] no galactic source class is energetic enough for the
production of particles at such high energies as discussed later in Section~\ref{acc_mech} and
\item[(b)] the particles' gyro-radius becomes too large and they escape from
  the galaxy already at lower energies. 
\item[(c)] In addition, at energies as high as $E_{CR}\sim 10^{20}$~eV, the particle diffusion is
low compared to the traveling length through the Galaxy~\cite{hillas}. The
observed particles point in this case back to their original source. The observed events
are isotropically distributed, which is only possible for traveling lengths
longer than the diameter of the galaxy.
\end{itemize}
This leaves extragalactic or exotic sources as the origin
of the highest energy events, for which
acceleration up to $E_{CR}\sim 10^{21}$~eV is possible. At the highest
observed energies, $E>5\cdot 10^{19}$~eV, nearby galaxies with distances smaller than
$\sim 50$~Mpc can still carry directional information. At these distances,
most galaxies are located in the supergalactic plane. Most recently, the {\sc
  Auger} experiment observed a first evidence for the correlation between the
V{\'e}ron-Cetty (V-C) catalog of Active Galactic Nuclei and Cosmic Rays of the highest
energies~\cite{auger_science2007}. In the analysis, three parameters have been optimized in order to
look for a correlation between the sources in the catalog and the events at
the highest energies: The maximum angular separation $\psi$ between Cosmic Ray
event and source, the maximum redshift $z_{\max}$ for which AGN in the catalog
are still considered and the threshold energy $E_{th}$, giving the lowest
energy at which Cosmic Ray events are still considered. These three parameters
ensure that
\begin{enumerate}
\item[(a)] magnetic deflection is considered to a certain amount ($\psi$
is larger than the point spread function of the detector), 
\item[(b)] only the local Universe is considered (the choice of $z_{\max}$
  leaves out distant galaxies), 
\item[(c)] only events with the highest energies are considered. Events of
  energies below $E_{th}$ are discarded, since lower-energy events are more
  sensitive to magnetic field deflections. 
\end{enumerate}
The final parameter set is
$(\psi,z_{\max},E_{th})=(3.1\deg,0.018,57$~EeV$)$,
and the sky plot is shown in Fig.~\ref{auger_skyplot}. The maximum redshift
of $z_{\max}=0.018$ corresponds to a distance of $75$~Mpc. The correlation is
a first evidence that sources following the structure of the supergalacitc
plane are the sources of the highest energy Cosmic Rays. The analysis does not
allow for the direct identification of the sources for different
reasons. Firstly, the catalog of AGN used in the analysis cannot considered to
be complete and is partly inhomogeneous, including different types of
AGN. Secondly, sources other than AGN, which also follow the local structure
of the Universe, may be responsible for the emission of UHECRs rather than AGN
themselves. Nonetheless, this result is a first step towards the
identification of the sources of extragalactic Cosmic Rays. 
%-------------------------------------------------------
\paragraph{The Greisen Zatsepin Kuzmin cutoff}
%-------------------------------------------------------
Protons effectively lose energy by interactions with the CMB on their way to Earth:
\be
p\,\gamma_{CMB}\rightarrow\left\{\begin{array}{l} \Delta^{+}\\p\,e^{+}\,e^{-}\,,\end{array}\right.\,.
\ee
The energy loss length for pair production at $\sim 10^{19}$~eV is about
$\sim500$~Mpc, while it is only $\sim 50$~Mpc for the production of a
$\Delta-$resonance. The latter process is therefore responsible for a rapid
decrease of the particle spectrum above $E_{CR}>5\cdot 10^{19}$~eV, if the
sources of Cosmic Rays are located at distances larger than $\sim 50$~Mpc~\cite{greisen,kuzmin_zatsepin68}. This effect is
named ''Greisen Zatsepin Kuzmin (GZK) cutoff''.

At the highest energies of $E_{CR}>5\cdot 10^{19}$~eV, a discrepancy between the
CR fluxes as observed by two
experiments, {\sc AGASA}\footnote{{\bf A}keno {\bf G}iant {\bf A}ir {\bf
    S}hower {\bf A}rray} and {\sc HiRes}\footnote{{\bf Hi}gh {\bf
    Res}olution Fly's Eye Detector} was found. {\sc AGASA} data
are represented by stars and {\sc HiRes} is shown as open boxes in Fig.~\ref{cr_spect}. While {\sc AGASA}
detected several events above $5\cdot 10^{19}$~eV, {\sc HiRes} observed a
decay of the spectrum. 
Given the prediction of the GZK cutoff, the result from {\sc HiRes} is expected,
while the {\sc AGASA} events need to be explained by exotic phenomena. A surface array for the measurement of the charged component of the
showers was used by {\sc AGASA}~\cite{agasa95} while {\sc HiRes}~\cite{hires} consisted of telescopes
measuring the emission of the showers' fluorescence light. At these
energies, there are quite large uncertainties in the calibration of the
spectra due to the low statistics and large systematic errors. Taking this
into account, it is possible to interpret the two results in a way which would
still fit a single theory, see~\cite{agasa_hires_auger}. The {\sc Auger} experiment is
being built to resolve the issue of the highest energy
events, see e.g.~\cite{dawson_tev}. With {\sc Auger}, the
two different techniques as used by {\sc AGASA} and {\sc HiRes} are combined in a hybrid
array. This allows for the investigation of systematic uncertainties. The
size of the array ensures sufficient statistics. {\sc Auger} has now
instrumented more than 85\% of a $3000$~km$^2$ surface array and all four
telescopes for fluorescence measurements are operating since March 2007. The hybrid array is
expected to be completed by the end of 2007~\cite{dawson_tev}. First results
from the {\sc Auger}
array show that a component beyond the GZK cutoff can be excluded: a powerlaw
behavior above $E=10^{19.6}$~eV can be excluded with a significance of
$6\,\sigma$~\cite{yamamoto_icrc2007}.
%----------------------------------------------
\subsubsection{Production of ultra high-energy Cosmic Rays \label{acc_mech}}
%----------------------------------------------
There are two basic scenarios for the production of charged Cosmic Rays at the
highest energies ($E_{CR}>3\cdot 10^{18}$~eV), referred to as the {\it
  top-down} and {\it bottom-up} scenarios. In the top-down scenario, UHECRs
come from the decays of superheavy particles with masses ranging from $10^{11}$~GeV
 up to the GUT scale, $M_{GUT}\sim 10^{24}$~eV. Such decays result in protons at the highest energies up to
$E_p\sim 10^{22}$~eV, and in this scenario, the GZK cutoff can be avoided,
since the protons are produced in the Earth's vicinity. Furthermore, the
proton signal is expected to be accompanied by a large flux of neutrinos and
high-energy photons, see e.g.~\cite{biermann_sigl}. The absence of such
signatures in combination with the confirmation of the GZK cutoff favors the production of protons in particle acceleration processes in
distant sources, referred to as the bottom-up scenario. 
\paragraph{Bottom-up}
Astrophysical environments are often characterized by the collision of
different plasmas. As an example, a remnants of supernova explosions are observed for
typically more than 1000 years, as a consequence of the supernova shell being
accelerated into the interstellar medium. A second example is the collision of two
galaxies due to gravitational interaction. A shock front is
produced, when a gas encounters other gas or a wall, with a velocity faster
than any signal velocity. Such phenomena are not only observed in
astrophysical environments, but also in other media, for instance in the atmosphere, such as supersonic
movement of planes or bullets in air. The plane or bullet moves faster than
the characteristic speed of the medium, the speed of sound, and produces a
shock wave, the Mach cone~\cite{mach_wentzel1884,mach_wentzel1885,mach1898}. In astrophysical shocks, the
characteristic speed of the plasma is the speed of magnetic waves.

The concept of stochastic particle acceleration has for the first time
been presented by Fermi~\cite{fermi49,fermi54}. In astrophysical shocks,
inhomogeneous magnetic fields are responsible for the
acceleration of the charged particles. A single particle
crosses the shock-front back and forth, gaining a
constant fraction of energy per encounter with a magnetic-field
inhomogeneity. The particle leaves the shock-front region when carrying
sufficient energy to escape. Stochastically, the acceleration of a whole
population of particle results in a powerlaw-behavior as observed for the
spectrum of charged Cosmic Rays. The theory of acceleration was refined in the
1970s by Bell~\cite{bell78a,bell78b}, Krymskii~\cite{krymskii77},
Blandford~\&~Ostriker~\cite{blandford_ostriker78} and
Axford, Leer~\&~Skadron~\cite{axford78}. While Bell worked out a microscopic approach, in which
individual particles are traced, Krymskii, Blandford~\&~Ostriker and Axford
worked on the macroscopic description of astrophysical shocks, neglecting any
individual movement of particles.
One important consequence of acceleration theory is that
the maximum energy depends on the magnetic field and the
size of the acceleration region as derived by Hillas~\cite{hillas}:
\be
E_{\max}^{18}=\beta_s\cdot Z\cdot B_{\mu G}\cdot R_{kpc}\,.
\label{emax:equ}
\ee
Here, $E_{\max}^{18}:=E_{\max}/(10^{18}$~eV) is the maximum energy which can
be achieved, $\beta_s=V_s/c$ is the shock velocity in terms of the speed of
light and $Z$ is the charge of the accelerated particle in
units of the charge of the electron, $e$. Furthermore, $B_{\mu G}:=B/(1\,\mu$G)
is the magnetic field of the acceleration region in units of $1\,\mu$G and $R_{kpc}:=R/(1$~kpc) is
the size of the acceleration region in units of $1$~kpc.
The spectral index of
the particle spectrum
depends on the conditions of the astrophysical shock, i.e.~the magnetic field
strength, the orientation of the shock-front towards the magnetic field, the
shock-front's velocity and the extension of the shock itself. 
%==============================================================
\subsubsection{Primary spectra and radiation fields\label{radiation:general_crs}}
%==============================================================
It is essential to study the correlation between accelerated primaries and secondary
radiation effects in order to connect photon
observations to Cosmic Rays and neutrinos. In this section, the focus lies on the synchrotron
emission, since its observation reveals the spectral behavior of the primaries
as discussed in the following.
%....................................................
\paragraph{Total synchrotron power}
%....................................................
The total radiated
  power $P_{tot}$ is proportional to $ m^{-4}$ with $m$ as the mass of the particle. Since the electron-proton mass ratio is $m_e/m_p\sim 5\cdot 10^{-4}$,
  the radiated energy from electrons is a factor $\sim 10^{13}$ higher than
  for protons. Protons only lose energy to synchrotron radiation for extremely
  high energies and large magnetic fields, since the power increases with the
  squared product of the external magnetic field $B$ and particle's energy $\ep$, $P_{tot}\propto
  (B\cdot \ep)^2$. Electrons undergo synchrotron
  losses at moderate energies already. 
%....................................................
\paragraph{Synchrotron radiation and the spectral shape}
%....................................................
For non-thermal spectra, the spectral index of the primary shock-accelerated particles can
  be expressed in terms of the synchrotron spectral index of a source. Electrons
  and protons follow the same distribution, i.e.~the spectral index of the
  electrons is the same as for the protons, $\alpha_p$. 
The power $P(\omega)$ per unit frequency $d\omega$ for a particle accelerated
  by an external magnetic field $B$ can be written in terms of a 
  function $F(x)$, which only depends on the dimensionless variable
  $x:=\omega/\omega_c$ (see~\cite{rybicki} for details):
\begin{equation}
P(\omega)\propto F(\frac{\omega}{\omega_c})\,.
\end{equation}
The critical frequency $\omega_c$ of the synchrotron spectrum is given as
\begin{equation}
\omega_c=\frac{3q\,B\,\sin\alpha}{2\,m\,c}\cdot \gamma^2\,,
\label{omegac}
\end{equation}
representing a measure for the maximum frequency of acceleration for the particle spectrum.
In this expression, $q$ is the charge, $m$ the mass and $\gamma$ is the
boost factor of the accelerated
particles. The latter can be expressed in terms of the energy, 
\be
\ep=\gamma\,m\,c^2\,.
\label{epgamma}
\ee
Since shock-accelerated primaries follow a powerlaw distribution,
\begin{equation}
\frac{dN_p}{d\ep}\,dE_p \propto {\ep}^{-\alpha_p} \,d\ep\,,
\end{equation}
the total radiated power $P_{tot}(\omega)$ can be expressed as
\begin{equation}
P_{tot}(\omega)\propto \int_{{\ep}^1}^{{\ep}^2}
P(\omega)\,\ep^{-\alpha_p}\,d\ep\propto
\int_{{\ep}^1}^{{\ep}^2}F(\frac{\omega}{\omega_c})\,\ep^{-\alpha_p}\,d\ep \,.
\end{equation}
With $\ep\propto\gamma\propto \omega_{c}^{1/2}$ (see Equ.~(\ref{omegac}) and Equ.~(\ref{epgamma})), $\gamma$
can be substituted for
$x:=\omega/\omega_c(\gamma)$ and the total power can be written as 
\begin{equation}
P_{tot}\propto \omega^{-(\alpha_p-1)/2}\cdot \int_{x_1}^{x_2} F(x)\cdot x^{(\alpha_p-3)/2}\,dx\,.
\end{equation}
Since the integral does not depend on $\omega$, the frequency dependency is
given as
\begin{equation}
P_{tot}\propto \omega^{-(\alpha_p-1)/2}\,.
\end{equation}
The total synchrotron spectrum therefore follows a powerlaw with a spectral
index $s$
\begin{equation}
P_{tot}\propto \omega^{-s}\,.
\end{equation}
This leads to a linear correlation between synchrotron and particle spectral
index,
\begin{equation}
s=\frac{\alpha_p-1}{2}\,.
\label{indices}
\end{equation}
The differential spectral index which will be used in the
following is $\alpha_{\gamma}=-s-1$ with $dN_{\gamma}/d\eg\propto {\eg}^{\alpha_{\gamma}}$.
The flatness of the synchrotron spectra is limited by the theory of
synchrotron radiation. Since a single
electron produces a spectrum with $\alpha_{\gamma}=-2/3$, the total synchrotron
spectrum of an electron population cannot be flatter than~$\alpha_{\gamma}\leq -2/3$. This phenomenon is referred to as the {\it line of
  death}\footnote{In the integral representation, $S\propto \nu^{-s}$, the
  spectrum behaves as $S\propto \nu^{1/3}$. While in GRB physics, it is more
  common to use the differential representation, the integral form is more
  common to use in
  the case of AGN spectra.}. 

The connection between synchrotron and particle spectral index becomes important, for instance, in the case of Gamma Ray Burst (GRB)
spectra, which are typically explained by synchrotron emission of
electrons. In some cases, the burst spectra are flatter than the maximum
values, which indicates that other phenomena like absorption due to high
optical depth need to play a role in the
radiative processes in GRBs as well. 
\paragraph{Electron cooling}
The calculations above assume that the dynamical timescale of the system is
much shorter than the cooling time of electrons due to radiation
losses. This
is called the {\em slow cooling regime}, see~\cite{kardashev62}. In the {\em fast
cooling regime} of long dynamical timescales compared to radiative cooling, the photon spectrum is flatter by $1/2$, 
\begin{equation}
s=\frac{p}{2}\,.
\end{equation}
For the prompt emission in GRBs, for example, the dynamical time scale is
short compared to the cooling time and slow cooling has to be considered. 
%....................................................
\paragraph{Further radiation effects}
%....................................................
Note that the synchrotron spectrum is in many cases altered by further
radiation effects. The synchrotron field can interact with the electron
population, leading to the {\it Inverse Compton} (IC) effect, which boost photons
to higher energies. This scenario is called {\it Synchrotron Self
  Compton} (SSC). Processes
like optical depth effects, extinction by dust, pair production or
bremsstrahlung can additionally alter the observed spectrum. When external
photon fields are present, Inverse Compton scattering does not need to be
induced by the synchrotron field. Such a process is referred to {\it External
  Inverse Compton} (EC) scattering. At the highest
energies ($\eg\sim$~TeV),
the decay of $\pi^0$ particles resulting from proton-proton and proton-photon
interactions can also dominate the spectrum. The last process competes with
the SSC model. TeV emission in such hadronic models are referred to as {\it
  Proton-Induced Cascades} (PIC), see~\cite{rachen00} for a discussion of the
emission features. The question whether high-energy photon
signals originate from hadronic ($\pi^0$-decays) or leptonic (SSC/EC) processes
is one of the most striking these days. In some cases, TeV photon emission can
also be explained by proton synchrotron radiation.\\
%==============================================
\subsection{Sources of high-energy photons}
%==============================================
From the radiation processes described above, it is clear that the emission of
high-energy photons is usually connected to the acceleration of electrons or
protons in astrophysical sources. In this section, the most energetic sources in the sky are discussed
with respect to their observation in photons, and the possible contribution to
the spectrum of charged Cosmic Rays which is displayed in Fig.~\ref{cr_spect}. The observed photon spectra are
essential for the prediction of neutrino fluxes, since photons are the only
messengers giving direct evidence on the properties of the sources. Galactic
sources are supernova remnants (SNRs) as likely sources for the
production of Cosmic Rays up to the knee as well as X-Ray Binaries (XRBs), in particular
microquasars, and pulsars which are candidates for the production of Cosmic
Rays above $100$~TeV~\cite{gaisser}.
The most energetic, extragalactic sources are Active Galactic Nuclei (AGN) as permanent
sources in the sky as well as Gamma Ray Bursts (GRBs) as transient
eruptions. Considering the abundance of the different source types and their
individual electromagnetic output, leads to the expectation that galactic
sources can produce the Cosmic Ray spectrum up to the ankle while
extragalactic sources are responsible for the CR flux above the ankle. The
power of electromagnetic output mirrors the power in Cosmic Rays, since
electromagnetic radiation originates from the charged particles in the
source.  Table~\ref{sources_cr_spect} lists source classes with their
intrinsic luminosity and possible
contribution to the Cosmic Ray spectrum.
\begin{table}[h!]
\centering{
\begin{tabular}{lll|ll}\hline
Source class& typical em.&life&energy range&Ref\\ 
            &                  output&time&&            \\\hline \hline
{\sc Galactic}&{\sc sources}&\\ \hline
SNR& $10^{42}$~erg/s&1000 yr&$10^{10}$~eV$<\ep<10^{15}$~eV&\cite{ginzburg1964}\\
SNR-wind&$10^{44}$~erg/s&1000 yr&$10^{10}$~eV$<\ep<10^{18}$~eV&\cite{voelk_biermann1988}\\
X-ray binaries&$10^{38}~$erg/s&$10^{5}-10^{6}$~yr&$10^{14}$~eV$<\ep<10^{18}$~eV&\cite{gaisser}\\ 
% z.B.:
% @ARTICLE{2005ApJ...621L.109I,
%   author = {{Ivanova}, N. and {Rasio}, F.~A. and {Lombardi}, Jr., J.~C. and 
%	{Dooley}, K.~L. and {Proulx}, Z.~F.},
%    title = "{Formation of Ultracompact X-Ray Binaries in Dense Star Clusters}",
%  journal = {\apjl},
%   eprint = {arXiv:astro-ph/0501617},
%     year = 2005,
%    month = mar,
%   volume = 621,
%    pages = {L109-L112},
%      doi = {10.1086/429220},
%   adsurl = {http://adsabs.harvard.edu/abs/2005ApJ...621L.109I},
%  adsnote = {Provided by the Smithsonian/NASA Astrophysics Data System}
%}
Pulsars&$10^{37}$~erg/s&$10^{6}$~yr&$10^{14}$~eV$<\ep<10^{18}$~eV&\cite{gaisser}\\ \hline\hline
% mean pulsar lifetime = 2.5*1e6yr, see Arshakyan, Astrophysics 37, vol 2 (1994)
{\sc Extragalactic}&{\sc sources}&&\\\hline
Galaxy clusters&$\sim 10^{44}$~erg/s&$10^{7}$ yr&$3\cdot 10^{18}$~eV$<\ep< 10^{21}$~eV&\cite{kang1997}\\ 
AGN&$10^{44}-10^{47}$~erg/s&$10^{7}$~yr&$3\cdot 10^{18}$~eV$<\ep< 10^{21}$~eV&\cite{biermann_strittmatter87}\\ 
% AGN lifetime from: http://arxiv.org/pdf/0709.2167
GRBs&$10^{49}-10^{51}$~erg/s&$>1-100$~s&$3\cdot 10^{18}$~eV$<\ep< 10^{21}$~eV&\cite{vietri95,wb97}\\\hline
\end{tabular}
\caption[Possible source classes of the Cosmic Ray spectrum]{Source
  classes possibly contributing to the Cosmic Ray spectrum. The values for the
typical electromagnetic emission and the life time are estimates based on the
objects' observations. It should be noted that the luminosity and lifetime
distributions can scatter, and there are objects within the given classes
which have values deviating from the given typical ones In the case of GRBs, a lower limit is given, the actual
value depends on whether or not GRB afterglow emission contributes to
UHECRs. The energy ranges are estimates from the given references.}
\label{sources_cr_spect}
}
\end{table}
%----------------------------------------------
\subsubsection{Active Galactic Nuclei}
%----------------------------------------------
A class of galaxies with a particularly bright core was detected for
the first time in 1962. An object, appearing star-like in the sky, showed
extreme radio-emission features and could therefore not be classified as a
star. The interpretation that this object, today known as 3C 273, was indeed a distant galaxy with a
very bright core, was suggested for the first time one year after the detection
by Maarten Schmidt~\cite{schmidt63}. This class of objects was referred to as
Quasi Stellar Objects (QSOs). 

\begin{figure}[h!]
\begin{center}
\epsfig{file=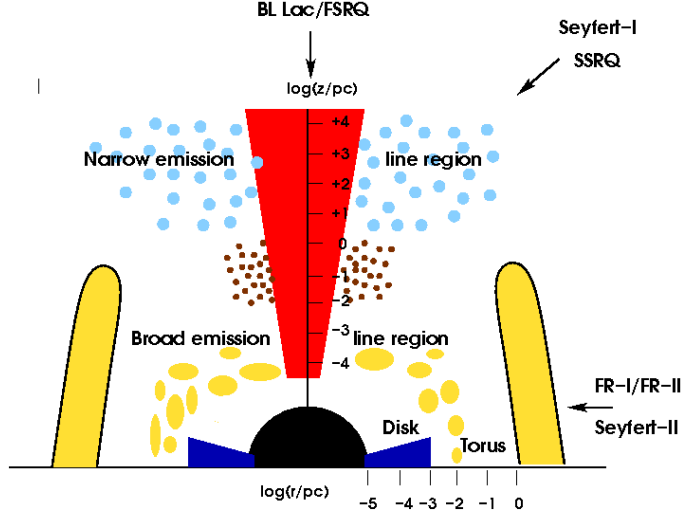,width =9cm}
\end{center}
%\vspace{-1cm}
\caption[AGN scheme.]{Scheme of a cylindrically symmetric AGN shown in the
  r-z-plane, both axes logarithmically scaled to 1~pc. It is indicated which
  objects are believed to be seen from which direction.
% For the definition of
%  the detailed AGN class scheme indicated in the figure, see~\cite{zier2}.
Figure after~\cite{zier2}.
}
\label{agnscheme}
\end{figure}

\begin{figure}
\centering{
\epsfig{file=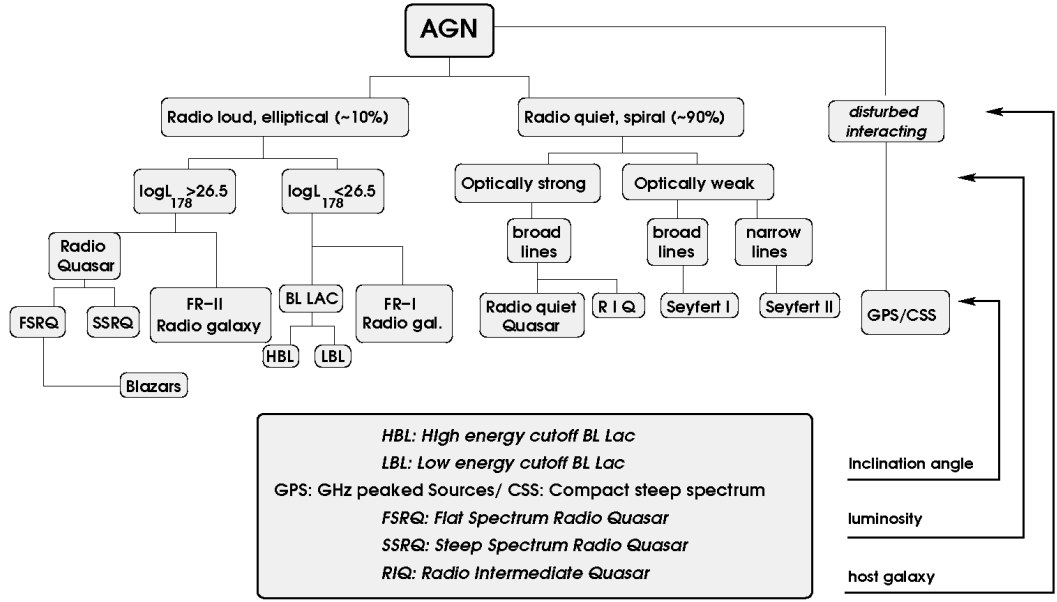,width=\linewidth}
\caption[AGN classification scheme]{AGN classification scheme, figure after~\cite{andreas}.}
\label{agn_scheme}
}
\end{figure}

Today, it is known that QSOs fit into the general classification scheme of
Active Galactic Nuclei (AGN), objects which are believed to be powered by a
rotating supermassive black hole in the center of the galaxy. A schematic view
of the general picture of AGN is shown in Fig.~\ref{agnscheme}. The core is
``active'' due to the accretion disk which forms around the central black hole
and radiates strongly at optical frequencies. The disk is fed by matter from a
dust torus. Perpendicular to the accretion disk, two relativistic jets are
emitted, transporting matter in form of lobes. Knots and hot spots along the jets emit radio
emission, leading to the strong observed radio signal of AGN. It is expected
that these knots and hot spots represent shock environments in which particles
are accelerated to high energies, in the case of hadrons up to proton energies
of $\ep\sim
10^{21}$~eV, see~\cite{biermann_strittmatter87}. In this
section, a general classification scheme for AGN is presented as well as
spectral and temporal properties of the sources.
%----------------------------------------------
\paragraph{AGN unification scheme}
%----------------------------------------------
Three main criteria can be used for the unification scheme of Active
Galactic Nuclei which is indicated schematically in
Fig.~\ref{agn_scheme}:
\begin{enumerate}
\item The activity of the source at radio wavelengths yields a
  division into radio loud and radio weak objects. About 90\% of all
  AGN are radio weak and are usually hosted in spiral galaxies, while
  radio loud nuclei are located in the centers of elliptic galaxies.
\item The luminosity of the object is a further classification
  criterion. Radio weak sources are subdivided into optically strong
  and optically weak sources, which can be distinguished by
  considering the features of the emission lines. Optically strong
  sources usually lack narrow emission lines which are present in the
  optically weak case. Both source types appear to have broad emission
  lines. Radio loud sources with extended jets ($\sim 100$~kpc) are
  subdivided at radio wavelengths into low luminosity and high
  luminosity objects at a critical luminosity of $L_{\nu}=2.5\cdot
  10^{26}$~W/Hz. The jets of compact objects such as GHz-Peaked
  Sources (GPS) and Compact Steep Sources (CSS) are believed to get
  stuck in matter.
\item The third classification criterion is the orientation of the AGN
  towards the observer. AGN are axisymmetric along the jet axis. In
  the branch of radio loud AGN, an object is classified as a blazar if
  one of the jets is pointed directly towards the observer. Flat
  Spectrum Radio Quasars (FSRQ) are the high luminosity population of
  the blazars while BL Lacs form the corresponding low luminosity
  population. So called Faranoff Riley (FR) galaxies are being looked
  at from the side, so that jets and torus are usually clearly
  visible. The high luminosity FR-II galaxies show a very strong radio
  emission at the outermost end of the jets, while the radio emission
  of the low luminosity class FR-I happens in knots throughout the
  jet.\\
For radio weak AGN, the objects are called radio weak quasars in the
optically strong case and Seyfert-I galaxies in the optically weak
case when looked at the gap between jet and AGN torus. The radio weak
equivalent to FR galaxies are Radio Intermediate Quasars (RIQ) and
Seyfert-II galaxies, where the observer's view is directed towards the
torus.
\end{enumerate}

\begin{figure}
\centering{
\epsfig{file=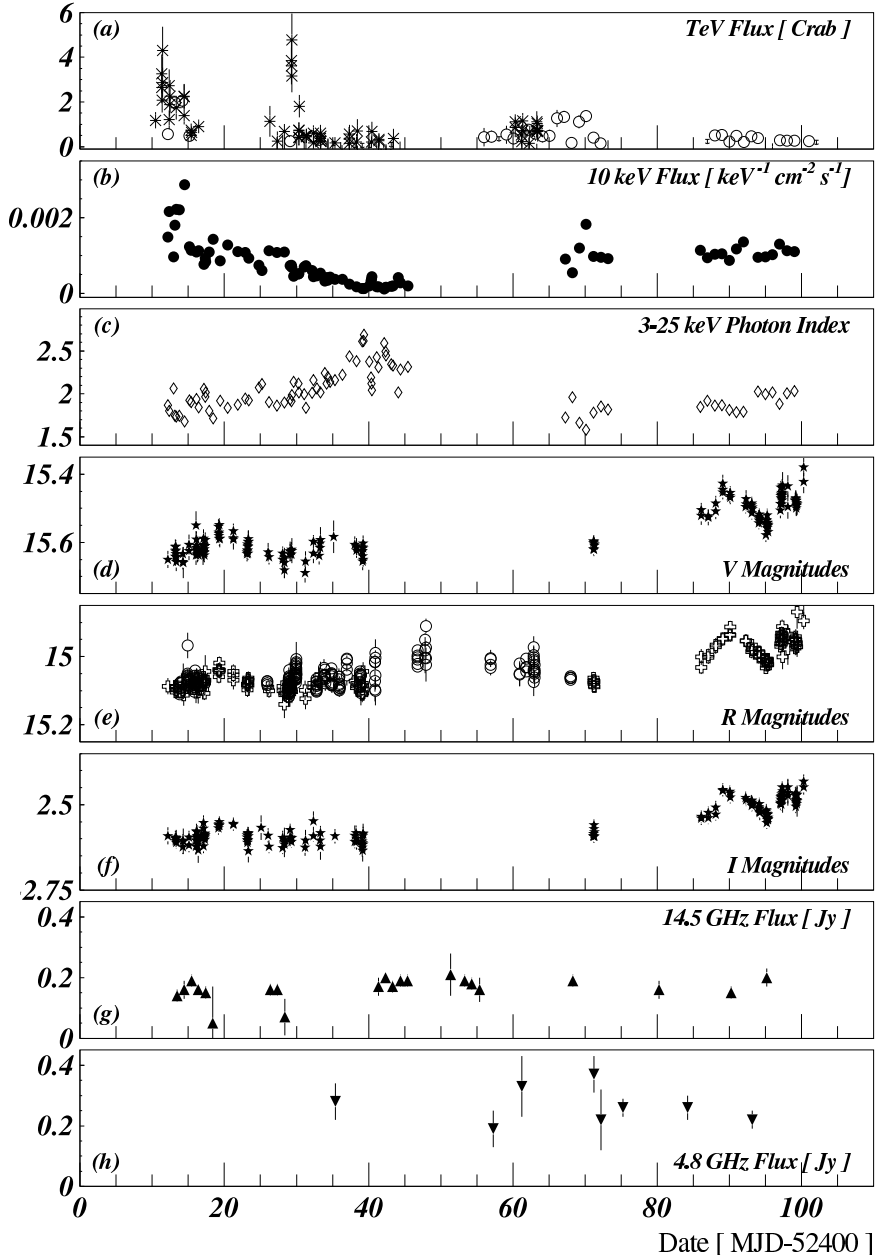,width=11.5cm}
\caption[Lightcurve of 1ES 1959+650 between May 18, and August 14,
2002]{Lightcurve of 1ES 1959+650 between May 18, and August 14, 2002. Figure
  from~\cite{1es1959_orphan}, with permission from the author; Data (from the
top): (a) TeV detection by {\sc Whipple} (stars) and {\sc HEGRA} (circles); X-ray measurements from {\sc RXTE},
(b): $10$~keV, (c): spectral photon index between $(3-25)$~keV;
Optical results, (d), (e) and (f): $V$~and $I$ magnitudes from {\sc Boltwood} and $R$ magnitude
from {\sc Boltwood} (Swiss crosses) and {\sc
Abastunami} (circles); (g) and (h): radio measurements at frequencies of $14.5$~GHz and
$4.8$~GHz from {\sc UMRAO}.}
\label{1es1959_variability}
}
\end{figure}

%==============================================
\paragraph{Multiwavelength observations of AGN}
%==============================================
AGN have been observed in all frequency bands, ranging from radio observations
up to TeV measurements. As an example, the lightcurves of the BL Lac object
1ES~1959+650 have been investigated in a
multiwavelength campaign  
between May 18, and August 14, 2002~\cite{1es1959_orphan}. The results are
shown in Fig.~\ref{1es1959_variability}. The bandpasses shown are TeV emission as
detected by {\sc Whipple} (stars) and {\sc HEGRA}\footnote{{\bf H}igh {\bf
    E}nergy {\bf G}amma-{\bf R}ay {\bf A}stronomy} (circles), X-ray emission
as measured by {\sc RXTE}\footnote{{\bf R}ossi {\bf X}-ray {\bf T}iming {\bf
    E}xplorer}, optical emission in the Violett, Red and Infrared band as well as radio
measurements from {\sc UMRAO}\footnote{{\bf U}niversity of {\bf M}ichigan {\bf
  R}adio {\bf A}strophysical {\bf O}bservatory} at frequencies of $14.5$~GHz and $4.8$~GHz. As can be seen from
the example, AGN are highly variable objects at all wavelengths. In
multiwavelength campaigns as the presented one, the correlation between the
temporal behavior in the different bandpasses is examined. Such campaigns are
relevant to determine the origin of the radiation. In the case of the observation of 1ES~1959+650, a
so-called {\it orphan flare} was observed, representing a rapid increase
in intensity only at TeV energies. Since SSC models necessarily predict the
correlated emission of TeV photons and X-rays, this scenario can be excluded
for the observed flare, while an EC scenario is still possible. A hadronic scenario in which the TeV photons come from
the decay of $\pi^{0}$ particles produced in proton-photon interactions does
not require the connection between X-rays and high-energy photons. Orphan
flares are therefore very interesting in the context of
neutrino emission.

Recent results from~\cite{hess_m87} could restrict the temporal variability at TeV
energies to less than $2$~days. This implies scales of the order of the
Schwarzschild radius of M~87, indicating that the TeV signal originates from
the core of the object and not from the jets. 

\begin{figure}
\centering{
\epsfig{file=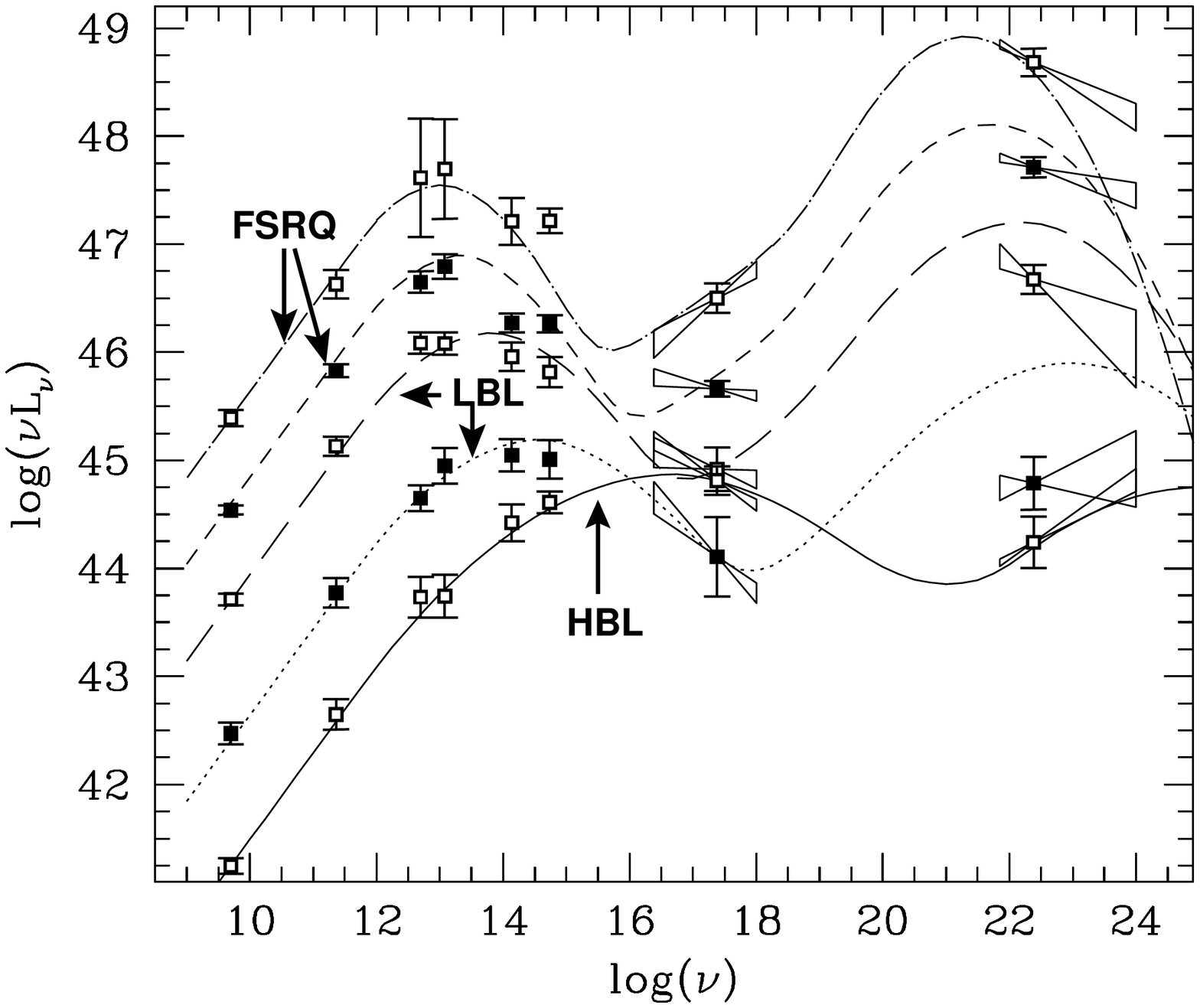,width=12cm}
\caption[Averaged SED for FSRQs, LBLs and HBLs]{Averaged SED for FSRQs, LBLs
  and HBLs. Reprint from~\cite{fossati1998}, with permission from MNRAS.}
\label{blazar_sed_comments}
}
\end{figure}

Most recently, the detection of 3C~279 by {\sc MAGIC} at energies of $\eg>$
was announced~\cite{prandini_icrc2007}. The detection this distant source
$(z=0.538)$ was unexpected, since very high-energy photons are believed to be
absorbed by the interaction with the infrared (IR) background, see~also
Section~\ref{ebl}. The flux at energies of above $\eg>100$~GeV is therefore
expected to be absorbed completely for sources with distances above $z\gtrsim 0.2$.
%==============================================
%\paragraph{Spectral energy distribution}
%==============================================
The spectral energy distribution (SED) of AGN typically shows two main bumps,
see Fig.~\ref{blazar_sed_comments},
apart from dust radiation, which can also lead to a bump in the spectrum, see e.g.~\cite{chini89}. The lower energy hump is believed to arise from synchrotron
radiation of electrons. Different radiation processes can be responsible for
the second hump at higher energies (GeV-TeV): in the Synchrotron Self
  Compton scenario, the synchrotron photons are up-scattered to high energies
by the primary electrons by the Inverse Compton effect. This implies the
direct correlation of the two humps in the case of an intensity variation. A
second possible scenario is the production of TeV photons in
$\pi^{0}$-decays. $\pi^{0}$ particles are produced in hadronic interactions
with photon fields or with each other in the source. Such a hadronic scenario
leads to the coincident production of high-energy neutrinos. It does not
necessarily imply the coincident variation of the lower and higher energy hump, since the
two emission signatures are not directly linked. A third component which can
contribute at TeV energies is the synchrotron radiation of protons. The latter
radiate in the case of high magnetic fields in the acceleration region.
Depending on the energy range of the second hump in the SED, BL Lac objects can be divided into a further sub-class of {\it High-peaked BL
  Lacs (HBLs)}, {\it Low-peaked BL Lacs
  (LBLs)} and {\it Flat Spectrum Radio Quasars (FSRQs)}. If the peak
  occurs at TeV energies, sources are called HBLs, while they are referred to
  as LBLs at peak energies in the GeV range. The source is called FSRQ for
  even lower peak energies. The source classes can be identified by taking
  their radio emission at $5$~GHz as a
  measure. Figure~\ref{blazar_sed_comments} shows the SED for a sample of
  blazars presented in~\cite{fossati1998}. The sample is divided into
  five sub-samples, selected via their radio luminosity at $5$~GHz, $\log(L_{radio}/$erg$)=\left\{<41,\,42-43,43-44,44-45,>45\right\}$. The
  overlaid curves are analytic approximations as described in~\cite{fossati1998}.
%..................................................................
\paragraph{TeV photons from AGN and the extragalactic background light\label{ebl}}
%..................................................................
Since 2003, the second generation of Imaging Air Cherenkov Telescopes (IACTs) is taking data,
with leading results from {\sc H.E.S.S.}\footnote{{\bf H}igh {\bf E}nergy {\bf S}tereoscopic {\bf
    S}ystem} and {\sc
  MAGIC}\footnote{{\bf M}ajor {\bf A}tmospheric {\bf
    G}amma {\bf I}maging {\bf C}herenkov Telescope}, and most recently also from {\sc VERITAS}\footnote{{\bf V}erv {\bf E}nergetic {\bf R}adiation {\bf I}maging {\bf
T}elescope {\bf A}rray {\bf S}ystem}. As of December 2007, 17
HBLs have been detected. 
Most recently, the first LBL, BL Lacertae, has been identified at TeV
energies by the {\sc MAGIC} experiment~\cite{albert_bllacertae}.
In addition, M~87 as a misaligned blazar, or a FR-I galaxy, is identified at
energies of $\eg>100$~GeV. With 3C~279, the first FSRQ has now been identified
by {\sc MAGIC} 
at $\eg>100$~GeV. This source is by far the most distant object detected at
these high energies.
  The AGN identified at TeV energies and their
spectral properties are listed in table~\ref{tev_agn_table}. 
It is not possible to give a time-independent spectral
parameterization of these variable TeV sources, since both the spectral index and the normalization vary with
the intensity of the signal. In the table, intense flare events have been chosen. The
values represent the maximum output per second observed so far. The
parameterization for the spectra is taken as
\be
\frac{dN_{TeV}}{dE_{TeV}}=F_{TeV}\cdot\left(\frac{E_{TeV}}{1\mbox{TeV}}\right)^{\alpha_{TeV}}\,,
\ee
with the spectral parameters as measured at Earth.
The signal of TeV photons is (partially) absorbed
at the highest energies by the Extragalactic Background Light
(EBL), comprised
of the infrared background as the trace of star formation, the cosmic microwave
background and other photon fields in the Universe~\cite{stecker92}. For a lower energy
threshold of ${\eg}^{th}\sim 100$~GeV, it was expected that sources up to
$z\sim 0.2$ can be detected. Thus, the detection of 3C~279 at $z=0.538$ was unexpected and
makes it necessary to reconsider the basic physics. Two assumptions go into
the conclusion that sources farther away than $z\sim 0.2$ should be completely
absorbed at $\eg>100$~GeV: (a) the presence of the IR background from star
formation; (b) the intrinsic particle spectra are not flatter than
$dN_{TeV}/dE_{TeV}\propto E_{TeV}^{-1.5}$. With the detection of 3C~279, it is most
likely that one of the assumptions does not hold. The intensity of the EBL at
IR wavelengths is difficult to determine directly. However, star
formation unavoidably leads to the production of at
least a minimum component of IR radiation. The observation of such a
distant source is difficult to explain by altering the EBL model within the
standard models of cosmology and particle physics. Beyond the standard model
of particle physics,
it is possible to introduce a new boson as done in~\cite{angelis2007}.  Photons
are in oscillation with a light
boson of a mass $m<10^{-10}$~eV, leading to the significant reduction of the
ELB. An further approach is presented in~\cite{protheroe_meyer2000},
explaining the long attenuation length via the violation of the
Lorentz invariance at the highest energies.
An explanation within the standard model is that the photon spectra are indeed extremely flat intrinsically,
which can be the result of very flat primary particle spectra. The statement from~\cite{aharonian_elb2006} that the photon spectra can intrinsically not
become flatter than $E_{TeV}^{-1.5}$ is questioned
in~\cite{stecker_baring2007}. The basic assumption which produces the
$\eg^{-1.5}$ limit is that the primary particle spectra producing the Inverse
Compton component at GeV-TeV photon energies is not flatter than $\sim
E_{p}^{-2}$ as it was shown by~\cite{bednarz_ostrowski98,kirk00,ostrowski_bednarz02} for relativistic, parallel shocks in the small
angle scattering regime. It is pointed out in different papers, however, that
the assumed shock configuration is very specific and does not apply to most
sources~\cite{baring04,mbq_icrc2007,mbq2007}. In~\cite{stecker_baring2007}, it is shown that using larger
scattering angles, very flat spectra can be produced. This leads in turn to
flatter IC spectra, making it possible to extend the absorption cutoff to higher redshifts.
To resolve this issue, further observations of more distant object are
necessary, and the lowering of the threshold energy for photon detection to a few tens of GeV will
also help.

While IACTs are designed for the detection of point sources, surface water Cherenkov detectors
instrument a large area with photomultipliers, being able to observe both
photons, electrons and muons from electromagnetic showers. This technique
allows for the observation of $2\pi$~sr at once, however, with a relatively low
sensitivity and a high-energy threshold. {\sc MILAGRO} succeeded with the observation of Mkn~421 and the
Crab nebula. In addition, diffuse emission regions in the Cygnus region were
found, which is not possible with IACTs due to the small field of view. The
next generation water Cherenkov telescopes, including the planned {\sc
  HAWC}\footnote{{\bf H}igh {\bf A}ltitude {\bf W}ater {\bf
    C}herenkov} experiment~\cite{hawc,hawc_madison06}, are expected to give more information on diffuse emission due to a
higher sensitivity.
\begin{table}[h!]
\centering{
\begin{tabular}{l|lllll}
\hline
Source&type&$z$&$\alpha_{TeV}$&$F_{TeV}$&first TeV \\
      & && &&detection\\
&&&&$10^{-13}$/(TeV s cm$^2$)&Ref\\\hline\hline
M~87&FR-I &0.004&$-2.22$&$11.7$&{\sc HEGRA}\\
    &     &     &({\sc H.E.S.S.}~2005)&\cite{hess_m87}&\cite{aharonian_m87}\\ \hline 
Mkn~421&HBL&0.031&$-2.14$&$311$&{\sc Whipple}\\
       &   &     &({\sc Whipple}
       2001)&\cite{mkn421_flare_whipple}&\cite{punch1992} \\\hline
Mkn 501&HBL&0.034&$-1.92$&$1001$&{\sc Whipple}\\
       &   &     &({\sc HEGRA} 1997)&\cite{aharonian_mkn501}&\cite{quinn1996} \\\hline
1ES 2344+514&HBL&0.044&$-2.54$&$510$&{\sc Whipple}\\
            &   &     &({\sc Whipple}~1995)&\cite{1es2344_schroedter}&\cite{catanese_1es2344} \\\hline
Mkn 180&HBL&0.045&$-3.3$&$12.0$&{\sc MAGIC}\\
       &   &     &({\sc MAGIC} 2006)&\cite{albert_mkn180}&\cite{albert_mkn180} \\ \hline
1ES 1959+650&HBL&0.047&$-2.83$&$740$&{\sc Tel.~Array}\\
            &   &     &({\sc HEGRA} 2002)&\cite{aharonian_hegra_1es1959}&\cite{nishiyama1999} \\\hline
BL Lacertae&LBL&0.069&$-3.6$&$3.6$&{\sc
  MAGIC}\\
&&&({\sc MAGIC} 2006)&\cite{albert_bllacertae}&\cite{albert_bllacertae}\\ \hline
PKS 0548-322&HBL&0.069&n/a&n/a&{\sc H.E.S.S.}\\
&&&-&-&\cite{glast_symposium}\\\hline
PKS 2005-489&HBL&0.071&$-4.0$&$2$&{\sc H.E.S.S.}\\
&&&({\sc H.E.S.S.} 03/04)&\cite{aharonian_pks2005}&\cite{aharonian_pks2005}
\\\hline
PKS 2005-489&HBL&0.071&$-4.0$&$2$&{\sc H.E.S.S.}\\
&&&({\sc H.E.S.S.} 03/04)&\cite{aharonian_pks2005}&\cite{aharonian_pks2005}\\\hline
RGB J0152+017&HBL&0.080&n/a&n/a&{\sc H.E.S.S.}\\
&&&-&-&\cite{hess_rgb0152} \\\hline
H 1426+428&HBL&0.129&$-3.5$&$50.0$&{\sc Whipple}\\
&&&({\sc Whipple} 2001)&\cite{petry2002}&\cite{horan2001} \\\hline
1ES 0229+200&HBL&0.139&$-2.5$&$6$&{\sc H.E.S.S.}\\
&&&({\sc H.E.S.S.} 05/06)&\cite{aharonian_1es0229}&\cite{glast_symposium}\\ \hline
H 2356-309&HBL&0.165&$-3.0$&$3.00$&{\sc H.E.S.S.}\\
&&&({\sc H.E.S.S.} 2004)&\cite{aharonian_h2356}&\cite{aharonian_h2356} \\\hline
\end{tabular}
}
\end{table}
\begin{table}[h!]
\centering{
\begin{tabular}{l|lllll}
\hline
Source&type&$z$&$\alpha_{TeV}$&$F_{TeV}$&first TeV \\
      & && &&detection\\
&&&&$10^{-13}$/(TeV s cm$^2$)&Ref\\\hline\hline
1ES 1218+304&HBL&0.182&$-3.0$&$12.7$&{\sc MAGIC}\\
&&&({\sc MAGIC} 2005)&\cite{albert_1es1218}&\cite{albert_1es1218} \\ \hline
1ES 1101-232&HBL&0.186&$-2.88$&$4.44$&{\sc H.E.S.S.}\\
&&&({\sc H.E.S.S.} 2006)&\cite{aharonian_1es1101}&\cite{aharonian_1es1101} \\\hline
1ES 0347-121&HBL&0.188&$3.10$&$4.52$&{\sc H.E.S.S.}\\
&&&&\cite{aharonian_1es0347}&\cite{aharonian_1es0347}\\\hline
1ES~1011+496&HBL&0.212&$-4.0$&$3.2$&{\sc MAGIC}\\
&&&({\sc MAGIC} 2006)&\cite{albert_1es1011}&\cite{albert_1es1011}\\\hline
3C~279&FSRQ&0.538&n/a&n/a&{\sc MAGIC}\\
&&&-&-&\cite{prandini_icrc2007}\\\hline
PG 1553+113&HBL&?&&&{\sc
  H.E.S.S.}\\
&&&&&\cite{aharonian_pg1553}\\
&&&-4.2&2.1&{\sc MAGIC}\\
&&&({\sc MAGIC} 05/06)&\cite{albert_pg1553}&\cite{albert_pg1553}\\\hline
\end{tabular}
\label{tev_agn_table}
\caption[AGN with TeV photon emission]{AGN with TeV photon emission. The spectrum is $dN_{TeV}/d\eg=F_{TeV}\cdot\left(\eg/1\mbox{TeV}\right)^{\alpha_{TeV}}$ Sources
  are sorted by redshift. The redshift of PG 1553+113 is still unknown. The
  sources with no spectral index announced (n/a) are preliminary.}
}
\end{table}
\clearpage
%----------------------------------------------
\subsubsection{Gamma Ray Bursts\label{grbs_photons:general}}
%----------------------------------------------
Photon eruptions of unknown origin were detected in the 1960th by both American
and Soviet military satellites. While it was immediately clear that these events were not man made,
but originated from outer space, the publication of the first observation in 1967 did not
happen before 1973 in the case of the {\sc Vela Satellites}~\cite{vela} and only a few months later in the 
case of the Soviet {\sc
  Kosmos-461}, and the American {\sc
  OSO-7}\footnote{{\bf O}rbiting {\bf S}olar {\bf O}bservatory-{\bf 7}} and {\sc IMP-6}\footnote{{\bf
    I}nterplanetary {\bf M}onitoring {\bf P}latform-{\bf 6}}
satellites~\cite{wheaton73,cline73,mazets74}. Systematic studies of these Gamma Ray Bursts
(GRBs) were done with {\sc BATSE}\footnote{{\bf B}urst {\bf a}nd {\bf T}ransient
  {\bf S}ource {\bf E}xperiment} on board of the
{\sc CGRO}\footnote{{\bf C}ompton {\bf G}amma {\bf R}ay {\bf O}bservatory} which was taking data for 9 years, between April 1991 and
June 2000~\cite{batse_cat}. During that time, 2704 GRBs in the
energy range of $(20,\,2000)$~keV were detected\footnote{In the following,
  the detection energy range of all quoted instruments is given in the
  notation $(E_{\min},\,E_{\max})~keV$ for simplicity.}. 
\begin{figure}[h!]
\centering{
\epsfig{file=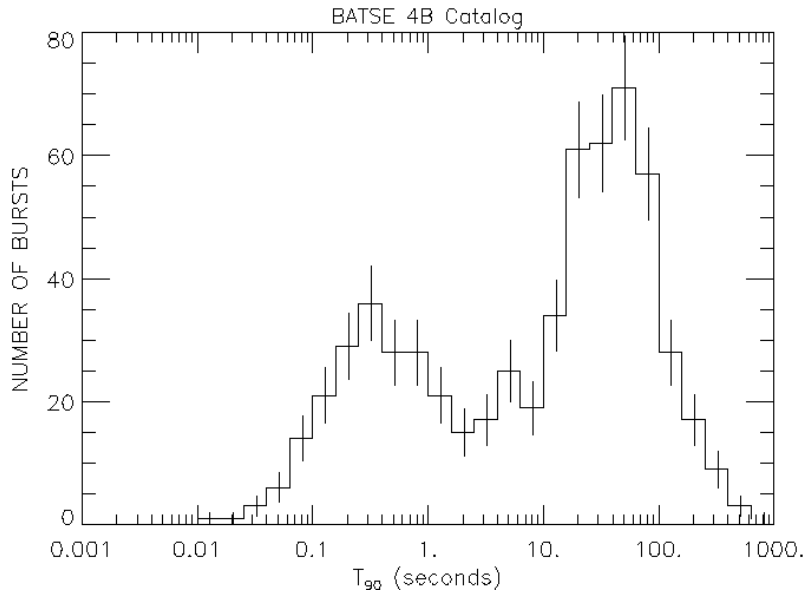,width=11cm}
\caption[Distribution of the duration $t_{90}$ for GRBs in the {\sc BATSE} 4B
  Catalog]{Distribution of $t_{90}$ for 1234 GRBs in the {\sc BATSE} 4B
  Catalog~\cite{batse_cat,batse_web}. Events with $t_{90}<2$~s are classified
  as short bursts, GRBs with $t_{90}>2$~s are called long bursts. Figure
  courtesy NASA/BATSE and NASA Gamma-Ray Astronomy group at National Space
  Science and Technology Center (NSSTC)~\cite{nasa_gammaray}.}
\label{batse_t90}
}
\end{figure}

\begin{figure}[ht]
\centering{
\epsfig{file=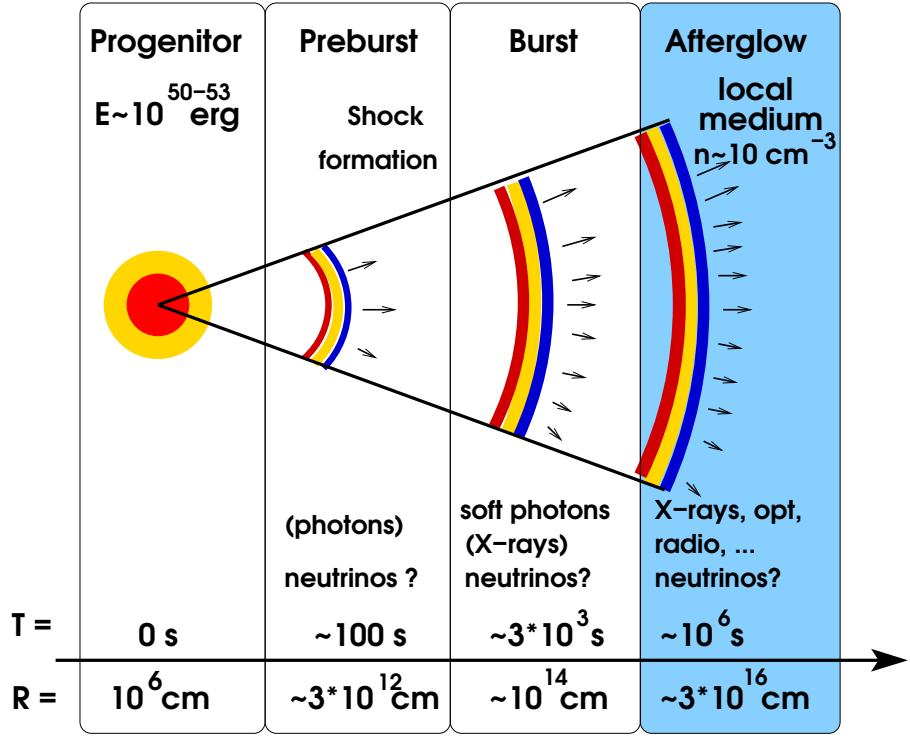,width=12cm}
\caption[Schematic view of the fireball scenario]{Schematic view of the fireball scenario.}
\label{fireball}
}
\end{figure}

Figure~\ref{batse_t90} shows
the distribution of the duration of the bursts $t_{90}$~\cite{batse_web,batse_cat},
defined in the way that 90\% of the signal was received during that time. Two
populations of bursts can be identified, classified as ``short''
($t_{90}<2$~s) and ``long'' ($t_{90}>2$~s) bursts. The spatial distribution of
GRBs in galactic coordinates as observed by the {\sc BATSE} reveals an
isotropic distribution with no visible clustering in the galactic plane or anywhere else. This indicates
an extragalactic origin of the events. However, scenarios of a galactic halo
with so far unknown sources of GRBs were also proposed, see
e.g.~\cite{galactic_halo_1,galactic_halo_2,galactic_halo_3}. A first
indication of a cosmological origin was given by the spatially non-Euclidean
distribution of the source luminosity. The final proof of the cosmological
distance of GRBs was possible in 1997 by the first afterglow observation by the {\sc BeppoSax}\footnote{{\bf
    Beppo} stands for Guiseppe Occhialini, and ``SAX'' is an acronym for
  {\bf S}atellite per {\bf A}stronomia {\bf X}} satellite, see e.g.~\cite{beppo_sax_afterglow}. While the prompt emission is
mainly detected in the keV-MeV band, the so-called
afterglow continues until long after the prompt
emission and is seen in basically all wavelength-bands, from the radio band up
to GeV-energies. From the afterglow-observation, host galaxies can be
identified, or absorption and emission lines can be measured to determine the
redshift at which the GRB occurred. These redshifts are cosmological, so that
GRBs are known to happen outside of our Galaxy. The reason for the intense
discussion of a galactic origin was that the photon fluence $F_{\gamma}$ in
the $(20,\, 2000)$~keV
energy band as measured by {\sc BATSE} scatters around
\begin{equation}
F_{\gamma}\sim 10^{-6}\mbox{ erg/cm}^2\,.
\end{equation}
The term ''fluence'' is used here as opposed to flux, since the units are
erg/cm$^2$, while a flux is typically measured per area and time interval.
For extragalactic distances, the total luminosity $L_{GRB}$ of a GRB event lies around
\begin{equation}
L_{GRB}\sim 10^{51}\mbox{ erg/s}
\end{equation}
for isotropic emission, but a little bit lower in the case of beamed emission
favored currently.
This tremendous output lies more than four orders
of magnitude higher than the typical output of AGN, the most luminous permanent source
class in the sky,
\begin{equation}
L_{AGN}\sim 10^{44}-10^{47}\mbox{erg/s}\,.
\end{equation}
While GRBs emit only for a short time of a few seconds, AGN are active over
long periods, so that the time integrated output is comparable, $\sim 10^{62}$~erg.
In the past decades, different models have been developed to explain the 
output from GRBs and AGN. Many of the classical arguments for the description of GRB physics are
borrowed from supernova remnants~\cite{woltjer72,cox72,cox74} or AGN
models~\cite{rees70}, since the phenomena are similar, although on different
spacetime-scales. Both AGN and GRBs show a variable time structure. supernova
remnants as well as long GRBs are produced in the explosion of stars. In all
three objects, shock fronts are responsible for particle acceleration and
therefore, the non-thermal electromagnetic spectrum can explained by
synchrotron radiation of electrons, Inverse Compton scattering and also by proton-photon interactions.
The favored model which is able to explain most of the phenomena
connected to a GRB is the fireball model, explaining the huge
electromagnetic emission by shock formation of relativistic plasma
shells, see e.g.~\cite{piran99,piran05,zhang04}. Alternatively, the Cannonball model tries to explain GRBs by colliding
plasma balls, see e.g.~\cite{dar_rujula,dar06}. However, the latter meets observations difficult to match the
predictions in this model, e.g.~the prediction of apparent motion for the
radio emission, which is
not observed for GRBs. Thus, only the fireball model will be discussed in
more detail in the following paragraph.
%---------------------------------------------------
\paragraph{Fireball model}
%---------------------------------------------------
A schematic view of the fireball model is shown in Fig.~\ref{fireball}. It
is based on the model of stellar outbursts as described in~\cite{sedov58} and
was enhanced including relativistic effects in order to match GRB observations. The
fireball model does not give any constraint on the progenitor. It yields a
phenomenological description of the actual burst observations. The basic idea is that a large amount of
mass is ejected within a short time interval by a central engine. The plasma
is ejected successively in shells. At some point, the outer shells slow down
and are caught by inner shells and a shock front is built up, accelerating
electrons and baryons in the plasma up to high energies. While protons can be
accelerated basically loss-free up to energies as high as
$10^{21}$~eV~\cite{vietri95,waxman_emax}, electrons lose their
energy to synchrotron radiation, escaping from the shocks as soon as the
region becomes optically thin. This is observed as prompt emission
from GRBs. Those shocks resulting from collisions of shells are called {\em
  internal shocks}. So-called {\em external
  shocks} result from collisions of
the shells with the interstellar medium leading to afterglow emission as
described in the following paragraph. While the prompt emission occurs mainly at energies
of $E_{\gamma}>100$~keV, afterglow emission is observed in almost all wavelength
bands.  Reviews on the
details of the underlying physics are given in
e.g.~\cite{piran99,piran05,zhang04}. 

%---------------------------------------------------
\paragraph{GRB experiments after BATSE}
%---------------------------------------------------
After the {\sc BATSE} era, many GRB satellites were taking data, each 
covering a much smaller field of view and thus providing much less statistics than
{\sc BATSE}. Additionally, many of the satellites were not able to give directional
information. For the determination of the GRB spatial origin, it is necessary
to have at least three detectors. {\sc BATSE} had four energy
channels and could localize GRBs. The precision was between approximately $10^{\circ}$ and a
few degrees. Other satellites have only one or two instruments on board. To
improve the localization of GRBs,
 the {\sc Interplanetary Network} was created already in the 1970s as an interconnection of
 all GRB satellites. The currently active third {\sc Interplanetary Network,
 IPN3}, was formed with the launch of {\sc Ulysses} in
 1990~\cite{ulysses,ulysses_web}. Since the beginning, more than 25 spacecraft
 missions have participated. The {\sc CGRO} joined in 1991 when it
 was launched. Today, {\sc HETE-II}\footnote{{\bf H}igh {\bf E}nergy {\bf
     T}ransient {\bf E}xplorer-II}~\cite{hete_classification,hete_web},
 {\sc INTEGRAL}\footnote{{\bf INTE}rnational {\bf G}amma-{\bf R}ay {\bf
     A}strophysics {\bf L}aboratory}~\cite{integral,integral_grb}, {\sc RHESSI}\footnote{{\bf R}amaty {\bf
     H}igh {\bf E}nergy {\bf S}olar {\bf S}pectroscopic {\bf
     I}mager} \cite{rhessi_web}, {\sc Mars
 Odyssey} \cite{mars_odyssey,mars_odyssey_web},
 {\sc Ulysses} \cite{ulysses,ulysses_web},
 {\sc Konus Wind} \cite{konus_short_cat,konus_web} and {\sc Swift} form the {\sc IPN3}. The most important GRB experiments with some of
 their individual properties are listed in
 table~\ref{grb_experiments}. With information of more than two instruments, GRB positions of a accuracies up to several
 square-arcminutes can be
 reconstructed. Before the launch of {\sc BeppoSax} in 1996, this was the only
 possibility of arcminute precision measurements. {\sc HETE-II} and {\sc INTEGRAL} are able to localize GRBs
 without additional information from {\sc IPN3}. {\sc Konus} is very sensitive to
 short GRBs and a catalog of $130$ short GRBs is examined in the context
 of neutrino emission, see Section~\ref{nus_grbs}.
\begin{table}[h!]
\centering{
\begin{footnotesize}
\begin{tabular}{l|llll|l}\hline
GRB Sat.&Launch-Demise&FoV$^{***}$&E-range&Localiz.&Reference\\ 
        &            &   &[keV]&Precision& \\\hline\hline
{\sc Vela 5B}&05/1969-06/1979&$6.1^{\circ}\times6.1^{\circ}$&(3, 750)&-&\cite{vela} \\\hline
{\sc Kosmos-641}&12/1971-09/1972&$\sim 2\,\pi$~sr&(28,
1000)&-&\cite{mazets74},\\
 &&&&&\cite{val_priv},\\
                &            &                &          & &\cite{mazets_atmos}\\\hline
{\sc Ulysses}$^{*}$&10/1990-now&$<75^{\circ}$&(5, 150)&-&\cite{ulysses},\\
                   &           &             &        & &\cite{ulysses_web} \\\hline
{\sc BATSE}&04/1991-06/2000&$\sim 2\pi$~sr&(20,
        2000)&degree&\cite{batse_cat},\\
          &                &              & & &\cite{batse_web}\\\hline
{\sc BeppoSax}&04/1996-04/2002&$0.5^{\circ}-1.3^{\circ}$&(0.1, 300)&arcmin&\cite{boella},\\
              &               &                         &          &      &\cite{beppo_sax_web}\\\hline
{\sc Mars Od.}$^{*}$&04/2001-now&$62^{\circ}$&(50, 10000)&-&\cite{mars_odyssey},\\
                    &           &            &           & &\cite{mars_odyssey_web}\\\hline
{\sc NEAR}&02/1996-02/2001&$60^{\circ}$&(1, 10000)&-&\cite{near_web} \\\hline
{\sc Konus W.}$^{*}$&11/1994-now&$\sim 2\pi$~sr&(10, 10000)&-&\cite{konus_web}\\\hline
{\sc RHESSI}$^{*}$&02/2002-now&$1^{\circ}$&(3, 20000)&-&\cite{rhessi_web}\\\hline
{\sc INTEGRAL}$^{*}$&10/2002-now& & &  &\cite{integral_grb}\\
SPI& &$16^{\circ}$&(18,8000)&-&\\
IBIS& &$9^{\circ}\times9^{\circ}$&(15, 10000)&arcmin&\\\hline
{\sc HETE-II}$^{*}$&10/2000-now &$1.5-3$~sr&(0.5, 400)&arcmin&\cite{hete_web},\\
                   &            &          &          &      &\cite{hete_classification} \\\hline
{\sc Swift}$^{*}$&11/2004-now&&&&\cite{swift_proc},\\
BAT& &$2$~sr&(15, 150)&arcmin&\cite{swift_web}\\
UVOT$^{**}$& &$17'\times 17'$&(170, 650)&arcsec& \\
XRT& &$23.6'\times 23.6'$&(0.2, 10)&arcsec& \\\hline
\end{tabular}
\end{footnotesize}
\caption[GRB satellites]{Some GRB Satellites and the basic properties of the main instruments
  for the detection of prompt emission. FoV= Field of View.\quad /$^{*}:$ Current
  member of IPN3.\quad/$^{**}$: the energy range for UVOT is given in terms of
  wavelength, in units of nano-meters (nm).\quad/$^{***}$: FoV for {\sc BATSE}
  and {\sc Kosmos-641} is actually $4\,\pi$~sr, but here it is assumed that
  $\sim$ half a hemisphere is occulted by the Earth.}
\label{grb_experiments}
}
\end{table}
\clearpage
The {\sc Swift} satellite was launched in November 2004 - for a
review see\linebreak e.g.~\cite{swift_proc} and references therein. {\sc Swift} is a
dedicated GRB satellite with four instruments on board. The main purpose of the
BAT\footnote{{\bf B}urst {\bf A}lert {\bf T}elescope} detector is the discovery of prompt
emission from GRBs. The main sensitivity is in the energy range of $(15,\,150)$~keV and
the field of view is about $2$~sr. About 100 GRBs per
year are detected with BAT. The XRT\footnote{{\bf X}-{\bf R}ay {\bf T}elescope} covers an energy range of $(0.2,\,10)$~keV and
serves afterglow observations. The UVOT\footnote{{\bf UV}/{\bf O}ptical {\bf T}elescope} is an
instrument for the detection of the optical afterglow at $(170,\,650)$~nm
wavelengths. 

\begin{figure}[hb]
\centering{
\epsfig{file=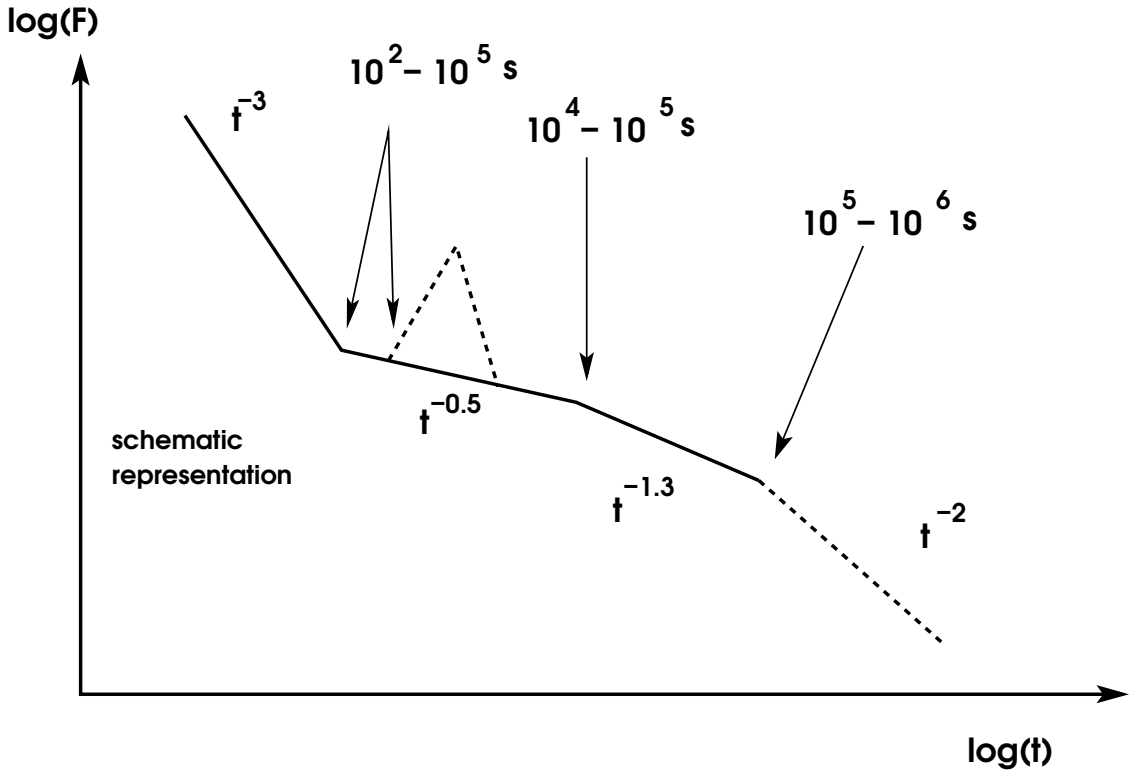,width=15.cm}
\caption[Early afterglow - temporal intensity behavior]{The figure shows the
  early afterglow flux decay with time. The scale is double
  logarithmic and the temporal behavior follows $t^{\alpha_t}$ with different
  decay indices $\alpha_t$ for different time periods as indicated in the
  figure. Figure after~\cite{meszaros_early_afterglow}. After a steep decay of
  $t^{-3}$, a shallow decay phase follows, $\sim t^{-0.5}$. In this second
  phase, afterglow flares can occur in the burst. The observation of this
  early afterglow up to about $10^{3}$~s after the burst has for the first
  time been observed by the {\sc Swift} experiment. The phase at later times,
  $t>10^{4}$~s was known since {\sc BATSE} times. A spectral behavior of
  $t^{-1.3}$ is observed at first, with a break in the slope at $\sim
  10^{5}-10^{6}$~s to a steeper behavior of $\sim t^{-2}$.}
\label{early_afterglow}
}
\end{figure}

The advantage of a satellite carrying both prompt emission and
afterglow instruments is that the afterglow can be followed almost starting
from the prompt emission phase. This has already lead to the discovery of
unexpected temporal behavior directly after the prompt emission, see
e.g.~\cite{meszaros_early_afterglow} for a review. The basic features of the
early afterglow are shown in
Fig.~\ref{early_afterglow}. While the afterglow appearance at $t\sim 10^{4}$~s after
the prompt emission had been known before, the temporal behavior at earlier
times was unexplored until the launch of {\sc Swift}. Two main features are
found. Firstly, a break in the temporal decay structure
$t^{\alpha_t}$ is observed. The decay index
changes from $\alpha_t \sim -3$ to $\sim -0.5$ at around $100$~s~ to~$1000$~s after the prompt
emission and it changes back to the previously observed $\alpha_t \sim -1.3$ behavior at
$10^{4}$~s~to~$10^{5}$~s. Secondly, X-ray flares are detected during the early afterglow, indicated by the dashed triangle in the
curve. X-ray flares occur only in a fraction of the observed
bursts. Different models to explain both the changes in the decay index and
the X-ray flares have been developed which are reviewed in
e.~g.~\cite{meszaros_early_afterglow}.
%================================================
\paragraph{Observed Redshifts from GRBs and GRB progenitors}
%================================================
\begin{figure}[ht]
\centering{
\epsfig{file=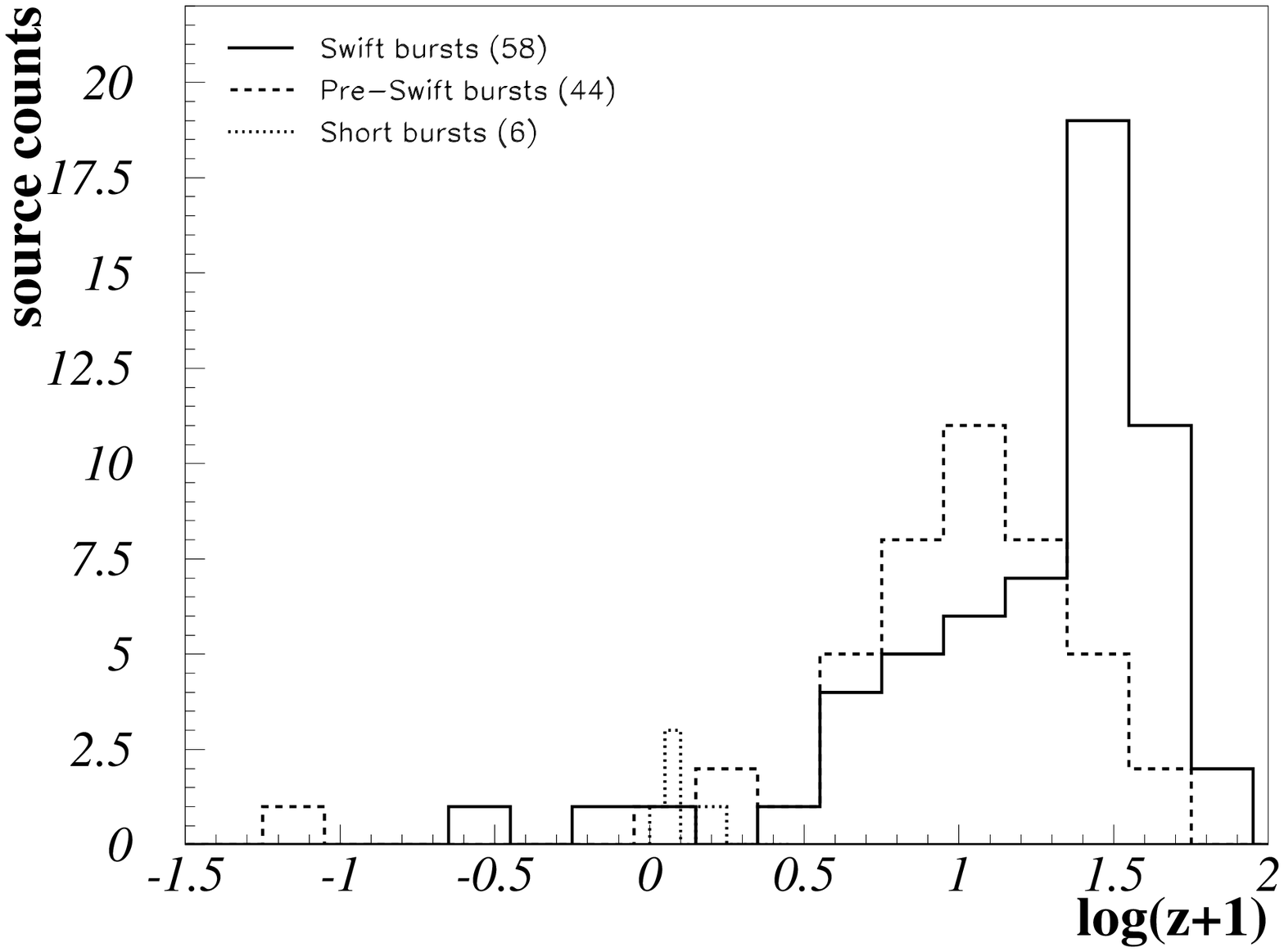 ,width=12cm}
\caption[Observed bursts with measured redshifts]{Observed bursts with measured redshifts. The solid line
  represents long bursts with redshifts from the {\sc Swift} time, the dashed line shows long GRBs in the pre-{\sc Swift} era. There is
  a clear shift of the distributions. Bursts in the pre-{\sc Swift} era
  (mostly {\sc BATSE} measurements) have lower redshifts on average. The
  dotted line shows the distribution of short bursts with measured redshifts
  in the {\sc Swift} era. There were no redshift detections of short bursts before the launch
  of {\sc Swift}.}
\label{redshifts}
}
\end{figure}
The first afterglow observation by {\sc BeppoSax} for GRB970228 also implied the first
measurement of the redshift of a GRB. Between 1997 and November 2004 - the time of
the launch of {\sc Swift} - redshifts of 44 GRBs have been detected. Since the
launch of {\sc Swift}, 58 long and 6 short GRB redshifts were
measured as of March 12 2007. Figure~\ref{redshifts} shows the redshift distribution of those GRBs. The
solid line represents {\sc Swift} bursts while the dashed line shows pre-{\sc
  Swift} measurements of redshifts. The mean of the distributions is shifted,
{\sc Swift} observations show a higher contribution of very distant
bursts. One reason for this is the better sensitivity of {\sc Swift} compared
to pre-{\sc Swift} instruments. Bursts at high redshifts have typically a
weaker fluence and may have been missed by instruments of lower
sensitivity. 
The higher statistics of bursts with measured redshifts relies on the possibility of early afterglow
observation of the {\sc Swift} instruments XRT and
UVOT. 

It is known since 2003 that long GRBs are connected to supernova explosions of
type Ic, which follow the death of Wolf-Rayet stars \cite{grb030329}. 
Two scenarios of producing jets in exploding stars have been discussed:
\begin{enumerate}
\item the core collapse from the center of a flattened, fast rotating star,
  see~\cite{sakashita71,moellenhoff76};
\item the magneto-rotational core collapse with an initial dipole magnetic
  field configuration as suggested in~\cite{bisnovatyi70,bisnovatyi71}.
\end{enumerate}
Short bursts have been proven in 2005 to originate from the merging of two neutron stars or a neutron
star and a black hole in a binary system~\cite{050509b,050709,050724}. The differences in the observed
events lie not only in the duration of the bursts, but are also seen in the
redshift distribution. While long GRBs are most likely to follow the star
formation rate and are located in starforming regions, short bursts happen in
regions of rather low star formation rate and at small redshifts $(z\sim
0.1)$. 

New results from the {\sc Swift} satellite show, however, that this scheme is
still too simple: there are exceptional bursts which do not fit into this
scheme (e.g.\ GRB060218, GRB060614)~\cite{campana,gehrels06}. Also, {\sc Swift} does not see the strong distinction
between short hard and long soft bursts as it was observed by {\sc
  BATSE}~\cite{sakamoto}. This indicates that the classification scheme is
more complex than it can be determined yet and needs to be refined in the
future.
%---------------------------------------------------
\paragraph{Classification of GRBs}
%---------------------------------------------------
In order to explain the observation of the prompt emission in GRBs, a boost
factor of $\Gamma > 100$ for the shock fronts is necessary. For lower boost
factors, the shock region is optically thick to pair production processes. On the other hand, the boost factor
must not exceed $\Gamma\sim 1000$, since protons would lose most of their
energy due to synchrotron radiation in that case~\cite{hh_02}. The peak
energy of the GRB, $E_{peak}$ is directly correlated to the boost factor and
lies around $E_{peak}\sim(100,\,1000)$~keV for regular GRBs. Typically, a boost factor
of $\Gamma=300$ is assumed. The two main sub-classes of GRBs are long and
short ones. The differences do not only lie in the duration of the events, but
also in the hardness of the spectra. Short bursts typically have much harder
spectra than long GRBs which is why they are usually referred to as Short Hard
Bursts (SHBs).

A measure for the hardness is the ratio of soft to hard emission,
\begin{equation}
\mbox{hardness ratio}:=H:=\frac{S(\mbox{hard})}{S(\mbox{soft})}\,.
\end{equation}
with $S$ as the photon flux. In the case of {\sc BATSE}, which had four energy
channels, channel 3 with an energy range of $(100,\,300)$~keV and channel 4, $(50,\,100)$~keV were used to
determine the hardness ratio. 
Bursts with $t_{90}<2$~s are
generally harder than long bursts ($t_{90}>2$~s). 
It should be noted, though,
that follow-up experiments like {\sc HETE-II}, {\sc Konus} and {\sc Swift} do not see
equally hard short bursts. 
For more details, see~\cite{sakamoto}.

\begin{figure}
\centering{
\epsfig{file=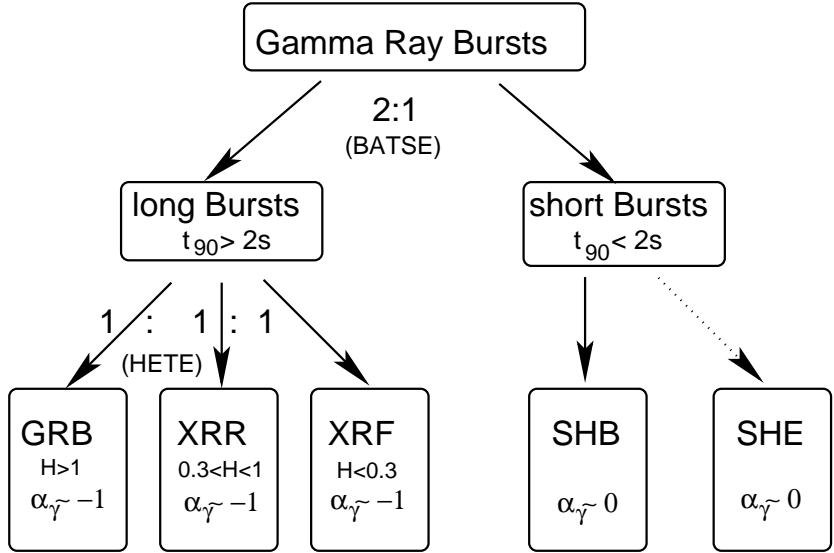,width=11cm}
\caption[Classification scheme of GRBs]{Classification scheme of GRBs.\label{grb_classification:fig}}
}
\end{figure}

Detailed studies by {\sc HETE-II} have shown that apart from the regular, long GRBs,
there are bursts having peak energies in the X-ray
regime~\cite{hete_classification}. For the classification  of events in terms
of the energy band of emission, the hardness ratio for the flux at
$(30,\,400)$~keV and $(2,\,30)$~keV was examined:
\begin{equation}
H_{HETE}=\frac{S(30,\,400)}{S(2,\,30)}\,.
\end{equation}
Regular GRBs have $H>1$. About 2/3 of the 45 {\sc HETE-II} bursts have, however,
$H<1$. This class of GRBs peaking in the X-ray regime, has further been
subdivided into X-Ray Rich bursts (XRRs) with X-ray and soft $\gamma$-ray
emission, $0.3<H<1$, and X-Ray Flashes
(XRFs) with only X-ray emission, $H<0.3$. The ratio between these three burst
classes as observed by {\sc HETE-II} is 
\begin{equation}
\mbox{(GRB:XRR:XRF)}=(1:1:1)\,.
\end{equation}
A schematic view of the classification of GRBs into long and short events and
the further subdivision of the long events into regular GRBs, XRRs and XRFs is
shown in Fig.~\ref{grb_classification:fig}. The ratio of long to short GRBs is
indicated of $2:1$ as detected by {\sc BATSE}~\cite{batse_cat}. However, {\sc HETE-II}
rather detects a ratio\linebreak of $3:1$. The detection ratio should be considered
as dependent on the instrument properties\footnote{{\sc Swift} for instance detects
only one short GRB for 18 long ones. This is also due to the fact that {\sc Swift}
detects at relatively low energies $E<150$~keV, while the emission for short
GRB rather happens at higher energies as discussed before.}. For the analysis of the GRB-XRR-XRF
relation, however, it is stated that not many GRBs are missed relative to XRR
and XRF events due to any observational effects~\cite{hete_classification}, so
that a 1:1:1 ratio seems to be reasonable.

One interpretation of the strong variation of the peak energy is based on variations in
the baryonic load of the shocks, which is an important parameter for the development
of the boost factor~\cite{loaded_fireball}. With a high baryonic load in the evolving jet, the system
cannot be accelerated to high energies due to the high mass of the
baryons. Thus, a this {\em dirty fireball} has
boost factors of an order of magnitude less than a regular GRB, $\Gamma \sim
10$. Since the peak energy evolves with $\Gamma$, the typical values for
dirty fireballs are $E_{peak}\sim (1,\,100)$~keV and XRRs or
XRFs are observed. GRBs as typically observed by {\sc BATSE} are classified as {\em 
 regular fireballs}. {\em Clean fireballs} have a very low baryonic load and thus, a
high gamma of around $\Gamma\sim 1000$, leading to peak energies of
$E_{peak}>$~MeV. The duration of such short high-energy bursts (SHE) would be small, since the
lack of heavy baryons enables the jet to evolve more rapidly, so that
$t_{90}\sim 0.1$~s. Such events have, however, not been observed yet. With the
launch of {\sc GLAST}\footnote{{\bf G}amma-Ray {\bf L}arge {\bf A}rea
  {\bf S}pace {\bf T}elescope}, the detection of such phenomena will be possible~\cite{glast}. A summary
of the different burst types and their basic parameters is given in table~\ref{grb_classification:tab}. The
fact that most bursts occur as regular fireballs shows that a certain fraction
of baryons needs to be present in the jet which inevitably leads to the
production of neutrinos in proton-photon interactions. The question about the
intensity of such a neutrino signal has still to be solved.
\begin{table}[h!]
\centering{
\begin{tabular}{l|lll}\hline
parameter&clean&dirty&regular\\
         & &(XRR/XRF)&(GRB)\\\hline\hline
$\Gamma$&$\sim 3000$&$\sim 30$&$\sim 300$\\
$E_{peak}$ [keV]&$>1000$&$1-100$&$100-1000$\\
$t_{90}$ [s] &0.1 &$\sim 10$&$\sim 10$ \\\hline
\end{tabular}
\caption[Basic parameters of clean/dirty and regular fireballs]{Basic
  parameters of clean/dirty and regular fireballs.}
\label{grb_classification:tab}
}
\end{table}

The classification scheme as presented here needs to be refined including a
more detailed view on the matter in the future: today, there are several
bursts which fall out of the scheme, e.g.~GRB060218 and GRB060614, and {\sc
  Swift} seems to detect a different sub-class of short bursts as compared to
{\sc BATSE}~\cite{sakamoto}. A summary of the observation of the different
sub-classes (XRF/GRB/short) with {\sc Swift} is given
in~\cite{zhang07}. In~\cite{zhang07}, the detection of the shallow decay of
the early afterglow as recently observed in many {\sc Swift} bursts as energy
injection features. Such a scenario disfavors the interpretation of XRFs as
low$-\Gamma$ GRBs. {\sc Swift} will help
to improve the current classification scheme. For now, the scheme as presented above
is still useful, since it can describe the majority of bursts.
%----------------------------------------------
\subsubsection{Galactic sources}
%----------------------------------------------
The electromagnetic output of galactic sources of non-thermal
emission can be used to estimate their contribution to the Cosmic Ray
spectrum. 

\begin{figure}[h!]
\centering{
\epsfig{file=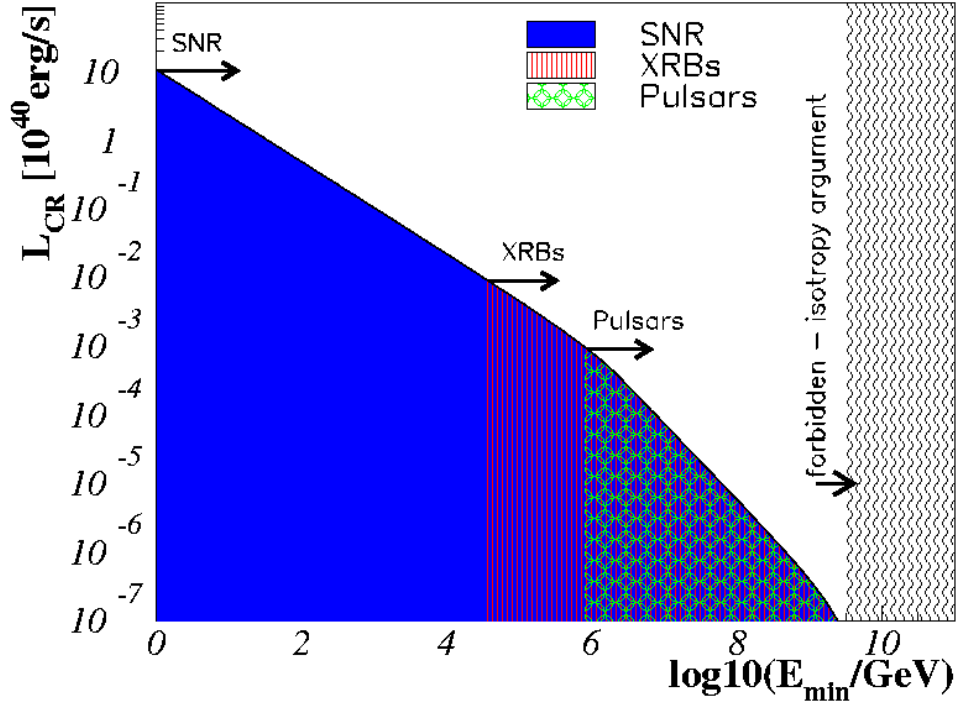,width=\linewidth}
\caption[Cosmic Ray luminosity from the Milky Way]{Cosmic Ray luminosity
  from the Milky Way and possible sources given
  their electromagnetic output. The upper bound of the curve indicates the Cosmic Ray
  luminosity for energies $\ep>E_{\min}$. Cosmic Rays above the ankle,
  i.e.~$\ep>3\cdot 10^{18}$~eV, cannot have a galactic origin, since the
  observed signal is too isotropic. Since supernova remnants in our Galaxy can
  produce a luminosity of $L_{SNR}\sim 10^{42}$~erg/s, they can be responsible
  for Cosmic Ray production starting at the lowest energies, i.e.~$\ep\sim
  1$~GeV (blue filled area). X-ray binaries ($L_{XRB}\sim 10^{38}$~erg/s) and
  pulsars ($L_{pulsar}\sim 10^{37}$~erg/s) can only be responsible for the
  Cosmic Ray flux above $10^{4.5}$~GeV (XRBs, red hatched area, lines)
  resp.~$10^{5.8}$~GeV (pulsars, green hatched area, circles). Alternatively,
  supernova winds can be responsible for the contribution above the knee,
  since higher energies can be reached in winds as opposed to SN explosions
  into the interstellar medium.}
\label{cr_lumi}
}
\end{figure}

The Cosmic Ray luminosity  can be written as
\be
L_{CR}=\frac{4\,\pi}{c}\cdot j_E(E_{\min})\cdot\frac{V_{GD}}{\tau_R}
\ee
assuming the production of Cosmic Rays in the Milky Way with a residence time
$\tau_R\approx 6\cdot 10^{6}$~yr
in the volume of the galactic disk $V_{GD}\approx 10^{67}$~cm$^3$.  Here, 
\be
j_E(E_{\min})=\int_{E_{\min}} \frac{dN_{p}}{d\ep}\,\ep\,d\ep
\ee
is the Cosmic Ray energy flux. The Cosmic Ray luminosity depends on the minimum
energy which is produced.
The calculation can only be
valid at energies below the ankle, $E_{CR}<3\cdot 10^{18}$~eV: at the highest
energies\linebreak \mbox{($E_{CR}>3\cdot 10^{18}$~eV)}, the observed spectrum is too isotropic to be of galactic
origin. Figure~\ref{cr_lumi} shows the
Cosmic Ray luminosity versus minimum energy. Three potential source classes
are indicated as filled/hatched areas below the curve. In the following, the
different source classes will be reviewed with their potential contribution to
the Cosmic Ray luminosity. Here, it is assumed that the
luminosity in Cosmic Rays produced by a source class must be less than the
total electromagnetic output of the same class, $L_{CR}^{class}<L_{em}^{class}$.
%..............................................
\paragraph{Supernova remnants}
%..............................................
The CR spectrum at energies below the knee is commonly believed to be produced
by the shock fronts in expanding shells of supernova remnants (blue filled area). When a star
perishes in a supernova (SN) explosion, the emitted material encounters the
interstellar medium (ISM), building a shock front of typical velocities $V_{S}\sim
10^{5}$~m/s. With the typical rate of SN explosions in a galaxy,
$dN_{SN}/dt\sim 3/(100\mbox{ yr})$, and a mean ejected mass of $M_{SN}\sim
10\,M_{\odot}$ per SN, the supernova remnant's
shock front is active for about $\Delta t_{SNR}\sim 1000$~yr. The luminosity of a single
SNR, $L_{SNR}\sim 10^{41}$~erg/s, can then be converted into the total
luminosity of SNRs in the Milky Way,
\begin{equation}
L_{tot}^{SNR}\approx \frac{dN_{SN}}{dt}\cdot L_{SNR}\cdot \Delta
t_{SNR}\approx 3\cdot 10^{42}\mbox{ erg/s}\,.
\end{equation}
Given the integral luminosity required for the production of Cosmic
Rays at $\ep>10^{9}$~eV, 
\begin{equation}
L_{CR}(\ep>1\mbox{ GeV})\approx 10^{41}\mbox{ erg/s}\,,
\end{equation}
SNRs are good candidates for the production of Cosmic Rays. It is, however,
difficult to explain the break in the spectrum at $\ep\sim 10^{15}$~eV. One
possibility is that leakage of particles out of the Milky Way becomes important,
leaving only heavy elements at the higher energies. This can lead to a
steepening of the spectrum. Another possibility is that SN explosions into
their own winds can be able to accelerate particles to higher energies, since
higher mass numbers (Helium up to iron) are produced. SN explosions of type Ib and Ic lose
their hydrogen (for Ic also the helium) envelope before collapsing. This leads
to a higher density of particles when the shock forms and thus to different shock
conditions. Regular SNRs can in this scenario produce CRs up to the knee and
SNR-Winds are responsible for the spectrum between the knee and the ankle.
%..............................................
\paragraph{Pulsars, X-ray binaries and microquasars}
%..............................................
As an alternative explanation for the contribution above the knee, systems
including neutron stars or black holes are considered.

Neutron stars can be observed due to their emission of electromagnetic
radiation along the magnetic field axis. Since the rotational axis of the
objects does not align with the magnetic field axis, they are observed as
pulsars: the emission is only seen when the particle jet points towards
Earth. Pulsars have periodic signals ranging from several seconds down to
milliseconds. The Crab as the most prominent, since most luminous, example is a millisecond pulsar. It is a neutron
star which was presumably produced in a SN explosion observed on July 04,
1054~\footnote{For a summary of historical SNe including SN~1054, see~\cite{sn1054_crab}.}. The
 pulsar wind nebula has
been observed at all wavelengths. From radio up to X-ray energies, the
supernova remnant is seen, while the pulsar itself is visible at X-ray and
higher energies. The observed TeV signal from the Crab is non-thermal,
an indication for particle acceleration in a shock environment, see e.g.~\cite{hess_crab}. The main reason why pulsars are good candidates for
particle acceleration are the very high magnetic fields of around $B\sim
10^{12}$~G. Pulsars have spin-down luminosities around $10^{37}$~erg/s and can
therefore not be responsible for the CR flux below the knee. They can, however,
contribute to the region between knee and ankle (see
Fig.~\ref{cr_lumi}, green hatched area, circles).

Magnetars, which are observed as Anomalous X-ray Pulsars
(AXPs) or Soft Gamma Repeaters
(SGRs), represent pulsars with even higher
magnetic fields of $B\sim 10^{15}$~G. For a summary on AXPs and SGRs see
e.g.~\cite{woods}. Five SGR candidates have
been observed in the Milky Way so far. These reveal themselves by randomly emitting
$\gamma$ radiation from time to time. The smaller eruptions have usually
thermal spectra, while giant bursts of non-thermal emission are observed from
time to time. Famous events are the outbursts of SGR~1806-20
on January 7, 1979 and on December 27, 2004 as well as the giant emission of
SGR~1900+14 on August 27, 1998.  Flare luminosities of SGRs
range from $\sim 10^{38}$~erg/s up to $\sim 10^{44}$~erg/s. AXPs show similar
phenomena, but with emission at X-ray energies rather than $\gamma$ rays. The
original definition of AXPs was the steady emission of X-rays. About half of
the AXPs today are
known to be variable in X-rays, though. Eight
AXP candidates have been observed in the Milky Way so far. The eruptions from magnetars are believed to come from
star quakes, exciting the surface of the magnetar which leads to the emission
of high-energy radiation. The energy output from AXPs lies at around
$10^{33}-10^{36}$~erg/s.

Binary systems including a neutron star or a black hole are good candidates
for shock acceleration as well. X-ray binaries consist of a compact object and
companion star.
In the case of Low-Mass X-ray Binaries (LMXBs), the non-degenerate star has
about a few solar masses. High Mass X-ray
binaries (HMXBs) are fed by a blue star (O/B star). 
Once the companion exceeds the
Roche volume of the binary system, it starts to feed the compact object with
matter. For HMXBs, accretion can also occur through accretion from stellar
winds. The neutron star or black hole in turn emits the gained energy  in form
of X-rays and sometimes in a jet along the magnetic axis. Such systems can lead to particle
acceleration up to the ankle at most. The typical electromagnetic energy of
such X-ray binary systems is $\sim 10^{38}$~erg/s. This is much lower than the
total energy required to explain the CR luminosity for
$\ep>10^{10}$~eV. But considering only events at higher energies,
i.e.~\mbox{$\ep>10^{13.5}$~eV} leads to a much lower CR luminosity due to the steeply
falling spectrum (see Fig.~\ref{cr_lumi}, red hatched area, lines). For a
summary of the XRB and CR connection, see e.g.~\cite{gaisser}.

A special case of an X-ray binary is the so-called microquasar, which are
X-ray binary systems with photon emission along two co-linear jets. 
 Such a phenomenon has for instance been observed in 
  GRS~1915+105~\cite{grs1915}. It is believed to be caused by instabilities
  in the accretion disk of the system. A burst is caused every time a
  particularly large amount of matter is accreted from the accompanying
  star. The development of the burst can be traced by the observation of the
  electromagnetic emission. Enhanced X-ray emission is seen close to the
  accretion disk, and radio to optical emission can be observed along the jet. For
  a recent summary of microquasar physics, see~\cite{mirabel_microq}.

For binary pulsar systems, a heavy
Be-star or O star is accompanied by a neutron star. Periodic emission of
high-energy photons can be observed in this case: Be-stars are massive stars with a circumstellar disk. The
path of the neutron star around the Be-star is highly elliptic. This can lead to
the accretion of matter by the neutron star from the massive star only near
the periastron, the point of closest distance of the two stars as discussed in~\cite{mirabel2006}. A periodic emission of TeV photons was
observed from three systems in the Milky Way, LS~5039 and PSR~B1259-63 were detected by
the {\sc H.E.S.S.} experiment~\cite{hess_ls5039,hess_psr}, while LS~I~61+303 was seen by
{\sc MAGIC}~\cite{magic_lsi61}. 
%\clearpage
%==============================================
\subsection{High-energy neutrinos from astrophysical sources \label{general:nu_models}}
\begin{figure}
\centering{
%Fig from Muecke et al, astro-ph/9903478
%\epsfig{file=figs/fig12c.eps,width=\linewidth}
\epsfig{file=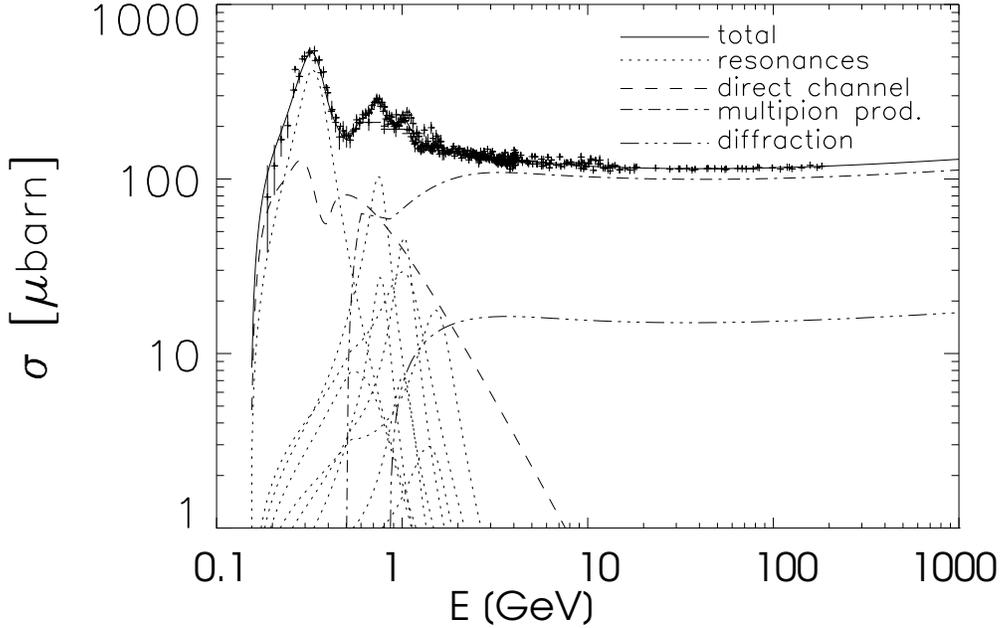,width=\linewidth}
}
\vspace{0cm}
\caption{Cross section for proton photon interactions in the center of mass
  system. Dotted lines represent baryonic resonances. Direct pion production
  processes are shown as the dashed line, while non-diffractive and
  diffractive multipion production are
  shown by the dash-dotted, upper and the dash-three-dotted, lower line,
  respectively. More information on the comparison between the data and the
  model is given in~\cite{sophia2000}. Reprinted from~\cite{sophia2000}, with permission from Elsevier.\label{multipion_cross_sec}}
\end{figure}

%==============================================
Neutrinos are produced in astrophysical shock fronts in proton-photon and/or
proton-proton interactions via
pion-production. The dominant channels are
\begin{eqnarray}
p\, \gamma &\longrightarrow&
\Delta^{+}\longrightarrow
\left\{
\begin{array}{lll}
p\,\pi^{0} &&\mbox{, fraction }2/3 \\ n\,  \pi^{+}&&\mbox{, fraction }1/3
\end{array}\right. \\
p\, p &\longrightarrow&\left\{
\begin{array}{lll}
p\, p\,  \pi^{0}&&\mbox{, fraction }2/3\\
p\,n\,\pi^{+}&&\mbox{, fraction }1/3\,.\end{array}\right.
\end{eqnarray}
The same processes occur for incident neutrons instead of protons, leading to
the production of $\pi^{-}$ particles. At higher energies, kaons can also
contribute to the spectrum~\cite{rachen00}.
Higher order processes are usually referred to as multipion production
processes. The total cross section for proton photon interactions in the
center of mass system is shown in Fig.~\ref{multipion_cross_sec}.

While the resulting neutrons are likely to interact before decaying, charged
pions decay and produce neutrinos,
\begin{eqnarray}
\pi^{+}\rightarrow\mu^{+}\,\nu_{\mu}&\rightarrow&
e^{+}\,\nu_{e}\,\overline{\nu}_{\mu}\,\nu_{\mu}\\
\pi^{-}\rightarrow\mu^{-}\,\overline{\nu}_{\mu}&\rightarrow& e^{-}\,\overline{\nu}_{e}\,\nu_{\mu}\,\overline{\nu}_{\mu}\,.
\end{eqnarray}

\subsubsection{Neutrino oscillations}
Assuming that pions of negative and positive charge occur equally, the
production flavor ratio of neutrinos at the source is
\begin{equation}
(\nu_{e}:\nu_{\mu}:\nu_{\tau})=(\overline{\nu}_{e}:\overline{\nu}_{\mu}:\overline{\nu}_{\tau})=(1:2:0)\,.
\end{equation}
This implies that tau neutrinos are not produced in astrophysical
sources. However, since neutrinos have non-vanishing mass eigenvalues, they
oscillate on their way to Earth. Neutrino flavor-eigenstates $|\nu_{\alpha}>$, $\alpha=e,\mu,\tau$, are connected to
neutrino mass-eigenstates $|\nu_{j}>$, $j=1,2,3$ via
\begin{equation}
|\nu_{\alpha}(t)>=\sum_{j=1,2,3} U_{\alpha j}\,\exp(-i\,E_{j} t)|\nu_{j}>\,.
\end{equation}
Here, $E_{j}$ is the eigenvalue for the Hamiltonian operator in vacuum, and
thus the energy of the mass-eigenstate. The mixing matrix $(U_{\alpha j})$
depends on three angles, $\theta_{12}, \theta_{13},$~and~$\theta_{23}$, and a
phase $\delta$. The matrix is
given as
\begin{equation}
U=
\left(
\begin{array}{lll}
c_{12}\,c_{13} & s_{12}\,c_{13}&s_{13}\,e^{-i\delta}\\
-s_{12}\,c_{23}-c_{12}\,s_{23}\,s_{13}\,e^{i\delta}&c_{12}\,c_{23}-s_{12}s_{23}s_{13}e^{i\delta}&s_{23}\,c_{13}\\
s_{12}\,s_{23}-c_{12}\,c_{23}\,s_{13}e^{i\delta}&-c_{12}\,s_{23}-s_{12}\,c_{23}\,s_{13}e^{i\delta}&c_{23}\,c_{13}
\end{array}
\right)\,.
\end{equation}
Here, $c_{jk}:=\cos(\theta_{jk})$ and $s_{jk}:=\sin(\theta_{jk})$, $j=1,2$
and $k=1,2,3$.
The probability for a neutrino to oscillate from a flavor state $\alpha$ to a
flavor state $\beta$ in a time $t$ starting from the emission of the neutrino
at the source, $t=0$, is given as
\begin{eqnarray}
P_{\nu_{\alpha}\rightarrow
  \nu_{\beta}}&=&\left|<\nu_{\beta}(t)|\nu_{\alpha}(t=0)>\right|^{2}\\
&=&\delta_{\alpha\beta}-4\,\sum_{j>i}U_{\alpha i}\,U_{\beta i}\,U_{\alpha
  j}\,U_{\beta j}\,\sin^2\left(\frac{\delta m_{ij}^{2}\,L}{4\,\en} \right)
\end{eqnarray}
for units of $c=\hbar=1$. Here, $\delta m_{ij}^{2}$ is the difference of the
squared masses of two flavors $i$ and $j$. 
While solar neutrino detection basically provides measurements for
$\theta_{12}$, i.e.~the oscillation from electron neutrinos to muon neutrinos,
atmospheric neutrinos provide $\theta_{23}$, where muon neutrinos oscillate
into tau neutrinos. The third angle $\theta_{13}$ appears to be the smallest
and limits are provided by reactor experiments. The parameter set is
approximately given as~\cite{sno04,k2k}
\begin{eqnarray}
\theta_{12}&\approx&\pi/6\\
\theta_{23}&\approx& \pi/4\\
\theta_{13}&\approx& 0\,.
\end{eqnarray}
Thus, the mixing matrix can be written as
\begin{equation}
U=\left(
\begin{array}{lll}
\frac{\sqrt{3}}{2}&\frac{1}{2}&0 \\
-\frac{1}{2\sqrt{2}}&\frac{\sqrt{3}}{2\sqrt{2}}&\frac{1}{\sqrt{2}}\\
\frac{1}{2\sqrt{2}}&-\frac{\sqrt{3}}{2\sqrt{2}}&\frac{1}{\sqrt{2}}
\end{array}
 \right)\,.
\end{equation}
%for flavor change and
%\begin{equation}
%P_{\nu_{\alpha}\rightarrow
%  \nu_{\alpha}}&=&1-\left|<\nu_{\beta}(t)|\nu_{\alpha}(t=0)>\right|^{2}\\
%\end{equation}
%for flavor conservation, using $\alpha\neq \beta$.

For a non-monochromatic neutrino beam, the probability has to be averaged over
the energy spectrum. Therefore, the $\sin$ term in the oscillation probability
is averaged to $\sim 0.5$ for large distances $L$~\cite{learned_pakvasa1995}. Therefore, the oscillation
probability is independent of time for distances larger than the size of the
solar system. The probability matrix for a neutrino flavor vector of
$(\nu_{e}^{source},\nu_{\mu}^{source},\nu_{\tau}^{source})$ changing to a
flavor vector $(\nu_{e}^{Earth},\nu_{\mu}^{Earth},\nu_{\tau}^{Earth})$ is
given as
\begin{equation}
\left(\begin{array}{l}
\nu_{e}^{source}\\
\nu_{\mu}^{source}\\
\nu_{\tau}^{source}\\
\end{array}\right)=\frac{1}{18}\cdot 
\begin{pmatrix}
10&&&4&&&&4\\
4&&&7&&&&7\\
4&&&7&&&&7\\
\end{pmatrix}
\cdot \left(\begin{array}{l}
\nu_{e}^{Earth}\\
\nu_{\mu}^{Earth}\\
\nu_{\tau}^{Earth}\\
\end{array}\right)
\label{flavor_ratios:equ}
\end{equation}
%\begin{equation}
%<P_{\nu_{\alpha}\rightarrow
%  \nu_{\beta}}>=\frac{1}{\int_{E_{\min}}^{E_{\max}} \frac{dN_{\nu}}{dE_{\nu}}\,\en\,d\en}\int_{E_{\min}}^{E_{\max}}
%  P_{\nu_{\alpha}\rightarrow \nu_{\beta}}\,\frac{dN_{\nu}}{d\en}\cdot \en\,d\en\,.
%\end{equation}
%In the case of astrophysical high-energy neutrinos, the energy spectrum is
%close to $dN_{\nu}/d\en\propto {\en}^{-2}$. In the simplified view of a two-neutrino framework, the
%oscillation probability for a baseline $L$ is given as
%\begin{equation}
%P_{\nu_{\alpha}\rightarrow
%  \nu_{\beta}}=\sin^{2}(2\theta)\sin^{2}\left(\frac{\Delta m^2\,L}{4\,\en}\right)\,.
%\end{equation}
%Here, $\Delta m^2$ is the difference of the squared masses of the two flavors
%and $\theta$ is the mixing angle between the states. The minimum and maximum
%energy of the neutrino spectrum can be chosen as $E_{\min}\approx 10$~GeV and
%$E_{\max}\approx 10^{10}$~GeV. 
At the source, the flavor ratio is
\begin{equation}
\left(\begin{array}{l}\nu_{e}^{source}\\\nu_{\mu}^{source}\\\nu_{\tau}^{source}\end{array}\right)=\left(\begin{array}{l}1\\2\\0\end{array}\right)\,.
\end{equation}
Applying Equ.~(\ref{flavor_ratios:equ}), the flavor vector approaches
\begin{equation}
\left(\begin{array}{l}\nu_{e}^{Earth}\\\nu_{\mu}^{Earth}\\\nu_{\tau}^{Earth}\end{array}\right)=\left(\begin{array}{l}1\\1\\1\end{array}\right)
\end{equation}
for path lengths exceeding the size of the solar system.

Thus, equal numbers of electron, muon and tau neutrinos are expected to be
observed at Earth.
\subsubsection{Normalization of the spectrum}
Given that the existence of high-energetic protons in astrophysical
environments is bound to neutrino production, the observed flux of charged
Cosmic Rays can be used to estimate the neutrino intensity expected from the
sources of Cosmic Rays, see e.g.~\cite{francis_barcelona} and references therein. The Cosmic Ray energy density above the ankle
$j_E(E_{\min}=3\cdot 10^{18}$~eV$)=\int_{E_{\min}=3\cdot 10^{18}\mbox{\tiny eV}} dN_p/d\ep\,\ep\,d\ep$, is expected to
be proportional to the neutrino energy density,
\begin{equation}
j_E(E_{\min})=x_{\nu/p}\cdot \int \frac{dN_{\nu}}{d\en}\,\en\,d\en\,.
\end{equation}
The constant of proportionality $x_{\nu/p}$ depends on the optical thickness of
the source to proton-photon interactions and on the fraction of energy
transferred to the charged pion. The fraction of energy going into the neutron
does not contribute to neutrino production. The average energy which is transfered
from the proton to the neutrino derives from the fraction of energy going into
the charged pion,
$\left<x_{p\rightarrow\pi} \right>\approx 0.2$.
The four leptons resulting from the pion decay are assumed to carry an equal
amount of energy. Consequently, each neutrino which is produced carries 1/4 of
the pion's original energy. The relation between neutrino and proton energy is
therefore
$\en=\ep/20$.

Due to the second channel in which $\pi^{0}$ particles are produced, the
neutrino flux is expected to be accompanied by a high-energy photon flux:
\begin{equation}
\pi^{0}\rightarrow \gamma\,\gamma\,.
\end{equation}
The resulting photons are produced at $>$~TeV energies. Thus, optically thin
sources emit TeV photons in coincidence with high-energy neutrinos and the
energy densities are proportional:
\begin{equation}
\int_{E_{\min}} \frac{dN_{\gamma}}{d\eg}\,\eg\,d\eg=x_{\nu/\gamma}\cdot
\int_{E_{\min}} \frac{dN_{\nu}}{d\en}\,\en\,d\en\,.
\label{x_nu_gamma}
\end{equation}
The constant of proportionality $x_{\nu/\gamma}$ depends on the fraction of
energy going into pion production. For optically thin sources, in the case of
$p\,p$ interactions, $1/3$ of the proton energy goes into each pion flavor and
the energy in $\nu_{\mu}+\overline{\nu}_{\mu}$ corresponds to the energy in
photons, $x_{\nu/\gamma}\sim 1$. For $p\,\gamma$ interactions,
$x_{\nu/\gamma}\sim 1/4$. If the source of neutrino emission is optically
thick to TeV photons, the TeV signal avalanches to lower energies until it
reaches a level at which it can escape the source. Thus, sources of GeV to
sub-MeV emission can be sources of neutrino production as well, assuming an
optically thick environment. Equation~(\ref{x_nu_gamma}) holds with different
values for $x_{\nu/\gamma}$ and modified integration limits.

In conclusion, the same sources which have been described as the potential
origin of Cosmic Rays and which show signs of non-thermal photon emission
are selected when discussing neutrino emission. In the following sections, the
possibility of neutrino emission from galactic sources like supernova
remnants, pulsars and X-ray binaries will be mentioned, as well as the
predictions of neutrino production in the decay or interaction of exotic
particles like WIMPs or monopoles.
Furthermore, extragalactic neutrino emission, i.e.~neutrinos from AGN, starbursts and
GRBs is discussed. 
\clearpage
%==============================================================================
\section{High-energy neutrino detection
  methods\label{nu_detection_methods}}\vspace{-0.5cm}
%==============================================================================
When a neutrino $\nu_l=\nu_e,\,\nu_{\mu},\,\nu_{\tau}$ interacts with a nucleon $N$ via charged current interactions, a lepton $l=e,\,\mu,\,\tau$ is 
produced,
\begin{equation}
N\,\nu_{l}\rightarrow X\,l\,.
\end{equation}
Here, $X$ indicates the hadronic product of the interaction, leading to a
hadronic cascade.
Charged particles in a medium emit Cherenkov light if traveling faster than the speed of
light in the same medium. The Cherenkov effect is described in detail
in~\cite{jackson}. The blue Cherenkov light can be detected by photomultiplier
tubes and both the
  incidental direction and the neutrino energy are reconstructible. The
  signatures for electrons, muons and tauons are very different. Electrons and
  tauons produce electromagnetic and hadronic cascades within the
  detector. About 20\% of the total energy go into the hadronic cascade, which
  arises from the nuclear recoil. The remaining 80\% of the energy are carried
  by the electromagnetic cascade, produced by the charged lepton interacting
  with the electrons of the medium~\cite{spencer2004}. In contrast, muons only undergo radiation
  losses and leave track-like signatures. The track of the muon can be reconstructed, since the Cherenkov
  signal is emitted in form of a directed cone. Tauons have very distinct signatures. A first cascade is produced in the first $\nu_{\tau}\,N$
  interaction, where a $\tau$ and a hadronic cascade is produced. The $\tau$
  in turn decays, producing a second cascade in which a further $\tau$ is
  generated. Thus, tau neutrinos have the unique
  property of regenerating themselves.

In the case of neutral current interactions,
\begin{equation}
N\,\nu_{l}\rightarrow X\,\nu_{l}
\end{equation}
and a hadronic cascade $X$ can be observed.\vspace{-0.5cm}
\subsection{Signatures}
\vspace{-0.5cm}
The primary detection technique for Cherenkov arrays is the measurement of neutrino-induced muon
tracks in water or ice. To guarantee that the observed muons are neutrino-induced, the Earth is used as
  a filter. While muons produced in the atmosphere are absorbed by the Earth,
  neutrinos traverse the Earth and the signature is unique. After the filtering of
atmospheric muons, the remaining signal mainly consists of neutrinos produced in hadronic showers in the atmosphere. Various analysis methods have been developed to separate this background from 
a potential signal from extraterrestrial sources. In addition
to the neutrino-induced lepton, a cascade produced
at the interaction vertex contributes to the signal. 

While the muon leaves an
optical track, the cascade events produce a radio and an acoustic signal in
addition to the optical emission.
The Landau-Pomeranchuk-Migdal (LPM) effect becomes important  at the highest
energies, $\en>10^{18}$~eV, where the cross sections for electron
bremsstrahung and photon electron-positron pair production are reduced. As
deduced in~\cite{spencer2004}, the electromagnetic shower lengthens to about
$100$~m at $10^{20}$~eV. Most of the signal is radiated at the end of the
cascade for each of the three detection methods - optical, radio and acoustic.
This is why a two-component signal can be expected at the highest energies.
%The three different detection methods - optical,
%radio and acoustic detection - are reviewed here after having discussed the
%background of atmospheric neutrinos. 

In the case of tau neutrinos, instead of a single cascade,
  multiple cascades are produced due to the regeneration of the tau. With {\sc IceCube}, so-called double-bang
  signatures are expected, where two cascades are observed within the
  instrumented volume. If one of the cascades happens outside the instrumented
  volume, the signature is called lollipop. A single cascade is observed
  together with a muon bundle, so that the cascade finally looks like a
  lollipop. The lower energy detection threshold is given by the fact that the
  cascade-like events need to be larger than the spacing of the {\sc IceCube}
  strings, which is about $125$~m. The assumption that the extension of the
  cascade event needs to be larger than about $\sim 200$~m leads to a lower
  detection threshold of $E_{\min}\sim $~PeV, see~\cite{beacom2003}. 

The reproduction of tauons renders possible the signal propagation through the
Earth. Therefore, tau neutrino cascades are suitable for the investigation of
  horizontal or Earth skimming neutrino air showers. The cascades are reproduced within the
  Earth until the shower comes out of the Earth and can be detected by either
  an Imaging Air Cherenkov Telescope or an air-shower surface array. Regular
  hadronic or electromagnetic air showers do not traverse the high column
  density and the neutrino signature is unique. This technique is for instance
  used by the {\sc Auger} experiment~\cite{auger_neutrinos2007} in the case of
  surface arrays and by the {\sc MAGIC} experiment in the case of
  IACTs~\cite{magicnus_santafe2007}.
\vspace{-0.5cm}
%---------------------------------------------
\subsection{Atmospheric neutrino background}
%---------------------------------------------
The main background in the search for extraterrestrial neutrinos are atmospheric muons and neutrinos. When primary CRs interact in
the upper atmosphere, a shower develops. If the primary is a hadron, e.g.~a
proton, secondary mesons and muons are produced while the shower propagates
through the atmosphere. Secondary mesons like charged pions and kaons produce
neutrinos when decaying. The neutrino spectrum is close to an $E^{-3.7}$
behavior in the energy range of $10^{3}$~GeV$<\en<10^{6}$. This is about one power steeper than the primary spectrum, $\sim
E^{-2.7}$, since a considerable fraction of pions and kaons ($\tau\sim
10^{-8}$~s) interact again before decaying. Neutrinos from pion and kaon
decays are usually referred to as {\it conventional atmospheric neutrinos} as
opposed to {\it prompt atmospheric neutrinos} from heavier hadrons.
%.....................................................................
\subsubsection{Conventional atmospheric neutrinos} 
%.....................................................................
An analytic description of the
muon neutrino spectrum at $\en>100$~GeV is given in~\cite{volkova80} as
\begin{footnotesize}
\be
\frac{dN}{dE_\nu d\Omega}\bigg|_{\nu_\mu}\bigg.(E_\nu,\theta)=\left\{
\begin{array}{ll}
0.0285\cdot E_{\nu}^{-2.69}\cdot&\\
\left[\frac{1}{1+6E_{\nu}/E_{\pi}(\theta)}\right.&\\
\left.+\frac{0.213}{1+1.44E_{\nu}/E_{K^{\pm}}(\theta)} \right]& 100\leq\frac{E_{\nu}}{\mbox{GeV}}<5.4\cdot 10^5\\
&\\
0.48\cdot E_\nu^{-4.04}\cdot&\\
\left[E_{\pi}(\theta)+0.89E_{K^{\pm}}(\theta)\right]&E_\nu\geq5.4\cdot 10^5\mbox{ GeV} 
\end{array} 
\right. 
\ee
\end{footnotesize}
and the electron neutrino spectrum is given as
\begin{footnotesize}
\be
\frac{dN}{dE_\nu d\Omega}\bigg|_{\nu_e}\bigg.(E_\nu,\theta)=\left\{
\begin{array}{ll}
0.0024\cdot E_{\nu}^{-2.69}\cdot&\\[0.1cm]
\left[\frac{0.05}{1+1.5E_{\nu}/E_{K^{\pm}}(\theta)}\right.&\\[0.1cm]
+\frac{0.185}{1+1.44E_{\nu}/E_{K^0}(\theta)}&\\[0.1cm]
+\left.\frac{11.4E^{\zeta(\theta)}}{1+1.21E_{\nu}/E_{\pi}(\theta)}
\right]& 100\leq\frac{E_{\nu}}{\mbox{GeV}}<3.7\cdot 10^5\\[0.1cm]
&\\
0.0071\cdot E_\nu^{-4.045}\cdot&\\[0.1cm]
\left[E_{K^{\pm}}(\theta)+3.7E_{K^0}(\theta)\right]&E_\nu\geq3.7\cdot 10^5\mbox{ GeV}\,. 
\end{array} 
\right. 
\ee
\end{footnotesize}
Here, $E_\pi$, $E_{K^{\pm}}$ and $E_{K^{0}}$ are angle distribution
parameters, depending on the zenith angle $\theta$. The factor $\zeta$ is energy
dependent as 
\be
\zeta(\theta)=a(\theta)+b(\theta)\cdot \log (E_{\nu})\,. 
\ee
The values of $E_\pi$, $E_{K^{\pm}}$, $E_{K^0}$, $a(\theta)$ and $b(\theta)$
are listed in table~\ref{angleparam}. At lower energies,
$\en<100$~GeV, the description is given numerically by~\cite{honda95}.
\begin{table}[h]
\centering{
\begin{footnotesize}
\begin{tabular}{l|llllllll}
\hline
$\cos(\theta)$&1&0.6&0.4&0.3&0.2&0.1&0.05&0\\ \hline\hline
$E_\pi(\theta)$~[GeV]&121&202&298&392&572&886&1060&1190\\ 
$E_{K^{\pm}}(\theta)$~[GeV]&897&1500&2190&2900&4220&6540&7820&8760\\
$E_{K^0}(\theta)$~[GeV]&$194$&$324$&$473$&$628$&$915$&$1410$&$1690$&$1890$ \\
$a(\theta)$&$-1$&$-0.355$&$-0.687$&$-0.619$&$-0.384$&$-0.095$&$0.0$&$0.083$\\
$b(\theta)$&$0.0$&$-0.23$&$-0.01$&$-0.007$&$-0.09$&$-0.165$&$-0.186$&$-0.215$\\ \hline
\end{tabular}
\end{footnotesize}
\caption{Angle distribution parameters for the atmospheric neutrino flux~\cite{volkova80}.}
\label{angleparam}
}
\end{table}

The ratio of neutrinos to anti-neutrinos is energy dependent. In~\cite{wei} it is
given as
\be
\begin{array}{l}
R_{atm}(E_\nu)=\left(dN_{\nu}/(dE_\nu d\Omega)\right)_\nu/\left(dN_{\nu}/(dE_\nu
  d\Omega)\right)_{\overline{\nu}}=\\
\left\{\begin{array}{lll}
1&&E_\nu\leq 1\mbox{ GeV}\\
1+0.3\cdot\log\left(E_{\nu}/GeV\right)&&1\mbox{ GeV}<E_\nu\leq
  100\mbox{ GeV}\\
0.6+0.5\cdot \log\left(E_\nu/GeV \right)&&100\mbox{
  GeV}<E_\nu\leq 1000\mbox{ GeV}\\
1.2+0.3\cdot \log\left(E_\nu/GeV \right)&&E_{\nu}>1000\mbox{ GeV}\,.
\end{array}
\right.
\end{array}
\ee
The spectrum is
one power steeper than the spectral behavior of the primaries, since the
probability that pions and
kaons interact before decaying increases with the energy~\cite{gaisser}. The intensity of the spectrum increases with the angle of
the neutrino towards the zenith angle. The larger the angle, the lower the
density gradient of the atmosphere and therefore, more pion decays are
possible for horizontal events as opposed to vertical events. 
%.....................................................
\subsubsection{Prompt atmospheric neutrinos}
\begin{figure}
\centering{
\epsfig{file=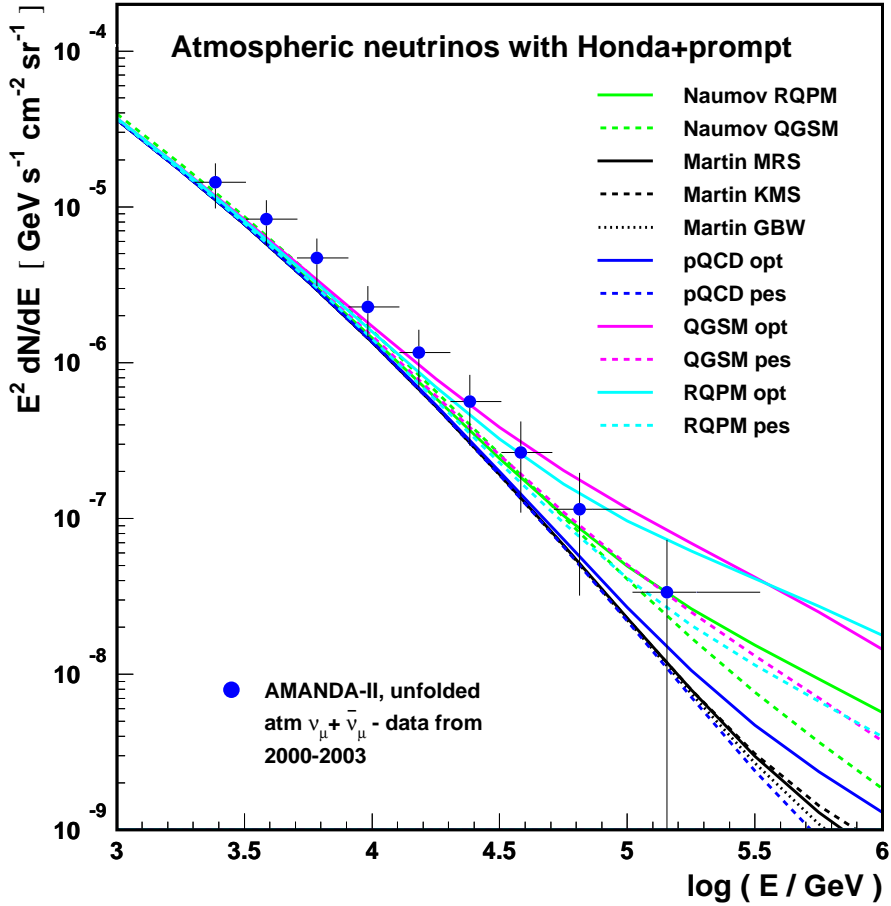,width=\linewidth}
}
\vspace{0cm}
\caption{Atmospheric neutrino flux models including the contribution from
  hadron decays with charm contents. The flux of conventional atmospheric
  neutrinos is given by Honda~\cite{honda_04}. The prompt neutrino flux
  predictions are models given in: Fiorentini et al.~\cite{naumov_RQPM_01,naumov_RQPM_89}
  ({\sl Naumov RQPM} and {\sl Naumov QGSM}); Martin et al.~\cite{martin_GBW}
  ({\sl Martin MRS},
  {\sl Martin KMS} and {\sl Martin GBW}); Costa~\cite{Costa} ({\sl pQCD opt},
  {\sl pQCD pes, QGSM
  opt, QGSM pes, RQPM opt, RQPM pes}). {\sl pQCD} models are determined using
  pertubative QCD in next to leading order. {\sl QGSM} models are half-empiric,
  using the {\bf q}uark {\bf g}luon {\bf s}tring {\bf m}odel pased on non-pertubative QCD
  calculations. The {\bf r}ecombination {\bf q}uark {\bf p}arton
  {\bf m}odel ({\sl RQPM}) is phenomenology-based and non-pertubative. Models marked {\sl
  opt} are optimistic parameterizations, {\sl pes} stands for pessimistic.  Figure courtesy Kirsten M\"unich, see also~\cite{kirsten_phd}.}
\label{honda_prompt_data_00_03}
\end{figure}

%.....................................................
When heavy hadrons with charm and beauty content are produced in the atmosphere, the
resulting neutrino spectrum appears flatter than the one of conventional
atmospheric neutrinos. Due to short lifetimes ( $\sim 10^{-12}$~s) these
hadrons do not interact before decaying for energies up to $\en<10^9$~GeV. Therefore,
the spectrum of the prompt neutrinos follows the primary index, $\propto
{\en}^{-2.7}$. The exact normalization and spectral behavior is difficult to
predict. In order to estimate the energy spectrum of neutrinos from hadrons
with charm contents, it is necessary to simulate interaction conditions at
extremely high energies ($\en>10^{5}$~GeV). Interaction models are based on
accelerator data reaching particle energies up to $\en\sim 10^{4}$~GeV at
most. The second challenge is the behavior of interactions at small
Bj\"orken$-x$. The Bj\"orken scale invariant $x$ is given as
\be
x=\frac{Q^2}{2\,M\,\nu}\,.
\ee 
Here, $Q^2$ is the negative four-momentum transfer, $M$ is the nucleon mass of
the interaction and $\nu$ is the energy transfer in the laboratory system, see
e.g.~\cite{schmitz_book,halzen_martin}. The
scale invariant is physically constrained to $0\leq x\leq 1$. Interactions
with small values of $x$ imply forward-scattered processes. Accelerator
experiments only yields measurements for $x\gtrsim 10^{-4}$, since the beam
pipe does not allow the detection of particles at smaller $x$. Therefore,
extrapolations to high energies and small $x$ are needed in order to calculate
the atmospheric charm flux. Figure~\ref{honda_prompt_data_00_03} shows the
atmospheric prediction including conventional neutrinos according
to~\cite{honda_04} together with different predictions of the prompt
atmospheric neutrino flux. At energies of around $\en\sim
10^{5}-10^{6}$~GeV, the prompt neutrino flux becomes stronger than the signal from
the conventional neutrinos. The uncertainties in the models are larger than
one order of magnitude. Large volume neutrino detectors can help to
determine the flux of prompt atmospheric neutrinos and thereby improve the
knowledge of particle interactions at the highest energies.

%---------------------------------------------
\subsection{General detection method}
%---------------------------------------------
An estimate of the detection rate of neutrino-induced muons/cascades $R$ can be given by folding the expected
neutrino flux at Earth with the probability of
the detection of the neutrino,
\begin{equation}
R(E_{\min},\theta)=\int_{E_{\min}}P_{\nu\rightarrow l}(E_{\nu},E_{\min})P_{shadow}(\theta,E_{\nu}) \frac{dN_{\nu}}{dE_{\nu}}dE_{\nu}\,.
\end{equation}
Here, $P_{\nu\rightarrow l}(E_{\nu},E_{\min})$ is the probability that a
neutrino interacts with a nucleus to produce either a muon - $l=\mu$ - or an
electromagnetic or hadronic
cascade - $l=$~cascade - which is detectable in a large volume neutrino
detector. It can be written as
\begin{equation}
P_{\nu\rightarrow
  l}=N_A\,\int_{E_{\min}}^{E_{\nu}}dE_{l}\,\frac{d\sigma}{dE_{l}}r_{l}(E_l,E_{\min})\,,
\label{p_numu:equ}
\end{equation}
where $N_A$ is Avogadro's constant. The range of the produced lepton within
detection range $r_l$ is determined by the geometric detection volume. For the
case of muon neutrinos, the resulting muon can be observed even if produced
outside the detector itself.
The average muon energy loss rate in a medium is given as~\cite{ghs1995}
\begin{equation}
\left<\frac{dE_\mu}{dX} \right>=-\alpha(E_\mu)-\beta(E_\mu)\cdot E_\mu \,.
\end{equation}
The first term occurs due to ionization losses of the muon. Monte Carlo
simulations show that $\alpha$
can approximately be regarded as a constant with a value of $\alpha\approx 2\mbox{ MeV}/(\mbox{g cm}^2)$ in media like rock.
The second term is due to bremsstrahlung, pair production and nuclear
interactions. The parameter $\beta$ is determined via Monte Carlo methods, with a value
which is approximately constant in a given medium. For  rock-like media,
\mbox{$\beta\approx 4\cdot 10^{-6}/(\mbox{g cm}^2)$}. For a discussion of
$\alpha$ and $\beta$ in the context of muon tracking in a medium using Monte
Carlo methods, see~\cite{mmc}.

The muon range is then given by the integral of the inverse of muon loss rate over the muon energy:
\begin{eqnarray}
r_{\mu}&=&\int_{E_{\mu}^{\min}}^{E_\mu} \frac{1}{\left<dE/dX \right>}
dE=-\int_{E_{\mu}^{\min}}^{E_\mu}\frac{1}{a+b\cdot E} \\
&=&\frac{1}{b}\log \frac{a/b+E_\mu}{a/b+E_{\mu}^{\min}}\,.
\end{eqnarray}

The differential cross section $d\sigma/dE_{l}$ includes charged and neutral
current interactions for cascade events,
$d\sigma/dE_{casc}=d\sigma/dE_{CC}+\sigma/dE_{NC}$. In the case of a muon
signal, only the charged current cross
section contributes, $d\sigma/dE_{casc}=d\sigma/dE_{CC}$. The energy threshold
of the detector is given as  $E_{\min}$. The cross section is determined by using the parton distribution
functions given by \cite{pdflib}. For the calculations in this review, the model of \cite{grv} is chosen.

The probability that a neutrino is absorbed by the Earth due to
neutrino-nucleon charged and neutral current interactions is given by the
shadow factor,
\be
P_{shadow}(X)=\exp(-X(\theta)/\lambda)\,,
\ee
in which $X$ is the column depth in units of cmwe and $\lambda$ is the mean path that the particle can survive without
interacting. The mean free path can be written in terms of the total cross section
$\sigma_{tot}$ as follows
\be
\lambda=\frac{1}{N_A\, \sigma_{tot}}\,.
\ee
Subsequently, the shadow factor can be expressed as
\begin{equation}
P_{shadow}(X)=\exp(-N_A\, \sigma_{tot} \, X)\,.
\end{equation}  
The column depth depends on the traveling distance
through Earth. It is also dependent on the Earth's density layers it has to
pass.

\begin{figure}
\centering{
\epsfig{file=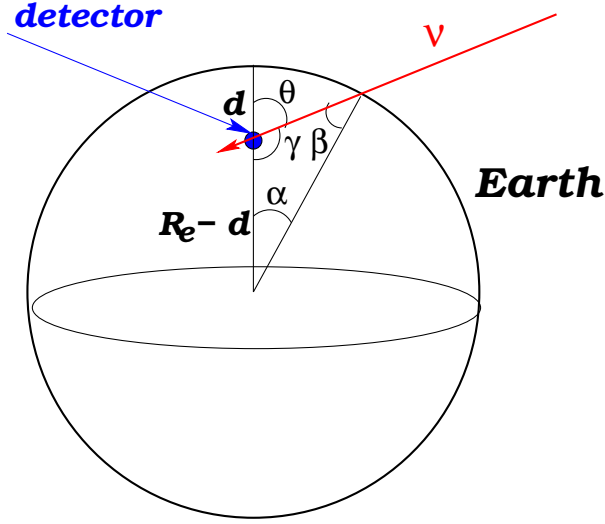,width=8cm}
\caption{Traveling distance of a neutrino through Earth.\label{earth_sdist}}
}
\vspace{0cm}
\end{figure}
In order to calculate the absorption probability the neutrino's path through the detector needs to be calculated.
Assuming a neutrino is entering the detector under a zenith angle $\theta$ (see
figure~\ref{earth_sdist}) and applying the sine theorem to the
triangle including $\gamma:=180^\circ-\theta$, $\beta$ and $\alpha$,
the relation
\begin{equation}
\frac{\sin \gamma}{\sin \beta}=\frac{R_{\oplus}}{R_{\oplus}-d}
\label{geometrie}
\end{equation}
is found. $R_{\oplus}$ is the radius of the Earth while $d$ is the depth at which
the detector is located under the Earth's surface. The angle $\alpha$ is determined as $\alpha=180^{\circ}-\gamma-\beta$. Applying the sine theorem
once more and using the equation~(\ref{geometrie}) with the approximation $d<<R_{\oplus}$, the distance $d_\nu$ which the
neutrino has to travel through Earth is given as\footnote{Note that this is only the case for
  $\sin \gamma\neq 0$. For $\sin \gamma=0$ and
  $\theta=0^\circ$, the traveling distance of the neutrino
  through Earth is equal to the depth of the detector under the
  Earth's surface $(d)$: $d_\nu(\theta=0)=d$. If, on the other hand, $\sin \gamma=0$ and
  $\theta=180^\circ$, the traveling distance is $d_\nu(\theta)=2\cdot
  R_{\oplus}-d$ since the neutrino arrives from the opposite direction.}
\begin{equation}
d_\nu=\frac{\sin\alpha}{\sin \gamma}R_{\oplus}\,.
\end{equation}

The column depth $X(\theta)$ is the product of the density
  of the Earth, $\rho_e$, with the traveling distance of the neutrino
  through Earth:
\begin{equation}
X(\theta)=\rho_e(\theta)\cdot d_\nu(\theta)\,.
\end{equation}

Depending on the zenith angle, the neutrino has to travel through different Earth layers which have different densities. A
model of the density profile of the Earth is given in~\cite{earthmodel},
\begin{footnotesize}
\begin{equation}
\rho(r) [g/cm^3]=\left\{ 
\begin{array}{lll}
13.0885-8.8381 \cdot x&&x<0.192\\[0.1cm]
12.5815-1.2638\cdot x-3.6426\cdot x^2-&\\
-5.5281\cdot x^3 & &0.192<x<0.546\\[0.1cm]
7.9565-6.4761\cdot x+5.5283\cdot x^2-&\\
-3.0807\cdot x^3&&0.546<x<0.895\\[0.1cm]
5.3197-1.4836\cdot x&&0.895<x<0.906\\[0.1cm]
11.2494-8.0298\cdot x&&0.906<x<0.937\\[0.1cm]
7.1089-3.8045\cdot x &&0.937<x<0.965\\[0.1cm]
2.691+0.6924\cdot x&&0.965<x<0.996\\[0.1cm]
2.9&&0.996<x<0.998\\[0.1cm]
2.6&&0.998<x<0.999\\[0.1cm]
1.02&& x\le 1\,.
\end{array}
\right.
\label{dens_profile}
\end{equation}
\end{footnotesize}
Here, $x$ is given as $x:=r/R_{\oplus}$ with $R_{\oplus}=6378$~km as the radius of the Earth.

\begin{figure}[h!]
\centering{
\epsfig{file=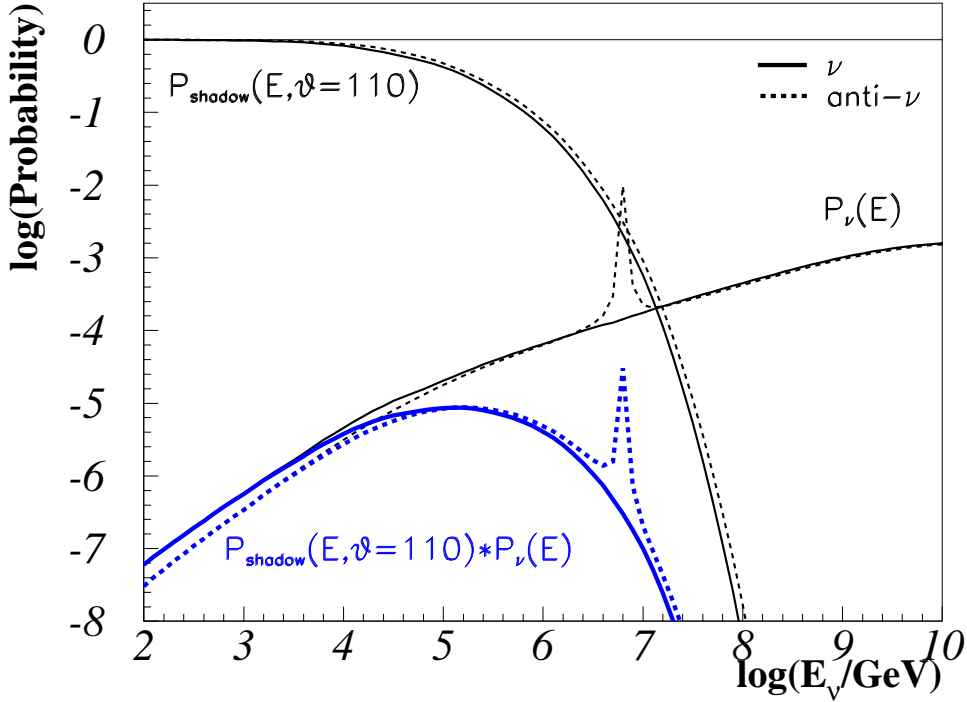,width=\linewidth}
\caption[Probability functions]{Probabilities as a function of energy for the
  case of electron neutrinos. The
  shadow factor is shown for a nadir angle of $\theta=110\deg$. The nadir
  angle is measured from the South Pole towards northern directions.
Higher inclination
  angles result in the shift of the cutoff towards lower energies. The
  interaction probability increases with energy. For anti-electron neutrinos,
  the Glashow resonance produces a peak at $10^{6.7}$~GeV. The product of the
  two probabilities (thick lines) is a measure for the detection probabilities (see text).}
\label{probabilities}
}
\end{figure}

The energy dependence of the shadow factor $P_{shadow}$ and interaction
probability is shown in Fig.~\ref{probabilities} for the case of electron
neutrinos. Solid lines represent neutrinos and dashed lines show
anti-neutrinos. The shadow factor is close to unity for low energies, while
the interaction probability increases with energy. The product of both
determines folded with the neutrino spectrum gives the event rate for a given
flux. 

Three main signatures can be detected from neutrino-nucleon interactions. The
main detection method is the optical detection of the Cherenkov light. Detectors in water and ice have been and are being built. The main differences
between the detector performance in ice and water lie in the scattering and
absorption lengths in water and ice. Ice absorbs only little of the
light, water has better scattering properties. In water, there are two
additional background components,
bioluminescence from bacteria populations in the sea water, and light from $^{40}$K decay. 

Neutrino-induced cascades can also be detected at radio
wavelengths and as acoustic pulses. These cascades are detectable over
long distances and therefore, radio and acoustic arrays do not need to be
instrumented as densely as optical detectors. On the other hand, larger volumes
need to be deployed, since the radio and acoustic detection methods become
effective at high threshold energies, $E_{\min}\sim 10^{7}-10^{8}$~GeV, at
which the expected neutrino flux is already very low.
%---------------------------------------------
\subsection{Optical detection in ice and water}
%---------------------------------------------
%......................................................
\subsubsection{Ice}
%......................................................
The largest currently operating, completed high-energy neutrino detector is\linebreak[4]
  {\sc AMANDA-II}\footnote{{\bf A}ntarctic {\bf M}uon {\bf A}nd {\bf N}eutrino
    {\bf D}etector {\bf A}rray-{\bf II}}. The low neutrino-nucleon cross section demands the
  instrumentation of large volumes to achieve a significant rate of
  events. {\sc AMANDA} is located in the Antarctic
ice, between 1500 m and 2000 m below the surface and covers a geometric volume
of $\sim$~0.016~km$^3$. The detector consists of 19 
strings with a total of 677 PhotoMultiplier Tubes (PMTs). {\sc AMANDA} can detect
  neutrinos with energies of a lower threshold energy of $E_{\min}\geq 50$~GeV.

{\sc AMANDA}'s successor {\sc IceCube} is currently being built around the\linebreak
{\sc AMANDA} experiment in 1500~m to 2500~m depth. The final configuration will
include a geometric volume of $1$~km$^3$ with $70-80$ strings, carrying
60 PMTs each. During the past four polar seasons (each year between October
and February the following year), a total of 40 strings has already been
deployed (1 string in 04/05, 8 strings in 05/06 and 13 strings in 06/07, 18
strings in 07/08).As half a cubic-kilometer is already instrumented,
completion of the full cubic-kilometer is expected in 2011. Since the spacing between the PMTs
and also between the strings is larger than in {\sc AMANDA}, the threshold
energy is slightly higher, $E_{\min}\sim 100$~GeV. With the increased
volume, higher energy events are easier to identify (the light deposition and
the path lengths are
increased with increasing energy) and better sensitivities can be achieved. 
%......................................................
\subsubsection{Water}
%......................................................
The optical detection of neutrinos in water via the Cherenkov effect has already been
probed by lower energy detectors like {\sc (Super)Kamiokande} and {\sc SNO} as
discussed above. In the case of high-energy neutrinos, natural water
reservoirs are used for detection. The first generation array {\sc Baikal}
detects neutrino-induced cascades and also muons in Lake Baikal~\cite{baikal,baikal_ralf06}. The advantage of this
lake is that it is deep enough to deploy $\sim 80$~m long strings in
depths of $\sim 1100$~m and that
the detector can be reached in winter due to the thick ice layer covering the
lake. {\sc Baikal} has been taking data with the {\sc NT200} detector since
1998. In April 2005, the detector was enlarged to {\sc NT200+}, enclosing
$0.005$~km$^3$ of water. The sensitivity to a diffuse flux for all three
neutrino flavors lies around $\ds \sim
2\cdot 10^{-7}\,\diffunits$~\cite{baikal_limits2006}.

Currently, different experiments are under construction in the Mediterranean
sea. In the {\sc ANTARES}\footnote{{\bf A}stronomy with a {\bf N}eutrino {\bf T}elescope and {\bf A}byss environmental {\bf RES}earch}~\cite{antares_prop,antares}  project, $300$~m long strings are being
deployed up to a depth of $\sim 2500$~m~\cite{antares_prop,antares_web}. As of
December 2007, 10 lines were deployed and the first muon tracks could already
be identified. The final detector configuration will consist of 12 strings
with an instrumented volume of $\sim 0.03$~km$^3$. A second array is being
constructed near the coast of Italy, the {\sc NEMO}\footnote{{\bf NE}utrino
  {\bf M}editerranean {\bf O}bservatory}
detector~\cite{nemo,nemo_web}, at a depth of $\sim 3500$~m. Such a high depth
has two main advantages: the flux of atmospheric muons from above is lower in
the detector, and also the bioluminescence background is low at high depths. The first tower
was deployed in 2006. In a second phase, the deployment of one or more
additional towers are planned.
{\sc NESTOR}\footnote{{\bf N}eutrino {\bf E}xtended {\bf S}ubmarine {\bf
    T}elescope with {\bf O}ceanographic {\bf R}esearch} is a third experiment,
being deployed near the coast of Greece at a depth of
$3800$~m~\cite{nestor,nestor_web}. The first phase of the detector was
deployed in March 2003, and first results are presented in form of the
measurement of horizontal atmospheric muons~\cite{nestor}.

The primary aim of the three arrays described above is to explore different
sites and techniques for neutrino detection in sea water in order to build a
$1$~km$^3$ detector, choosing the optimal location and design, namely {\sc
  KM3NeT}~\cite{km3net}. This detector will be complementary to {\sc IceCube}, detecting
mainly muon neutrinos from the southern hemisphere. With both arrays, a view
of the complete sky is possible. The view on the southern hemisphere enables
{\sc KM3NeT} the observation of the galactic center, which is only restrictedly visible from
the South Pole. 
%.............................................
\subsubsection{Analysis strategies}
%.............................................
In order to keep the analysis blinded to avoid experimenters bias, analysis
cuts are optimized using off-source samples created by scrambling the right
ascension of events or excluding the time window of transient emissions under
investigation. Diffuse flux analyses are optimized
on a low energy sample, where the signal is expected to be negligible.

The search for steady point sources is possible in two ways: the selection of
data according to specific source candidates and the search for spatial
clustering in the data independent of any source location.
The mean sensitivity in the northern hemisphere to an ${\en}^{-2}$ neutrino flux
from a point source, $\ps={\en}^{2}\cdot \left.dN/dE\right|_{limit}$, given by the {\sc AMANDA} experiment is~\cite{5yrs}
\begin{equation}
\ps=5.9\cdot 10^{-8}\,\pointunits
\end{equation}
for 5 years of data taking~\cite{5yrs}.

\begin{figure}[h!]
\centering{
\includegraphics[width=\linewidth]{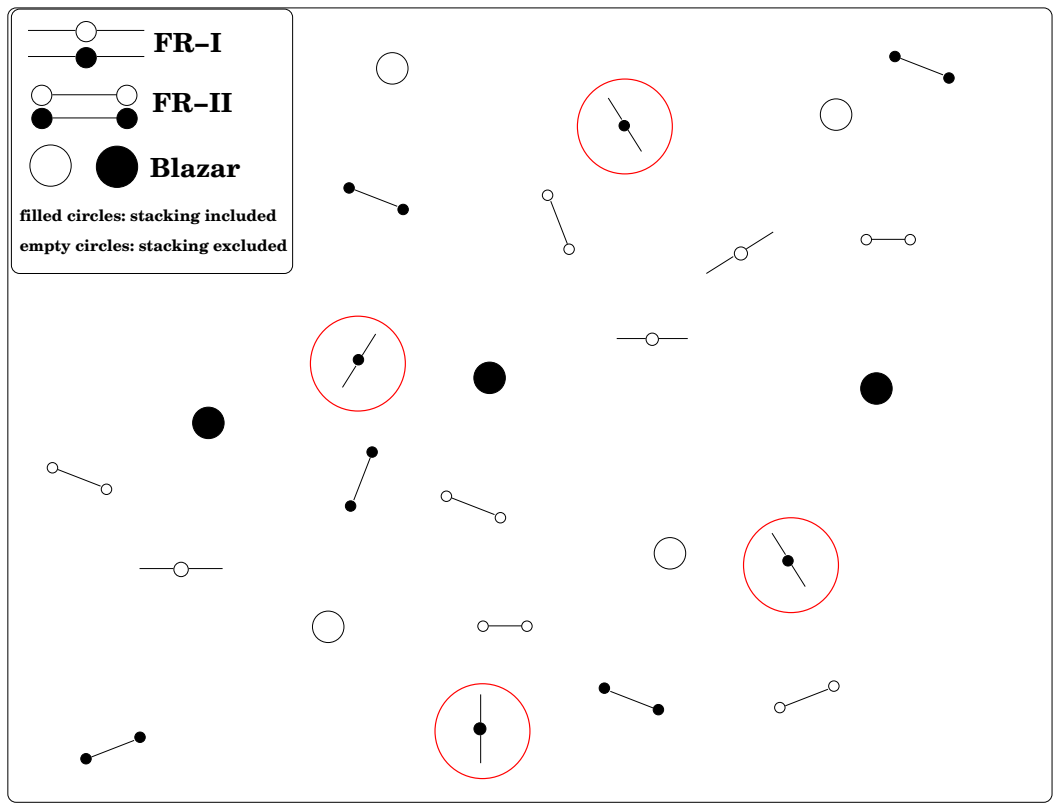}
\caption{Schematic figure of the principle of source stacking. Objects are
classified into source classes (here: FR-I and FR-II galaxies as well as
blazars). Under the signal hypothesis that photons in a certain bandpass are
directly correlated to neutrinos, the sources are selected according to their
strength in the corresponding bandpass. The most intense sources are then
stacked for an analysis of a neutrino signal from the source class. This is
indicated for the case of FR-I galaxies as circles around the objects. Only
the strongest sources are used in the stacking analysis (filled circles),
while weaker objects are not used (empty circles). Figure after~\cite{stacking_diffuse2007}.\label{stacking_diffuse_fig}}
}
\end{figure}

The search for single point sources can be complemented by stacking classes of
sources according to the direct correlation between the photon output and the
potential neutrino signal. This was done with {\sc AMANDA} data for 11 different AGN samples
that were selected at different wavelength bands, see~\cite{andreas}. The
optimum sensitivity was typically achieved by the stacking of the $\sim 10$
strongest sources in the catalog.  The principle of source stacking is indicated schematically in
  Fig.~\ref{stacking_diffuse_fig}. As an example, three source classes, FR-I
  and FR-II galaxies as well as blazars, are displayed. The stacking limits
  are obtained by stacking the most luminous sources of the same class in the
  sky, indicated by sources with filled circles. Weaker sources (empty
  circles) are not included. 

\begin{figure}
\includegraphics[width=14.5cm]{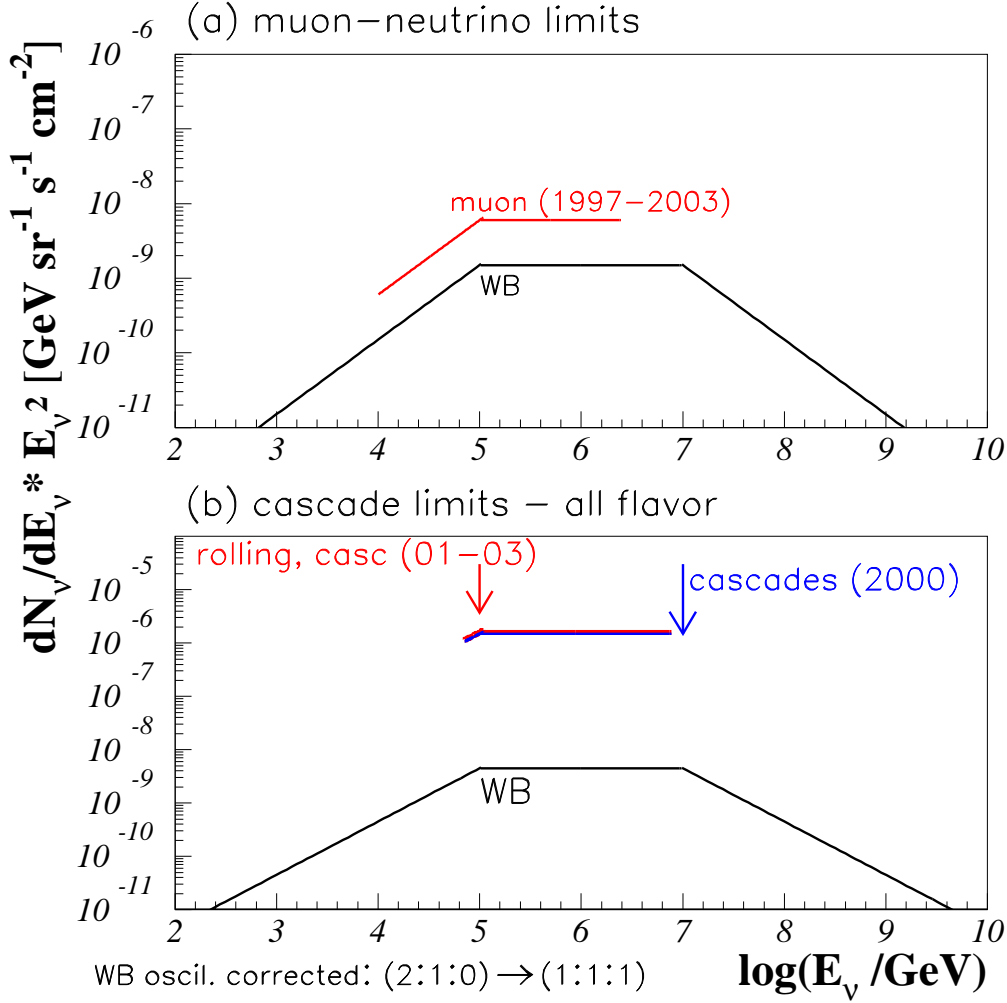}
\caption[GRB diffuse neutrino flux prediction and limits]{ Diffuse neutrino flux prediction by~\cite{wb97,wb99} and limits to the flux
prediction for different channels. (a) results based on the detection of
neutrino-induced muons~\cite{kyler_paper}; (b) cascade measurements~\cite{grbs_cascade07}.\label{grb_limits:fig}}
\end{figure}

The search for GRBs is almost background free, since the events are selected
according to the time and the direction of the GRB. All bursts in the northern hemisphere
during the operation time of {\sc AMANDA} have been 
analyzed by temporal and spatial selection in the 
data. This leaves a vanishing amount of atmospheric neutrinos in the sample. 
The background is estimated by choosing a 2 hour 
time window around the burst duration. A similar 
analysis was done for cascade events where also 
the southern hemisphere can be analyzed.
Additionally, {\sc AMANDA} searches for GRBs which are 
too weak to be detected in photons but have a 
potentially strong neutrino signal. Here, the data are 
analyzed according to density fluctuations in the 
temporal structure. No signal above the 
atmospheric neutrino background was detected in 
any of the searches. Limits on the neutrino flux from GRBs 
could be set for each of the searches as indicated in Fig.~\ref{grb_limits:fig}. A Waxman-Bahcall spectrum was assumed 
where each burst is parameterized the same 
way~\cite{wb97,wb99}. The theory of GRB neutrino emission is discussed in
detail in Section~\ref{nus_grbs}. The search for up-going muons gives limits on 
the muon neutrino flux (Fig.~\ref{grb_limits:fig}(a)),
 while the cascade analysis 
gives all-flavor limits (Fig.~\ref{grb_limits:fig}(b)).
Analyses are now underway 
which use individually 
parameterized spectra for 
each GRB. 
For more details on each of the GRB analysis methods of {\sc AMANDA}, see~\cite{icrc05_amanda}.
While the {\sc AMANDA} sensitivity is not high enough to reach the sensitivity of the
flux predictions, {\sc IceCube} is at the level of the diffuse signal
prediction. Also, for individual parameterizations of GRB spectra, the
strongest bursts would contribute with about $\sim1$~event/km$^3$. 

In the diffuse analysis, high-energy (HE) events from all directions are examined
with respect to the spectral energy behavior of the sample. A flattening of
the total neutrino spectrum at $\en\sim 10^{5}-10^{6}$~GeV is expected when a flat, astrophysical component
($dN_{\nu}/d\en\sim {\en}^{-2}$) overcomes the steep atmospheric background ($dN_{\nu}/d\en\sim
{\en}^{-3.7}$). The most restrictive limit $\dl={\en}^{2}\,dN_{\nu}/d\en$ is
given by the {\sc AMANDA} experiment, from the diffuse analysis for the years
2000 to 2003,
\begin{equation}
\dl=7.4\cdot 10^{-8}\,\diffunits
\end{equation}
in the energy range of $4.2<\log(\en/\mbox{GeV})<6.4$,
see~\cite{jess_diffuse}.
 
Diffuse neutrino flux limits are typically obtained by optimizing the analysis cuts on
${\en}^{-2}$ spectra. The dependency of the response function of the
detector to different spectra is considered in~\cite{jess_diffuse} and limits could be set for different
spectral shapes (e.g.~${\en}^{-3}$) or specific models.
Varying the spectral index in
the simulation shows that the event distribution simulated for {\sc AMANDA} peaks at
very different energies depending on the assumed spectral index. While, for an
${\en}^{-2}$ spectrum, $90\%$ of the signal lies between
$4.2<\log(\en/\mbox{GeV})<6.4$ for the analysis presented in~\cite{jess_diffuse}, an ${\en}^{-3}$ spectrum shows an
event distribution located about an order of magnitude lower in energy while
an ${\en}^{-1}$ spectrum shifts the sensitivity to higher energies. This shows
that it is useful and more accurate to model the spectra according to the
predicted shape. 
%---------------------------------------------
\subsection{Detection of extremely high-energy neutrinos}
%---------------------------------------------
Current neutrino experiments are able to measure the atmospheric neutrino
spectrum up to $\sim100$~TeV without observing a significant contribution from
extragalactic sources~\cite{jess_diffuse,kirsten_icrc07,kirsten_phd}. Successor experiments like {\sc IceCube} and {\sc KM3NeT} aim at
the detection of neutrinos at $100$~TeV up to $100$~PeV. The detection of a
signal at even higher energies is restrictedly also possible with {\sc IceCube}, but
to achieve a good detection probability of the cosmogenic neutrino
flux, new methods are being developed which are complementary to optical
detection. Acoustic and radio neutrino detection aims at the measurement of
neutrinos at Extremely High Energies (EHEs), i.e.\ $E_{\min}\sim
10^{8}$~GeV. In addition, horizontal air-showers or air-showers from rocks,
the sea or the Earth are neutrino signatures, since other particles are not
able to leave traces through such dense matter. 
Experiments like {\sc Auger}, {\sc HiRes}, {\sc MAGIC}, etc.~can
investigate the neutrino flux at energies around $\sim 10^8$~GeV by searching
for (sub-)horizontal events.
%...................................................
\subsubsection{Radio detection}
%...................................................
As it was discussed for the first time by Askaryan~\cite{askaryan62}, the
Cherenkov effect is not only present at optical and UV wavelengths, but has a
strong component in the radio. An electromagnetic
shower is caused by a neutrino, produced by interactions of the secondary
lepton with electrons from the molecules in the medium.
Positrons from pair production annihilate with electrons in the atoms which leads to a negative charge excess of approximately $30 \%$.
The shower-size is around $\sim 10$~cm and the power $P$ of Cherenkov
radiation is proportional to the
frequency $\nu$ and the bandwidth $\Delta \nu$:
\be
P \propto \nu \cdot \Delta \nu \,.
\ee
The refraction index for light in a medium changes back to $n \approx 1$ at frequencies above the UV light and the 
differential power decreases. Therefore, Cherenkov radiation is optically seen as blue light. The power is proportional to
the squared electrical field,
\be
P\sim \left|\vec{E} \right|^2
\ee
and for optical frequencies, $\left|\vec{E} \right| \sim \sqrt{N}$ with N as
the number of particles. Thus, the power is proportional
to $N$:
\be
P_{opt}\propto N\,.
\ee
For lower frequencies the corresponding wavelength is bigger than the size of the shower $\left(
\lambda \geq d\right)$ and the electromagnetic fields of the particles are in
phase. Consequently, the amount of the electric field scales with the number
of particles, $\left| \vec{E}\right|\propto N$. Therefore, 
\be
P_{rad}\propto \left|\vec{E} \right|^2 \propto  N^2 \,.
\ee
The ratio between radio power $P_{rad}$ and optical power $P_{opt}$ of the Cherenkov light is then
\be
\frac{P_{rad}}{P_{opt}}=N \left(\frac{\nu_{rad}}{\nu_{opt}}\right)^2 \,.
\ee
Typical frequencies are $\nu_{rad}=(100,\,500)$~MHz and $\nu_{opt}=75\cdot 
10^9$~MHz and thus,
\be
\frac{P_{rad}}{P_{opt}}\approx (1.8-44.4)\cdot 10^{-20}N \,.
\ee
A particle excess of $N\approx 10^{18}-10^{20}$ would be necessary to reach an
intensity equivalent to the optical intensity. However, lower efficiencies are
sufficient for the observation of the effect:
the Askaryan effect was observed for cascade-like events in
sand~\cite{david_sand}, salt~\cite{david_salt} and also in
ice~\cite{david_ice}. While cascade signals result in a significant radio
signal as detected,
single muons do not radiate enough to be observed.
It is used for the purpose of neutrino detection in the {\sc
  RICE}\footnote{{\bf R}adio {\bf I}ce {\bf C}herenkov {\bf E}xperiment} experiments which aims at the detection of the radio signal
from neutrino-induced cascades at energies of $(10^{15},\,10^{18})$~eV in the ice
below the geographic South Pole~\cite{rice}. A geometric volume of
$0.008$~km$^3$ is instrumented with dipole receivers at a depth of 100~m to 300~m
below the surface. The advantage of the high-energy threshold is that the
atmospheric background of muons and neutrinos can be neglected. The challenge
is the very small signal rate at such high energies. So far, no neutrino signal could be identified
yet. Most recent studies concerned the search for individual GRBs, with limits
to GRB afterglow neutrino emission models given in~\cite{rice_grbs07}. A second detector
trying to measure the radio signal from neutrino-induced cascades is the 
{\sc ANITA}\footnote{{\bf AN}tarctic {\bf I}mpulsive {\bf T}ransient {\bf
    A}ntenna Experiment} balloon-borne experiment~\cite{anita_limits06},
sensitive in the energy range of $(10^{18.5},\,10^{23.5})$~eV. Radio emission
arising from neutrino interactions with nucleons in the Antarctic ice sheet is
measured from above.
First limits
come from test-flights of {\sc ANITA-lite}, the first full {\sc ANITA} flight
was successfully done in December 2006, with 3.5 circles around Antarctica,
with no results published yet. {\sc ANITA} is
expected to investigate the model of GZK neutrinos, either confirming a
positive signal, or setting limits which will restrict source evolution
scenarios for the production of ultra high-energy protons, see
Section~\ref{gzk_nus:general}.

Next generation detectors of radio emission from neutrino-induced cascades are
planned in salt and in ice. In ice, the {\sc AURA}\footnote{{\bf A}skaryan
  {\bf U}nderice
  {\bf R}adio {\bf A}rray} experiment~\cite{aura} will be succeeding {\sc RICE}. The final aim is
the instrumentation of a surface of about $\sim 5-10$~km$^3$ in Antarctica. A new project using
radio antennas on top of the Ross Ice Shelf at the Antarctic coast is {\sc ARIANNA}\footnote{{\bf A}ntarctic {\bf R}oss
  {\bf I}ceshelf
  {\bf AN}tenna {\bf N}eutrino {\bf A}rray}~\cite{barwick_tev}, making use of
neutrino-induced radio emission in the ice shelf.
Salt is a good target material, since it has about
$2.3$ times the density of water and ice, and therefore also provides more than
2 times more nucleons within the same volume. The neutrino effective interaction
volume is hence increased. The {\sc SalSA}\footnote{{\bf Sal}t dome {\bf S}hower
  {\bf A}rray} experiment is
planned, where parts of a salt dome is proposed to be instrumented for the
purpose of neutrino detection. First tests are already being done in salt
domes~\cite{salsa05}.

Another method of detecting neutrino-induced radio emission is to observe the
moon with radio telescopes. The moon yields a good target for neutrino
interaction and therefore, an enhanced radio signal is expected from the
direction of the moon. Projects investigating this effect are planning to use
the {\sc  WSRT}\footnote{{\bf W}esterbork  {\bf S}ynthesis {\bf R}adio  {\bf T}elescope}
or {\sc SKA}\footnote{{\bf S}quare {\bf K}ilometre {\bf A}rray}. The latter is
planned to be used in the context of the {\sc LUNASKA}\footnote{{\bf L}unar
  {\bf U}ltra high-energy {\bf N}eutrino
  {\bf A}strophysics using {\bf SKA}} project~\cite{lunaska_web}. For an
overview of the status of neutrino radio detection methods, see e.g.~\cite{david_tev}.
%...................................................
\subsubsection{Acoustic detection}
%...................................................
Detectable acoustic signals are expected to accompany neutrino-induced
cascades. While cascades are compact and have high energy densities, which
produces the acoustic signal, single muons have too low energy density to be detected as acoustic
pulses. An estimate of the thermal energy deposit and the energy
densities is given in~\cite{learned1979}.
The  {\sc SAUND}\footnote{{\bf S}tudy of {\bf A}coustic {\bf
      U}ltra high-energy {\bf N}eutrino {\bf D}etection}~\cite{saund}
  experiment was the first array built for acoustic neutrino detection. It
  makes use of a military basis in the Bahamas, located between
  depths of $1570$~m and $1600$~m, spread over a surface of $\sim 250$~km$^2$. First results
  were presented in~\cite{saund}. The acoustic technique in the ocean is also
  tested at the {\sc ANTARES} site with the {\sc AMADEUS}\footnote{{\bf
      A}NTARES {\bf M}odules for {\bf A}coustic {\bf DE}tection {\bf U}nder
    the {\bf S}ea} detector, currently
  being deployed along with the {\sc ANTARES} strings~\cite{amadeus}. The main aim of these
  prototype detectors is the study of the background and general water
  properties. For the detection of an EeV neutrino signal, an extremely large
  array ($\sim 10$~km$^2$ surface area) needs to be deployed. In addition to water, ice and
  salt are good environments occurring naturally for the installation of a
  large volume acoustic neutrino detector. In the {\sc SPATS}\footnote{{\bf
      S}outh {\bf P}ole {\bf A}coustic {\bf T}est {\bf S}etup} project~\cite{boeser_arena,amanda_tev_proc}, first hydrophones were deployed on
  three {\sc IceCube} strings as of December 2007 in order to test the
  acoustic properties of the Antarctic ice. The final aim is again the
  construction of a $\sim 10$~km$^2$ array, in a
depth of $200-1500$~m, for the detection of EeV
  neutrinos. For a summary of the current status of acoustic neutrino
  detection arrays, see~\cite{vandenbroucke_acoustic}.\vspace{-0.5cm}
%...................................................
\subsubsection{Detection of neutrino-induced air showers}
%...................................................
The hadronic and electromagnetic cascades which are produced in
neutrino-nucleon interactions via the neutral and charged current can also be
observed as air showers, in particular in the case of tau
neutrinos. Neutrino-induced showers can be distinguished from regular Cosmic
Ray showers by choosing events which originate from close to or below the horizon. While
Cosmic Ray showers are absorbed by Earth, showers from {\it
  Earth-skimming neutrinos} are induced in the outer mantle of the Earth and
propagate further through the atmosphere. Neutrinos traveling through the core
of the Earth are attenuated and only nearly horizontal events can be observed
as inclined air showers, see e.g.~\cite{zas_nu_airshowers2005}. Searches for
neutrino-induced air showers are performed by the Cosmic Ray detectors,
e.g. by {\sc
  Auger}~\cite{auger_neutrinos2007} and  {\sc
  HiRes}~\cite{hires_neutrinos_leptonphoton2007}. A {\bf C}osmic {\bf R}ay
{\bf T}au {\bf N}eutrino {\bf T}elscope ({\sc CRTNT}), dedicated to the detection of
neutrino-induced air showers, is proposed
with possible locations in Tibet near the ARGO-YBJ array~\cite{crtnt_icrc2007}. Cherenkov telescopes with the primary aim of TeV photon showers
are also performing neutrino-shower analyses, see e.g.~{\sc
  MAGIC}~\cite{magic_nus_icrc2007}. The detection of the radio
signal from air-showers as planned by the {\sc LOFAR}\footnote{{\bf LO}w {\bf F}requency {\bf AR}ray} experiment~\cite{lofar} also gives
the possibility for neutrino detection with a threshold energy of
$\en>10^{22}$~eV~\cite{radio_neutrinos_icrc2007}. At these highest energies,
exotic particle decays, like topological defects, can be tested. The energy threshold is too high for
neutrino fluxes as produced in AGN or GRBs, since the maximum neutrino energy
lies below $10^{20}$~eV. The detection of neutrino-induced radio showers from
the moon has similar a detection threshold and is investigated by experiments
like {\sc FORTE}\footnote{{\bf F}ast {\bf O}n-orbit {\bf R}ecording of {\bf T}ransient
  {\bf E}vents}, {\sc GLUE}\footnote{{\bf G}oldstone {\bf L}unar {\bf
    U}ltra High-Energy Neutrino {\bf E}xperiment} and {\sc NuMoon}.
%%...................................................
\subsection{Hybrid arrays}
%%...................................................
The combination of different experimental techniques in one experiment or at
the same location has been proven to help
the understanding of the fundamental processes to be measured. The latest
example is the {\sc Auger} experiment which uses the complementary pieces of
information about the air shower from both fluorescence and surface
detectors. Events can be reconstructed with higher accuracy and systematic
errors can be reduced.  
A similar approach done with the combination of {\sc IceCube} and {\sc
  IceTop}~\cite{tom2006}. Surface stations are deployed together with the {\sc IceCube}
strings in order to measure the electromagnetic component of air showers at the surface in combination with
the high-energy muons in the air showers in {\sc IceCube}. High-energy muons
originate from the fragmentation region of hadronic showers, which is not well
known~\cite{tom2006}. The combination of a surface array with an underground
muon detector gives a good energy resolution, while the measurement of
low-energy muons at the surface provides a better mass resolution, see e.g.~\cite{tom2006}.

In the case of neutrino detection, the three techniques optical, acoustic and radio detection of neutrinos provide complementary information on the neutrino-induced showers.
Therefore, the best reconstruction of single neutrino events will be possible, if all three
detection methods are combined in a hybrid array. First simulations of a
combined array of acoustic, radio and optical array have been done for ice~\cite{acoustic_hybrid_icrc05}. Here, the {\sc
  IceCube} configuration is used together with a further outer ring of
strings for optical detection, as well as an acoustic and radio array with a
diameter of $10$~km around {\sc IceCube}. A scheme of the configuration and of
the effective volumes of the three arrays and the combined array is given
in~\cite{vandenbroucke_arena}.
\subsection{Current diffuse neutrino flux limits}
In order to optimize
the signal to noise ratio, it is useful to assume a certain spectral shape for
the potential extraterrestrial contribution. This way, specific cuts can be
applied in the data in order to reduce the background as much as possible
while keeping most of the signal. Extraterrestrial neutrinos are typically
believed to have a spectral behavior close to ${\en}^{-2}$, and thus, limits
are usually derived assuming an ${\en}^{-2}$
spectrum. Table~\ref{nu_limits:tab} lists current neutrino flux limits,
organized according to the lower value of the energy range.
All limits are given
for a $90\%$ confidence level, except for the {\sc RICE} results, where $95\%$
confidence level values are quoted. It is important to note that the different
experiments are sensitive to different neutrino flavors. It is typically
assumed that the flavor ratio for extraterrestrial neutrinos is
$(\nu_e:\nu_{\mu}:\nu_{\tau})=(1:1:1)$. Therefore, single-flavor limits can
generally be scaled to all-flavor limits by applying a factor of $3$. However, since
the flavor ratio of $1:1:1$ is not exact and can even change significantly at
energies above $1$~TeV~\cite{waxman_nuflavorratio2005}, the limits are quoted as given by the
different experiments, see references in the table. 
\begin{table}
\centering{
\begin{tabular}{ll|lll}
experiment&flavor&$E^2\,\dl$t&energy range&Ref\\ 
          &$(\nu+\overline{\nu})$&$10^{-7}\,\diffunits$&[GeV]&\\\hline\hline
{\sc Fr{\'e}jus}&$\nu_{\mu}$&$470$&$(1.5e3,\,1e4)$&\cite{frejus_limit,frejus_spectrum}\\
{\sc AMANDA}&$\nu_{\mu}$&$0.74$&$(1.6e4,\,2.5e6)$&\cite{jess_diffuse} \\
{\sc Baikal}&$\nu_{e}+\nu_{\mu}+\nu_{\tau}$&$8.1$&$(2e4,\,5e7)$&\cite{baikal_limits2006}\\
{\sc AMANDA}&$\nu_{\mu}$&$2.6$&$(1e5,\,3e5)$&\cite{kirsten_icrc05}\\
{\sc AMANDA}&$\nu_{e}+\nu_{\mu}+\nu_{\tau}$&$2.7$&$(2e5,\,1e9)$&\cite{lisa_3yrs}\\
{\sc RICE}&$\nu_{e}+\nu_{\mu}+\nu_{\tau}$&$10$&$(1e8,\,1e11)$&\cite{rice_limits2006}\\
{\sc
  Auger}&$\nu_{\tau}$&$1.3$&$(2e8,\,2e10)$&\cite{auger_neutrinos2007}\\
{\sc HiRes}&$\nu_{\tau}$&$4.2$&$(1e9,\,1e10)$&\cite{hires_neutrinos_leptonphoton2007}\\
{\sc ANITA-lite}&$\nu_{e}+\nu_{\mu}+\nu_{\tau}$&$16$&$(3e9,\,3e14)$&\cite{anita_limits06} \\
\end{tabular}
\caption{Limits to an ${\en}^{-2}-$shaped, diffuse neutrino flux. The results
  are organized according to the energy range.}
\label{nu_limits:tab}
}
\end{table}

Specific models can also be tested by changing the signal expectation from an
$\en^{-2}$ spectrum to a specific spectral shape. Such limits have been set by
e.g.~{\sc AMANDA} in the case of muon neutrinos~\cite{jess_diffuse} and all
flavors~\cite{lisa_3yrs} and {\sc Baikal}~\cite{baikal_limits2006}. It is also
possible to derive differential, model-\linebreak independent limits with methods as
described in~\cite{rice_limits2006}. These typically trace the energy-dependence of the
detector's sensitivity, giving best results at the maximum performance energy,
see e.g.~results from {\sc GLUE}~\cite{glue_nulimits2004}, {\sc
  FORTE}~\cite{forte_nulimit} and {\sc
  NuMoon}~\cite{radio_neutrinos_icrc2007}. 

During the past decade, neutrino limits could be improved from values of
$\dl=470\cdot 10^{-7}\,\diffunits$ at TeV energies ({\sc Fr{\'e}jus}) to four orders of magnitude lower values,
reaching limits as low as\linebreak $\dl=0.74\cdot 10^{-7}\,\diffunits$ at PeV energies ({\sc AMANDA}). With {\sc
  IceCube}, it will be possible to further improve this value by one order of
magnitude within only one year of full operation, and up to two orders of
magnitude in less than five years, giving the opportunity to detect the first
extraterrestrial neutrinos at $>$PeV energies.
\clearpage
%%%%%%%%%%%%%%%%%%%%%%%%%%%%%%%%%%%%%%%%%%%%%%%%%%%%%%%%%%%%%%%
\section{Galactic sources\label{galactic}}
%%%%%%%%%%%%%%%%%%%%%%%%%%%%%%%%%%%%%%%%%%%%%%%%%%%%%%%%%%%%%%%
The basic assumptions for neutrino production in galactic sources are the same
as for extragalactic neutrino sources: in the case of optically thin sources,
TeV photon emission can be directly correlated to neutrino emission in the case of a
hadronic acceleration scenario. In addition, the emission at lower energies
can be interpreted as an avalanched high-energy signal. One striking
difference between galactic and extragalactic neutrino spectra is the maximum
energy, since galactic sources produce maximum proton energies of \mbox{$E_p<3\cdot 10^{18}$~eV}
  which is mirrored in the neutrinos' relatively low maximum energy. 
Candidates for neutrino
  emission are the following source classes, which are also listed in table~\ref{galactic_nu_sources}.
\begin{itemize}
\item {\em Supernova Explosions}\\
Supernova Explosions release a large amount of their energy via the emission
of MeV neutrinos. This phenomenon was observed for the first time for
SN1987A. Today's neutrino detectors have created an early warning system for
such explosions in our galaxy, the {\bf S}uper{\bf N}ova {\bf E}arly {\bf
  W}arning {\bf S}ystem (SNEWS)~\cite{antonioli2004}. Since neutrinos are emitted prior to the
optical eruption, the SNEWS network can trigger optical, radio, X-ray and
other telescopes which would probably miss the explosion otherwise. On
average, a supernova explosion is believed to happen within our galaxy about
every thirty years.\\
Apart from the thermal neutrino emission prior to the optical outburst, a
TeV-neutrino flux is expected to arise from proton-nuclei interactions in Type-II supernova explosions
as described in~\cite{waxman_loeb2001}. Protons are believed to be accelerated
in shock-fronts up to proton energies of $\ep>20$~TeV, resulting in a neutrino flux
up to neutrino energies of $\en<10$~TeV. 
\item {\em Supernova remnants}\\
Young supernova remnants make up of the dominant part of the
charged Cosmic Ray spectrum~\cite{ginzburg1964}. Neutrinos are expected to be
produced in the shocks of both shell-type SNRs, see
e.g.~\cite{alvarez_muniz_halzen2002}, and plerions, see e.g.~\cite{guetta_amato2003}.\\
The observation of TeV emission from the shell
can be interpreted as a signal from $\pi^0-$decays. 
The spectral information from the shell-type 
SNRs and from plerions can be used to calculate the neutrino signal, see~\cite{kappes}. The event rates for the single sources are of the order of $1-10$
events in a cubic kilometer detector with a lower energy threshold of $E_{\min}=1$~TeV, at a rate of atmospheric neutrinos of the
same order or even higher. This implies that an effective background rejection
is necessary, possibly by performing a stacking analysis in order to observe a significant signal. 
\item {\em Microquasars}\\
Neutrino emission from jet-disk like microquasars was suggested\linebreak
in~\cite{bednarek_nus}. Here, it is proposed that nuclei being accelerated in
the jet of the system interact with the accretion disk of the compact object
or directly with the accompanying Wolf-Rayet star. While photons are absorbed
in the dense star or accretion disk, most neutrinos traverse the objects and
escape the production region without significant energy loss. The same kind of
processes can happen with B- and O-type stars instead of the Wolf-Rayet star.\\
The detection of a periodic TeV photon signal from the two microquasars LS~5039 and
LS~I~61+303 encourages models predicting high-energy neutrino emission from
massive star/compact companion-type binaries. If the TeV photons are produced in
photohadronic interactions, the neutrino signal is directly correlated to the
TeV photon signal as described in~\cite{torres_halzen_lsi}. A model presented
in~\cite{boettcher2006} predicts an anti-correlation between very high-energy
photons and neutrinos. While the region is optically thin in the high-state of high-energy
photon emission, it is optically thick in the low-state, allowing for the
production of high-energy neutrinos.
\item {\em Soft gamma repeaters and anomalous X-ray pulsars}\\
Most peculiar events in our galaxy are the so-called soft gamma repeaters or,
if the detection is in X-rays rather than in soft gamma-rays, the anomalous
X-ray pulsars. Their emission is irregular and does not follow a periodic
structure. In addition, the intensity of the flares varies
immensely. Low-intensity flares are typically thermal, while the most intense
flares have a non-thermal component. The latter are therefore good candidates
for neutrino emission. The neutrino emission from the giant flare of SGR~1806-20
was calculated in~\cite{francis_sgr}. This flare had a total energy
release of $\sim 10^{46}$~erg, which is several orders of magnitude above typical
flaring states. The non-thermal photon spectrum is extrapolated to TeV photon
energies in~\cite{francis_sgr}, assuming that the TeV signal is proportional
to the neutrino luminosity. A detection rate of a few events is expected in a cubic
kilometer array for such a giant flare.
\item {\em Molecular clouds and diffuse TeV emission}\\
 Molecular clouds have been observed by
  {\sc H.E.S.S.}~\cite{hess_clouds}. These are good candidates for neutrino emission: TeV
  emission is observed, since an astrophysical accelerator hides behind the
  clouds. The protons are absorbed by the dense clouds ($n_H\sim 250/$cm$^3$)
  and produce TeV photons and high-energy neutrinos through proton-proton and
  proton-photon interactions. Recently, \mbox{{\sc MILAGRO}}\footnote{{\bf M}ultiple {\bf I}nstitution {\bf L}os {\bf A}lamos {\bf G}amma
  {\bf R}ay {\bf O}bservatory}  detected diffuse TeV photon
  emission from the Cygnus region~\cite{goodman_tev}. This is a good candidate for a neutrino
  production site as well as pointed out in~\cite{francis_cygnus}.
\end{itemize}
So far, only the outer part of the galactic plane can be observed thoroughly,
since the most sensitive instruments, {\sc AMANDA} and {\sc IceCube} are
located in the southern hemisphere. Due to the detector location, the 
northern hemisphere is visible (see Section~\ref{nu_detection_methods}). With {\sc
  KM3NeT} to be built in the north, the southern hemisphere and
therefore galactic center will be
observable. The remaining challenge is the small maximum neutrino energy,
$\en<100$~TeV. For a dedicated summary of galactic neutrino sources,
see~\cite{bednarek_review}.
\begin{table}
\centering{
\begin{tabular}{l|llll}
source class&origin&normalization&$E_{\max}$&model(s)\\\hline\hline
SN explosion&&&&\\
shocks&$p\,n$&em.~luminosity&$10$~TeV&\cite{waxman_loeb2001}\\\hline
Shell-SNR&&&&\\
pulsars&$p\,\gamma,\,p\,p$&Cosmic Rays&$100$~TeV&\cite{berezinsky_smirnov1975,sato1977}\\
 diffuse&$p\,\gamma,\,p\,p$&TeV photons&$100$~TeV&\cite{kappes} \\
RX~J~1713.7-3946&$p\,\gamma,\,p\,p$&TeV photons&$10$~TeV&\cite{alvarez_muniz_halzen2002}\\\hline
Plerions&&&&\\
identified sources&$p\,p$&TeV photons&$100$~TeV&\cite{guetta_amato2003} \\
Crab&$n\,p$&TeV photons&$200$~TeV&\cite{bednarek_protheroe1997}\\\hline
Microquasars&&&&\\
jet&$p\,\gamma,\,p\,p$&jet-disk symbiosis&$100$~TeV&\cite{bednarek_nus}\\
LS~I~61+303&$p\,\gamma,\,p\,p$&TeV photons&$100$~TeV&\cite{torres_halzen_lsi,boettcher2006}\\\hline
SGRs\&AXPs&$p\,\gamma,\,p\,p$&TeV photons&$100$~TeV&\cite{francis_sgr}\\\hline
Cygnus-region&$p\,\gamma$&TeV photons&$100$~TeV&\cite{francis_cygnus}\\\hline
\end{tabular}
\caption{Examples for galactic sources of neutrino emission. Source classes
  for which neutrino emission is likely are listed. The origin of the neutrino
  flux is listed together with the electromagnetic or hadronic flux which is
  used to normalize the neutrino spectrum. In addition, the maximum neutrino
  energy is listed. The last column gives references of proposed models. A more detailed
  description can be found in e.g.~\cite{bednarek_review}.\label{galactic_nu_sources}} 
}
\end{table}

%%%%%%%%%%%%%%%%%%%%%%%%%%%%%%%%%%%%%%%%%%%%

%%%%%%%%%%%%%%%%%%%%%%%%%%%%%%%%%%%%%%%%%%%%
%AGN Neutrinos

%%%%%%%%%%%%%%%%%%%%%%%%%%%%%%%%%%%%%%%%%%%%%%%%%%%%%%%%%%%%%%%
\section{Neutrinos from Active Galactic Nuclei\label{nus_agn}}
%%%%%%%%%%%%%%%%%%%%%%%%%%%%%%%%%%%%%%%%%%%%%%%%%%%%%%%%%%%%%%%
The most luminous, persistent objects in the sky are Active Galactic
Nuclei.
In hadronic acceleration models,
it can be assumed that for each class of AGN, the electromagnetic emission is
correlated to a neutrino signal. Apart from individual normalization factors, the
corresponding cosmological integrations are basically mathematically
identical. In this section, the correlation between the emission of neutrinos
and photons at different wavelengths will be reviewed according to neutrino
flux models which are currently being discussed in the
literature. 
%=====================================================
\subsection{Calculation of a diffuse neutrino flux}
%=====================================================
The total diffuse neutrino spectrum from a certain source class at Earth is given
by the single source spectrum $d\Phi_{\nu}/dE_{\nu}$ convoluted with the redshift and luminosity
dependent source number. A factor $1/(4\pi\,d_{L}^{2}$) is applied to account
for the decrease of the source flux with the luminosity distance $d_L$. The
number of sources is given by the product of the luminosity function 
of AGN, $dn/dL/dV$, and the comoving volume $dV/dz$. Thus, the total diffuse
neutrino flux $dN_{\nu}/d\en$  can be written as
\begin{equation}
\frac{dN_{\nu}}{d\en}(E_{\nu}^{0})=\int_{z}
\int_{L}dz\,dL\,\frac{d\Phi_{\nu}}{dE_{\nu}}(E_{\nu}^{0},L,z)\cdot
\frac{dn}{dL\,dV}(L,z)\cdot \frac{dV}{dz}\cdot \frac{1}{4\pi\,d_{L}(z)^{2}}\,.
\end{equation}
A shift of the neutrino energy at the source $E_{\nu}$ to lower energies at
the detection site due to the expanding Universe, $E_{\nu}^{0}=E_{\nu}/(1+z)$
is taken into account. A simple cosmology of $\Omega_m=1$
and $\Omega_{\Lambda}=0$ is still used for many models, since most spatial distributions of the
source samples were given according to this cosmology at the time of the
publication of the models. The changes in the normalization of the result when
using the experimentally confirmed model of $\Omega_m\sim0.3$ and
$\Omega_{\Lambda}\sim 0.7$ \cite{wmap06} should be negligible, as pointed out
in~\cite{dunlop}.
%------------------------------------------------------------------
\subsubsection{Different classes, different normalization options}
\begin{figure}[h]
\centering{
\includegraphics[width=\linewidth]{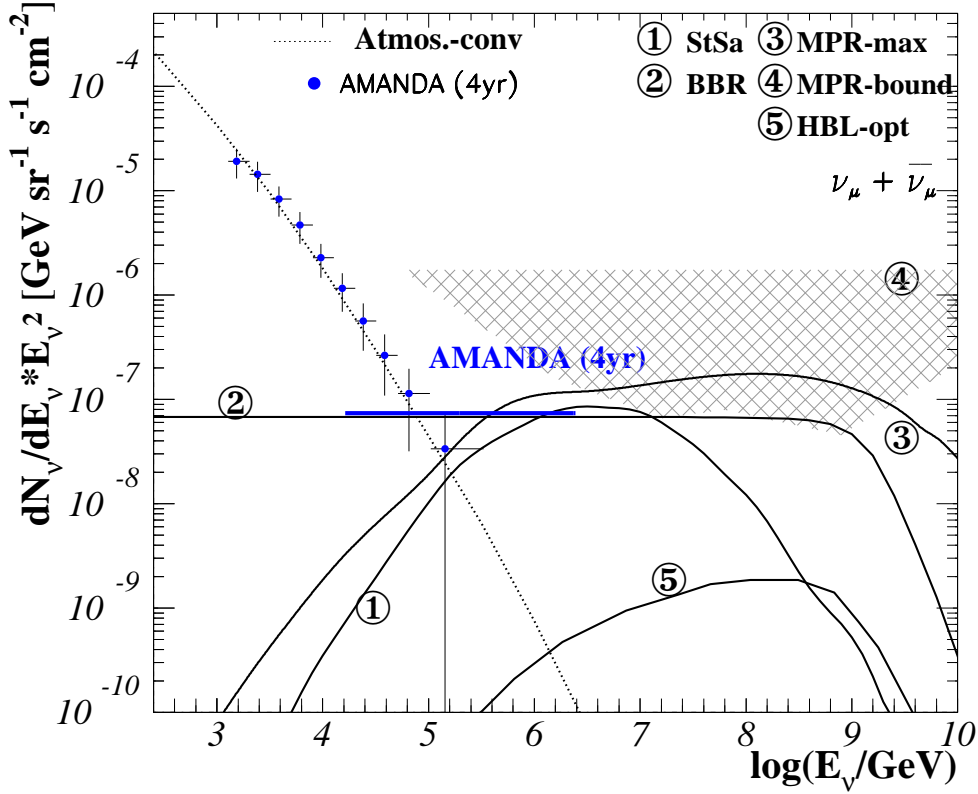}
\protect\caption[AGN neutrino flux models.]{AGN neutrino flux models: \ding{172}
  $\nu$-MeV correlation for blazars~\cite{stecker_mod}; \ding{173} $\nu$-radio
  correlation for FR-II galaxies \& FSQRs~\cite{bbr_05}, using an optical
  depth of $\tau=0.2$; \ding{174} maximum
  contribution from {\sc EGRET} sources~\cite{mpr}; \ding{175} upper bound
for $\nu$s from optically thick sources (upper, straight bound) and from
optically thin sources (lower, curved bound of shaded area)~\cite{mpr}; \ding{176}
Optimum prediction for High-peaked BL Lacs (HBL-opt) within the proton-blazar model~\cite{muecke}. The
atmospheric prediction is shown as the dotted line~\cite{volkova80}, data are
for four years of {\sc AMANDA-II} data~\cite{kirsten_icrc07,kirsten_phd}, as
well as the limit given in~\cite{jess_diffuse}.}
\label{review_agn_models}
}
\end{figure}

%------------------------------------------------------------------
The overall normalization of the neutrino spectra builds on the correlation
between the observed flux from photons or charged Cosmic Rays as discussed in
Section~\ref{multimessenger}. In the following section, the prevailing diffuse
neutrino flux models are discussed.
Table~\ref{nu_model:table} summarizes the different models with
their very basic assumptions. Figure~\ref{review_agn_models} shows a selection
of the models in comparison with current {\sc AMANDA} data and the atmospheric
prediction.

\begin{table}[h!]
\centering{
\begin{tabular}{l|llll}\hline
source class&Normalization&$\nu$ correlation&origin&model(s)\\
&wavelength&to wavelength&& \\\hline\hline
blazars&CRs&proton flux, responsible&jet&\cite{mpr}\\ 
       &   &for $p\,\gamma$ in source& &\cite{muecke}\\\hline
&$>100$~MeV&cascaded $\pi^{0}$ signal&jet&\cite{mpr}\\
&          &connected to $\pi^{+/-}$&&\cite{muecke}\\
&          &production ($\Rightarrow\,\nu$s) &&\cite{mannheimjet}\\\hline
&$>$~MeV&cascaded $\pi^{0}$ signal&jet&\cite{mannheimjet}\\
&       &connected to $\pi^{+/-}$&    &\cite{stecker_mod}\\\hline
FSRQs&radio&jet-disk correlation&jet&\cite{bbr_05}\\
     &     &radio $\sim$ total power& \\\hline
FR-II&radio&jet-disk correlation&jet&\cite{bbr_05}\\
     &     &radio $\sim$ total power& \\\hline
radio quiet AGN&X-ray&cascaded $\pi^{0}$
signal&disk&\cite{stecker96}\\
& & & &\cite{nellen}\\
& & & &\cite{alvarez_xrays}\\\hline
\end{tabular}
\caption[Neutrino models for AGN]{Neutrino models for Active Galactic Nuclei. The models are developed
  for different AGN sub-classes, using different signal hypotheses. Each model
  uses either charged Cosmic Rays or non-thermal photon emission from the
  given source class to normalize the neutrino spectrum. The CR flux
  gives evidence for proton acceleration. protons interact with the
  photon field to produce neutrinos. The correlation between neutrinos and
  MeV-GeV photons can be present if the photons arise from $\pi^{0}$-decays
  which implies the coincident production of charged pions, the latter decaying
  to leave neutrinos. The normalization to the radio signal from AGN can be
  used, since the radio power is connected to the total power of the AGN via
  the jet-disk model. A
  fraction of the total power goes into neutrinos. X-rays originate from the
  foot of AGN jets. In the cited models, these X-rays are assumed to be
  produced in $\pi^{0}$ decays, the signal cascading in an optically thick
  environment from TeV-energies to X-ray wavelengths.}
\label{nu_model:table}
}
\end{table}
%-------------------------------------------------------------
\subsubsection{TeV blazars}
%-------------------------------------------------------------
Sources of electromagnetic TeV emission can
be interpreted as optically thin to photon-neutron interactions,
$\tau_{\gamma\,n}\ll 1$, in hadronic acceleration models~\cite{muecke,mpr}. In such a
scenario, charged Cosmic Rays (CRs) are produced in the vicinity of the source
through the decay of the escaping neutrons. In this case, the resulting
neutrino energy density would be proportional to the extragalactic CR component measured
at Earth.
 The neutrino flux from High frequency peaked BL Lacs (HBLs) was calculated
using the connection between Cosmic Rays and neutrino emission,
see~\cite{muecke}. The neutrino emission peaks at very high energies,
i.e.~$\en\sim 10^{8}$~GeV. Due to the low intensity of the signal, it will be
difficult to detect such a contribution.
 A theoretical upper bound of such a contribution to the diffuse
neutrino flux was derived by Mannheim, Protheroe and Rachen ({\em MPR bound}), see \cite{mpr} and Fig.~\ref{review_agn_models}.

As discussed in Section~\ref{multimessenger}, the high-energy photon tail is
absorbed by the extragalactic background light. Therefore, only the local
Universe can be observed with TeV photons. The most distant source observed by
IACTs so far is 3C~279 at a redshift of $z=0.538$, detected with the {\sc
  MAGIC} telescope~\cite{prandini_icrc2007}, which is considered as the VHE
photon horizon here. In the following, it is assumed that the sources can be
observed to this maximum redshift. Neutrinos, being co-produced with
TeV photons in hadronic scenarios, offer
the possibility to observe the same source classes, but without any
losses. The spectra to be observed at Earth correspond to those emitted at the
sources, only shifted in energy by $1+z$. The neutrino flux, originating from
the local sources, which can be observed in coincidence with TeV photon
experiments, is a factor of $\eta(z_{\max})$ smaller than the total neutrino
flux. The absorption factor $\eta$ is determined by calculating the ratio of
the neutrino flux up to the first
formation of galaxies and the total neutrino flux up to  $z_{\max}$,
\begin{equation}
\eta=\frac{dN_{\nu}/d\en(total)}{dN_{\nu}/d\en(z_{\max})}\,.
\end{equation}
The total diffuse neutrino flux cannot exceed flux limits as determined from
{\sc AMANDA} data, the most stringent one being 
\be
\dl=7.4\cdot
10^{-8}\,\diffunits
\label{diff_limit_jess:equ}
\ee
as presented in~\cite{jess_diffuse}. Therefore, the flux which can be
observed simultaneously in TeV photons and neutrinos is restricted to
\be
{\en}^{2}\cdot \frac{dN_{\nu}}{d\en}<\eta(z_{\max})^{-1}\cdot \dl\,.
\ee 
The absorption factor $\eta$ is presented in~\cite{stacking_diffuse2007} with
values of $\eta(z_{\max}=0.5)=17$ and $\eta(z_{\max}=1)=5$. Assuming a horizon
for TeV photon detection of $z_{\max}=0.5$ or $z_{\max}=1$ results in a
maximum flux of
\be
{\en}^{2}\,\frac{dN_{\nu}}{d\en}(z_{\max}=0.5)<4.3\cdot 10^{-9}\,\diffunits
\ee
and
\be
{\en}^{2}\,\frac{dN_{\nu}}{d\en}(z_{\max}=1)<14.8\cdot 10^{-9}\,\diffunits\,,
\ee
respectively. This result implies that the diffuse fluxes to be observed in
both TeV photons and neutrinos are relatively small, even for {\sc IceCube}
prospects. It emphasized the importance of neutrino analyses independent of
photon measurements, such as the general diffuse search, see e.g.~\cite{jess_diffuse,kirsten_phd}.
%-------------------------------------------------------------
\subsubsection{EGRET and COMPTEL blazars}
%-------------------------------------------------------------
In the case of optically thick sources ($\tau_{\gamma\,n}>1$), the TeV photons
from $\pi^{0}-$decays interact with nucleons in
the source before escaping at lower energies leading to the emission of
sub-TeV photons. Therefore, the diffuse
extragalactic background measured by the {\sc EGRET}\footnote{{\bf E}nergetic {\bf
    G}amma {\bf R}ay {\bf E}xperiment {\bf T}elescope} 
experiment~\cite{egret_new} ($\eg>100$~MeV) can be interpreted as an avalanched TeV
signal from blazars and can thus be used to normalize the neutrino flux from
{\sc EGRET} or {\sc COMPTEL}\footnote{{\bf COM}pton {\bf TEL}escope}-type sources. Again, an upper bound to the contribution from such
sources is given in~\cite{mpr}, which is much less restrictive than the optically thin
case. Apart from the bound, a calculation of the maximum contribution from
blazars is given in~\cite{mpr}. 
In addition to the contribution from HBLs given in~\cite{muecke}, a contribution from the optically thick Low-frequency peaked BL Lacs (LBLs)
can be calculated using the {\sc EGRET} diffuse extragalactic background for a
normalization of the neutrino spectrum, see~\cite{muecke}.
A model of proton-photon interactions in AGN and
collisions of protons from the core with protons of the host galaxy is derived
in~\cite{mannheimjet}. For sources with an even higher optical thickness, only
photons below $100$~MeV escape. In this case, the neutrino signal can be
normalized to the diffuse extragalactic contribution measured by {\sc COMPTEL} at
energies in the range of $(0.8,\,30)$~MeV~\cite{comptel}. This would enhance the
contribution of neutrinos from proton-photon interactions by almost an order of magnitude as
shown in~\cite{mannheimjet}. 

A model by~\cite{stecker96} was originally using the diffuse cosmic X-ray background and has recently been modified in a way that it is using the
{\sc COMPTEL} diffuse background to normalize the neutrino
spectrum~\cite{stecker_mod}, see Fig.~\ref{review_agn_models}. This reduces
the formerly very high contribution by a factor of 10. In addition, oscillations have been taken
into account which leads to a further reduction by a factor of 2. 
%-------------------------------------------------------------
\subsubsection{X-ray detected AGN}
%-------------------------------------------------------------
The measurement of the diffuse extragalactic contribution in X-rays by {\sc
  RO\-SAT}\footnote{{\bf RO}entgen {\bf SAT}ellite}
has raised the question whether it is produced by radio-weak AGN. Assuming that the
X-ray emission comes from the foot of the jet, the X-ray
signal would be accompanied by a neutrino flux. A model by~\cite{nellen} and an approach by~\cite{stecker96} have
been presented, predicting a correlation between X-ray emitting AGN and
  neutrino emission from the same sources. In these models, $p\,\gamma$
  interactions produce high-energy photons, which cascade down to X-ray
  energies before escaping due to the optically thick environment. An alternative scenario for the
  explanation of the X-ray flux from AGN would be the up-scattering of thermal
electrons via the Inverse Compton effect. In that case, the X-ray component
would not be accompanied by a neutrino signal, see e.g~\cite{msr}. Until today, about $75\%$ of the diffuse
X-ray signal has been resolved by {\sc ROSAT}~\cite{xray_sources}, with the
help of {\sc Chandra} and {\sc XMM-Newton}\footnote{{\bf X}-ray {\bf M}ulti-{\bf
      M}irror Mission {\bf Newton}} data,
this number can be updated to $90\%$, see~\cite{xray_sources} and references therein.  More than $70\%$ of the diffuse background
are possibly connected to the X-ray emission of Active Galactic Nuclei most of
which are radio weak. The search for a diffuse, extraterrestrial neutrino flux with {\sc AMANDA}
did not yield a significant signal and restricts these models strongly. 
%------------------------------------------------------------------
\subsubsection{FR-II galaxies and FSRQs\label{agn:nus_permanent}}
%------------------------------------------------------------------
The diffuse neutrino flux from FR-II galaxies using average parameters has
been calculated in~\cite{bbr_05}. For the derivation of a source spectrum, it is assumed that the neutrino
spectrum follows the proton spectrum in first order approximation. Effects
from multi-pion production are neglected. The changes to the spectrum
due to this effect would be negligible with respect to other remaining
uncertainties in the calculation, such as the lack of knowledge of the sources'
optical depth. For the normalization of the spectrum, it is assumed that the
neutrino output of a single AGN is proportional to the total power of the jet given by the disk luminosity $L_{disk}$~\cite{bbr_05}.
The disk luminosity is linked to the radio output $L_{radio}$ via the \emph{jet-disk symbiosis}
approach which was developed in~\cite{falcke1,falcke,falcke2}.  The
spectral behavior is generically chosen as ${\en}^{-2}$, with a resulting
spectrum of
\be
{\en}^2\cdot \frac{dN_{\nu}}{d\en}=1.43\cdot \tau_{eff}\cdot 10^{-7}\,\diffunits
\ee
for one neutrino plus anti-neutrino flavor, including oscillations.
Here, $\tau_{eff}$ is the effective optical depth to $p\,\gamma$
interactions. 

With the currently most restrictive flux limit, see Equ.~(\ref{diff_limit_jess:equ}),
the optical depth has to be smaller than $\tau_{eff}<0.6$.  The maximum flux
for $\tau_{eff}=0.6$ is shown in Fig.~\ref{review_agn_models}. The limit,
however, only applies if the spectrum does not deviate strongly from the
${\en}^{-2}-$shape. It was pointed out by different authors,
e.g.~\cite{baring04,mbq2007}, that the spectral shape can vary
significantly under certain
conditions of relativistic shocks. While flatter spectra may still be excluded by the limit,
steeper spectra are more difficult to investigate. When interpreting the radio
spectrum as synchrotron radiation from electrons, the radio spectral index
$\overline{s}\sim 0.8$ is correlated to the primary index as $p=2\cdot s+1$.
The neutrino spectrum, following the proton spectrum, could therefore be as
steep as ${\en}^{-2.6}$.
%===================================================================
\subsection{Individual source spectra}
%===================================================================
For the calculation of the diffuse spectra discussed above, mean parameters
have been used. In this section, different predictions implementing the individual
variations of the single sources are discussed. 
%-------------------------------------------------------------------
\subsubsection{TeV emitters}
%-------------------------------------------------------------------
A simple estimate of the neutrino flux from TeV sources is to use the photon
flux at TeV energies under the assumption that it arises from photo-hadronic
interactions. It is known that the TeV photon flux is diminished by interactions with the
extragalactic background light. Thus, the TeV photon flux at Earth and its
spectral index are used here as a
lower limit for the photon emission from the blazars. The neutrino flux is given as
\be
\frac{dN_{\nu}}{d\en}=A_{\nu}'\cdot\left(\frac{\en}{\mbox{TeV}}\right)^{-\alpha_{\nu}'}
\ee
with the spectral index $\alpha_{\nu}'\approx\alpha_{TeV}$. 
The normalization factor is calculated by correlating the neutrino flux
luminosity with the TeV photon flux as
\be
\int \frac{dN_{\nu}}{d\en}\,\en\,d\en=K\cdot\int \frac{dN_{TeV}}{dE_{TeV}}\,E_{TeV}\,dE_{TeV}\,.
\ee
Here, $K=1$ for $p\,p$ interactions and $K=1/4$ for $p\,\gamma$
interactions, see e.g.~\cite{francis_barcelona}. Assuming that the spectral indices for neutrino and TeV photon
emission and the energy threshold ($E_{\min}\sim 100$~GeV) are about equal
translates the above equation into a direct correlation between the flux
normalizations,
\be
A_{\nu}'\approx K\cdot F_{TeV}\,.
\ee
Table~\ref{tev_sources_neutrinos} lists the normalization for
the neutrino spectrum of the 13 HBLs with given spectral information, the LBL
BL Lacertae and M~87. The values have been calculated using the parameters in
table~\ref{tev_agn_table} in Section~\ref{multimessenger}. The results are listed
for a $1:1$ correlation between photons and ($\nu_{\mu}+\overline{\nu}_{\mu}$)-neutrinos as it is the case for
proton-proton interactions. Dividing the values by 4 yields the flux in case
of $p\,\gamma$ interactions.
It needs to be considered that the  TeV photon flux from blazars is
highly variable. The intensity of the low state can be several orders of
magnitude below the high-state flux. Both difficulties and opportunities arise
from the strong variability.

The estimated neutrino signal from high-state photon states in AGN and point
source neutrino flux limits as given in~\cite{5yrs} for 6 of
the individual sources are listed in
table~\ref{tev_sources_neutrinos}. However, the
limits cannot directly be compared to the estimated flux: the calculated
values are only valid during the time of the flare, while the limits apply to
 the flux {\it averaged} over the five years of observation. The latter is 
 significantly lower and for many of the TeV sources, the low-emission state
 is not known. Low-state observations require long observation times. The duty cycle of Cherenkov telescopes
is quite low, since only dark nights can be used for measurements. 
Therefore, extensive long-term monitoring of the sources is not possible with the current
telescopes. Many flares are  missed and the average flux is difficult to determine. Small telescopes, dedicated to the
long-term monitoring of TeV blazars, are planned to be rebuilt from {\sc
  HEGRA} telescopes as described in~\cite{dwarf_icrc2007}. Currently, the {\sc Whipple}
telescope uses part of its time for blazar monitoring in order to get a better
estimate of the TeV lightcurve~\cite{whipple_monitor_icrc2007}. These attempts
may be an option to get a better understanding of the variability of the
sources. The small telescopes are, however, typically not sensitive to
low-state emission. This has to be examined by the more sensitive telescopes,
{\sc H.E.S.S., MAGIC, VERITAS} \& co.

The flaring states also provide an advantage for high-energy neutrino telescopes: for known flares like the ones listed in
table~\ref{tev_sources_neutrinos}, temporal cuts in the data of neutrino
telescopes can be used in addition to spatial cuts. If only the duration of
the flare is used for an analysis, the estimated flux values can be used as
reference values. This way, a much higher flux is expected, in addition to a
reduced background, which increases the probability of detecting a
signal. Analysis methods concerning AGN flare searches are being developed as presented in~\cite{elisa_tev}.

\begin{table}
\centering{
\begin{tabular}{l|lll}
\hline
Source&year&$A_{\nu}^{K=1}$ (flare)&Limit  $(\nu_{\mu}+\overline{\nu}_{\mu},\,2000-2004)$ \\
&(TeV obs)&$10^{-13}$TeV$^{-1}$ s$^{-1}$ cm$^{-2}$&$10^{-13}$TeV$^{-1}$ s$^{-1}$ cm$^{-2}$\\\hline\hline
M~87&2005&11.7&440\\\hline
Mkn~421&2001&311&370\\\hline
Mkn 501&1997&1001&740\\\hline
1ES 2344+514&1995&510&300\\\hline
Mkn 180&2006&12.0&-\\\hline
1ES 1959+650&2002&740&680\\\hline
BL Lacertae&2006&3.6&-\\\hline
PKS 2005-489&03/04&2&-\\\hline
PKS 2155-304&2006&2060&-\\\hline
H 1426+428&2001&50.0&470\\\hline
1 ES 0229+200&05/06&6&-\\\hline
H 2356-309&2004&3.00&-\\\hline
1ES 1218+304&2005&12.7&-\\\hline
1ES 1101-232&2006&4.44&-\\\hline
1ES 0347-121&2006&4.5&-\\\hline
1ES~1011+496&2006&3.2&-\\\hline
PG 1553+113&05/06&2.1&-\\\hline
\end{tabular}
\caption[Neutrino flux from AGN of TeV emission]{Neutrino flux ($\nu_{\mu}+\overline{\nu}_{\mu}$) from AGN of
  TeV emission. Limits are taken from~\cite{5yrs}, having been calculated from
  5 years of {\sc AMANDA} data.}
\label{tev_sources_neutrinos}
}
\end{table}
%-------------------------------------------------------------------
\subsubsection{FR-II galaxies and FSRQs}
\begin{figure}[h]
\centering{
\includegraphics[width=10.cm]{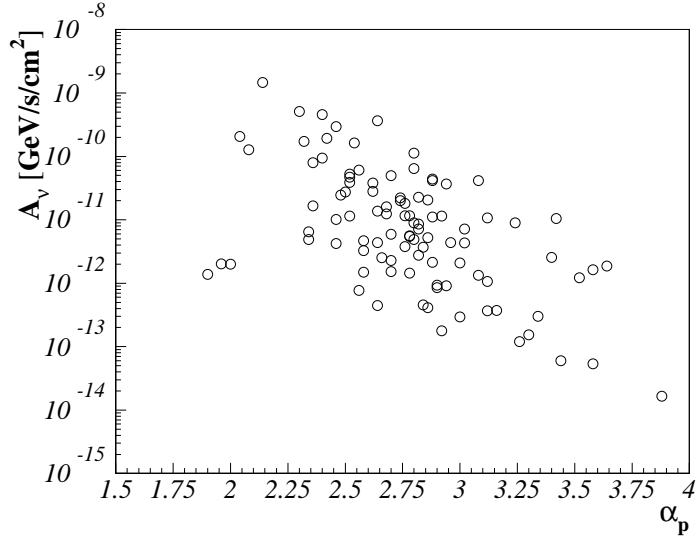}
\protect\caption[Neutrino flux normalization and index $\alpha_p$ (FR-II galaxies)]{Correlation between the neutrino flux normalization and the
  particle spectral index $\alpha_p$. While the majority of sources is
  concentrated around $A_{\nu}\sim 10^{-11}$~GeV/s/cm$^2$ and $\alpha_p\sim 2.6$, a strong
  correlation between $A_{\nu}$ and $\alpha_p$ is visible - bright sources are
typically flatter.}
\label{s_alpha}
}
\end{figure}

%-------------------------------------------------------------------
A sub-sample of 114
FR-II sources from the steep spectrum sample given in~\cite{willott} was used to analyze
the differences between an average prediction and a source property based
estimate, see~\cite{bbr_icrc05}. The radio flux $S$ is converted into the radio luminosity as
\begin{equation}
L_{radio}=4\,\pi\,d_{l}^{2}\cdot S
\end{equation}
and the spectral radio index $s$ is used to determine the proton spectral
index for each case individually, $\alpha_p=2\cdot s+1$. 
The single source normalization versus the
primary spectral index is shown in Fig.~\ref{s_alpha}. The normalization is
given at $\en=1$~GeV. There is a strong
correlation showing that flat neutrino spectra are stronger than steep ones. 
The coincidence spectrum is much flatter than the average prediction, 
\begin{equation}
\frac{dN_{\nu}}{dE_{\nu}}\propto E_{\nu}^{-2.2}
\end{equation}
Using the mean spectral index $\overline{s}=0.8$ results in a much steeper mean
spectrum due to the correlation between the normalization $A_{\nu}$ and the
spectral index $\alpha_p$.
%================================================
\subsection{Theoretical upper bounds on neutrino emission}
%================================================
An upper bound on the diffuse neutrino flux from extragalactic sources was
derived in~\cite{wb99}. Here, the observed spectrum of UHECRs was used in
order to calculate the maximum neutrino flux possible to be produced together
with the highest energy Cosmic Rays. The bound is given as
\begin{equation}
{\en}^2\,\frac{dN_{\nu}}{d\en}=1.5\cdot 10^{-8} \,\diffunits
\end{equation}
for muon and anti-muon neutrinos.
It is pointed out in~\cite{mpr} that this bound is, however, only valid for
sources optically thin to neutron-photon interactions. In those sources,
neutrons escape, decay and produce the flux of UHECRs. High-energy photons
from proton-photon interactions escape at TeV-PeV energies. The bound derived
in~\cite{wb99} does therefore not apply to those classes of AGN which are
optically thick to neutron-photon interactions. In addition, it is pointed out
in~\cite{mpr} that the calculations in~\cite{wb99} assume a strict
${\en}^{-2}-$spectrum. The neutrino spectrum, however, deviates from the
primary spectrum of charged particles, which needs to be taken into account in
the case of the derivation of an upper bound. Taking these facts into account,
a bound for transparent sources and optically thick sources was derived
in~\cite{mpr} as indicated in Fig.~\ref{review_agn_models}. The upper, straight line of the
shaded area represents the bound for optically thick sources, while the lower
line shows the upper bound concerning optically thin sources. 

It is generally very difficult to derive absolute upper bounds. There are
still large uncertainties in many of the parameters. Taking for instance the
flux of high energy Cosmic Rays, which is uncertain by at least a factor of
two, when comparing the results from {\sc HiRes} and {\sc Auger}, see
Fig.~\ref{cr_spect}. Further uncertainties come from the spectral index of the primary
flux and the neutrino flux, the redshift distribution of the different source
classes, as well as the factor which connects UHECRs or
photon emission with neutrinos.

It is shown
in~\cite{francis_barcelona} that the derived bound from~\cite{wb99} should
rather be interpreted as a flux estimate considering the connection between
UHECRs and neutrinos, with an intensity between
\begin{equation}
{\en}^2\,\frac{dN_{\nu}}{d\en}=(1-5)\cdot 10^{-8} \,\diffunits
\end{equation}
for muon and anti-muon neutrinos, using recent Cosmic Ray data.
%================================================
\subsection{North and south view - next generation neutrino detectors}
%================================================
Both for point source and diffuse searches, the orientation of the field of
view of the detector is crucial. The sources are not
distributed isotropically. Sky surveys can be biased by
the location of the instruments and the brightest source in a sample sometimes
makes up the dominant contribution from a source class. The latter is, for instance,
the case for FR-I galaxies, where M~87 gives most of the radio flux in the
total sample. The reason is that M~87 is by far the closest of the sources, at a
distance of only $\sim 16$~Mpc.

The measured diffuse flux depends on the location of the
detector. If the luminosity distribution is
steep, implying that nearby sources make up most of the contribution, the
signal is much stronger with the nearby sources in the field of view.

{\sc IceCube} is currently being built at the geographic South Pole with a
full view on the northern hemisphere in the muon-neutrino channel. The
complementary instrument, {\sc Km3NeT}, planned to be built in the
Mediterranean, will cover the southern hemisphere.  Depending on the source
class, optimal observation conditions are provided by either southern or
northern telescopes.
For instance, in the case of {\sc EGRET}-detected sources, the three
brightest objects are in the southern sky. The largest fraction of identified
FR-I and FR-II galaxies are located in the northern hemisphere, since the supergalactic plane is mostly located
in the northern hemisphere and there may also be a bias in the radio selection
of objects.  A detailed description of the most important catalogs is
given in~\cite{stacking_diffuse2007}.
%================================================
\subsection{Physics conclusions from current neutrino flux limits}\vspace{-0.5cm}
%================================================
The primary goal of large volume neutrino arrays is 
the detection of a signal from extraterrestrial sources. Different methods
have been developed in order to reduce the atmospheric background, using the
model predictions described above for an optimal selection of data. Here, energy cuts are applied for diffuse
searches and spatial and temporal cuts are used for point source
searches. The stacking of the most intense objects of a source class can
further improve the signal to background ratio. This was done for 11 AGN
classes with the {\sc AMANDA} detector, as described in~\cite{andreas,5yrs}. Here, typically the ten brightest sources in the
sample were used for the analysis. The point source limits $\stl$ can be
interpreted as diffuse limits, $\sdl$ as described in~\cite{stacking_diffuse2007},
\be
\sdl=\frac{\epsilon\cdot \xi}{2\,\pi\,\mbox{sr}}\,.
\ee
The factor $\epsilon$ includes identified sources contributing to the
diffuse flux, which have not been considered in the stacking
analysis\footnote{In many samples, most of the sources are too faint to be
  considered in the stacking analysis - including them would reduce the
  sensitivity. They still have a significant contribution to the
  diffuse flux.}. The second factor $\xi$ accounts for the diffuse background
of unidentified sources. This is only known for some of the
source classes and needs to be estimated in other cases. The factor
$1/(2\,\pi\,$~sr$)$ accounts for the field of view of the {\sc AMANDA} detector.

The best results are achieved for identified {\sc EGRET} blazars. The stacking
diffuse limit is given as
\be
\sdl_{EGRET}=7.25\cdot 10^{-8}\,\diffunits
\ee
in the energy range of $1$~TeV$<\en<1000$~TeV.
This is not a general diffuse limit, but it only applies for diffuse flux
predictions building on a correlation between the photon flux as measured by
{\sc EGRET} and neutrino emission. While the limit cannot restrict current
predictions, second generation neutrino telescopes like {\sc IceCube} and {\sc
  KM3NeT} will be sensitive to the
prediction of neutrinos from $p\,p$ interactions described
in~\cite{mannheimjet} at energies around $1$~TeV. The advantage of the
stacking method is that it is sensitive to neutrino fluxes at relatively low
energies, while the general diffuse search is restricted to energies of above
$10$~TeV due to the high atmospheric background.

The diffuse search, on the other hand, is more sensitive in most of the cases
at higher energies, since the stacking method loses sensitivity due to the
high diffusive factors $\xi$ which can reach values up to $\sim 10^5$ in the
worst case. With current neutrino flux limits, it is possible to exclude the
correlation between X-rays and neutrino
emission~\cite{stacking_diffuse2007}. The three predictions,
i.e.~\cite{stecker96,nellen,alvarez_xrays} show a violation of the limit by
about an order of magnitude. It is therefore excluded that neutrinos are
produced in coincidence with X-rays from radio-quite AGN.
The modified model of Stecker et al.~\cite{stecker_mod}, assuming a
correlation between neutrinos  and MeV photons rather than X-rays is a factor
of $20$ lower than the original prediction. The factor of $20$ arises from the
fact that the diffuse MeV flux is lower than the X-ray flux, which translates
directly into a lower neutrino flux.

A search for ultra high-energy neutrinos was performed
in~\cite{lisa_3yrs}. At energies around $\en\sim 1$~PeV, neutrinos are
effectively absorbed by the Earth and only reach the detector when coming from
close to the horizon. Events from above are difficult to distinguish
from atmospheric muons, which leaves horizontal neutrinos for the
analysis. The limit for three year of data taking (year 2000-2003) lies at
\be
\dl=2.7\cdot 10^{-7}\,\diffunits\,.
\ee
Limits have been derived according to different models. The limit lies a factor of $1.25$ below the prediction
of~\cite{mpr}, presenting the maximum contribution from optically
thin sources. The most optimistic parametrization can therefore be
excluded. 
With {\sc IceCube}, a sensitivity gain of about two orders of magnitude in
one year of observation is expected, providing more information on
the parameters of different flux models at high energies.
\clearpage
%%%%%%%%%%%%%%%%%%%%%%%%%%%%%%%%%%%%%%%%%%%%
\subsection{The Olbers' paradox for neutrinos}
%%%%%%%%%%%%%%%%%%%%%%%%%%%%%%%%%%%%%%%%%%%%
First predictions of high-energy neutrino emission were made in the 1990s with maximum
parameter settings in order to estimate what is possible. These models could
now be revised using more realistic parameters due to current neutrino flux
limits. While it is obviously the aim to {\em detect} high-energy neutrinos
from extraterrestrial sources, the limits which have so far resulted from the
searches are very valuable in the context of multimessenger astrophysics as
discussed above. These limits are in analogy to the {\em Olbers'
  paradox}. The paradox arose when Olbers calculated that - with a homogeneous
distribution of astrophysical objects in the sky, and with these objects
radiating isotropically - that the sky should be isotropically bright as the sun itself~\cite{olbers}. There was,
however, a very obvious, optical limit on the radiation during night-time. The
sky was, and still is, dark! The paradox could not be solved before the 1950s,
when different assumptions from Olbers' calculations fell: the sources are not
isotropically distributed in the sky, the Universe is not infinite and it is
not in thermal equilibrium. 

Similar considerations can be done for every other wavelength, where a
prediction is made which violates experimental limits. The limits give the
opportunity to constrain the parameters of the model. With
neutrino-astrophysics, we are still at the beginning of exploring these
predictions. Still, current neutrino flux limits start to become sensitive to
prevailing neutrino flux models and start to constrain them. Within a short
time, the next
generation of large volume detectors will be able to reach sensitivities more
than $1-2$ orders of magnitude better than current instruments and it can be
expected, that the first extraterrestrial neutrinos will be detected soon.

%%%%%%%%%%%%%%%%%%%%%%%%%%%%%%%%%%%%%%%%%%%%

%%%%%%%%%%%%%%%%%%%%%%%%%%%%%%%%%%%%%%%%%%%%
%GRB Neutrinos
%%%%%%%%%%%%%%%%%%%%%%%%%%%%%%%%%%%%%%%%%%%%%%%%%%%%%%%%%%%%%%%
\section{Gamma Ray Bursts as neutrino sources\label{nus_grbs}}
\begin{figure}[h!]
\centering{
\epsfig{file=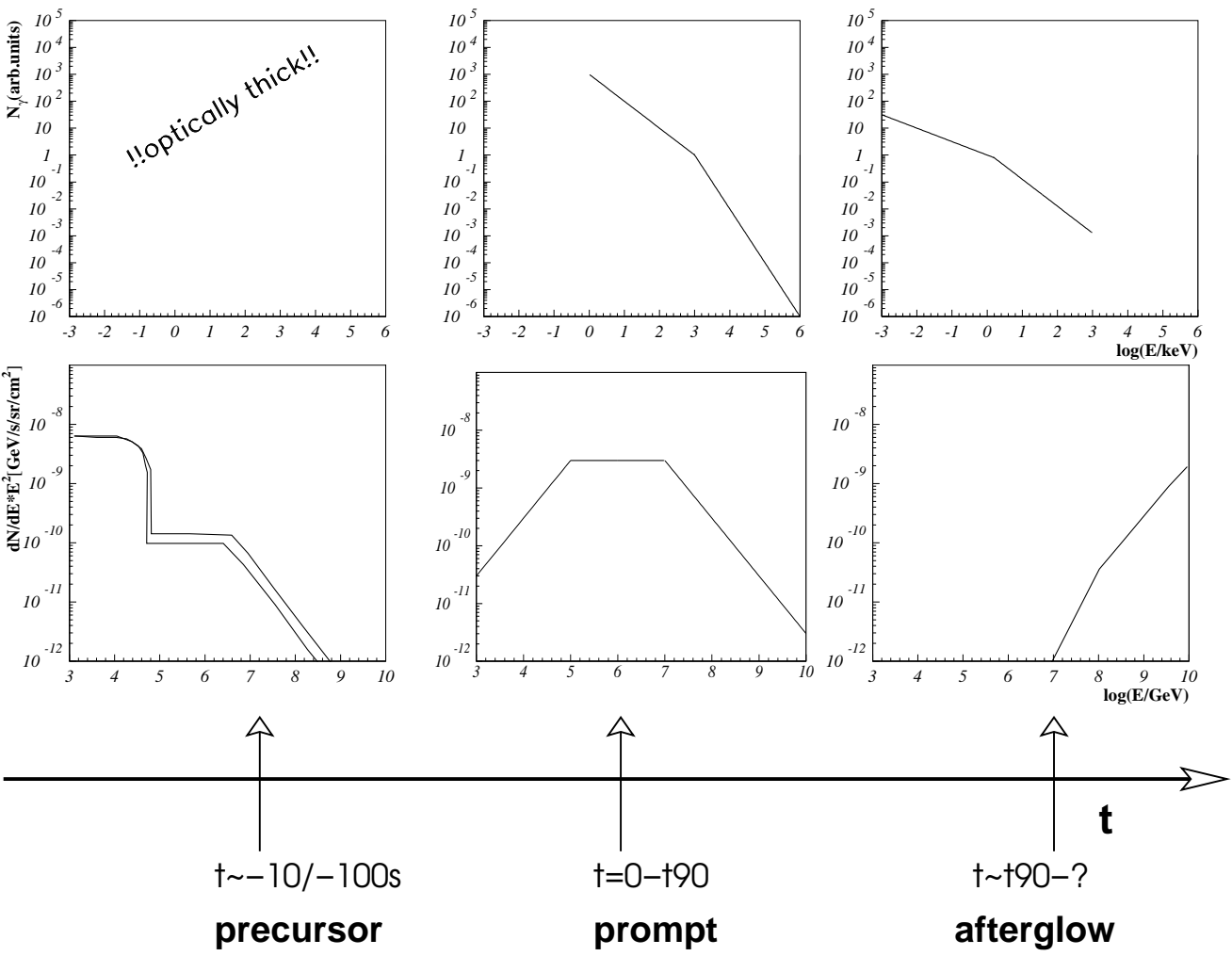,width=18.5cm,angle=90}
\caption[Overview of different neutrino production scenarios in GRBs]{Overview of different neutrino production scena\-rios during the three
  different phases of a GRB. The corresponding electromagnetic output is
  indicated schematically as well. The different flux models are described in
  the text.}
\label{grb_spectra:fig}
}
\end{figure}

\vspace{-0.5cm}
%%%%%%%%%%%%%%%%%%%%%%%%%%%%%%%%%%%%%%%%%%%%%%%%%%%%%%%%%%%%%%%
A summary of the photon and neutrino emission from GRBs is shown in
Fig.~\ref{grb_spectra:fig}. Three phases
of non-thermal emission are expected, i.e.~precursors hours prior to the
GRB~\cite{precursor_nus}, emission during the prompt phase~\cite{wb97,wb99} as well as
afterglow emission~\cite{afterglow_nus}. The lower row shows the neutrino
energy spectra for the three phases while the upper row displays the
corresponding electromagnetic output schematically.
\begin{itemize}
\item A {\em precursor model} has
been developed in \cite{precursor_nus}. The basic idea is that  a shock forms
when the pre-GRB matter collides with the wind of the central pulsar or the
SNR shell.
At this point, the
burst is still opaque to photon emission and the shock environment yields a good
target for neutrino production by shock-accelerated protons interacting with
thermal X-rays in the sub-stellar jet cavity. The shocks happen at smaller
radii than the prompt emission and at lower boost factors $\Gamma$. The
neutrino signal could also be accompanied by a signal in the far infrared. The low
energy part of the neutrino spectrum arises from neutrino production in $p\,p$
interactions and is ${\en}^{-2}$-shaped. The flux at energies $E_{\nu}>10^5$~GeV
originates from proton-photon interactions. The two lines in the lower left panel of
Fig.~\ref{grb_spectra:fig} represent different
shock/jet radii and envelope mass settings. 
\item The {\em prompt} photon {\em emission} from GRBs can be correlated to
  the observed flux of UHECRs, since protons are believed to be accelerated in
  the highly relativistic shocks~\cite{vietri95,wb97,wb99}. The acceleration
  of protons in turn implies the production of neutrinos through photon-hadronic interactions. The resulting flux is shown in the
lower middle graph of Fig.~\ref{grb_spectra:fig}.
\item  {\it Afterglow neutrinos} are produced when the internal shocks from the
  original fireball
hit the interstellar medium and produce external\linebreak shocks~\cite{afterglow_nus}. As for the prompt
emission, the synchrotron emission of electrons gives evidence for the
existence of relativistic charged particles which in turn implies neutrino
production by the baryonic component of the jet and the photon field. In the case
of afterglow emission, 
Waxman and Bahcall~\cite{afterglow_nus} predict the acceleration of
ultra high-energy protons ($E_p>10^{20}$~eV) in reverse, mildly relativistic shocks~\cite{meszaros_rees}. The acceleration of protons to such high energies implies
the production of neutrinos in proton-photon interactions in environments
optically thick to proton-photon or proton-proton interactions. Waxman and
Bahcall conclude in~\cite{afterglow_nus}, that a significant neutrino flux during the
afterglow phase is directly correlated to the electromagnetic afterglow emission.
\end{itemize}
The three different emission phases are discussed in more detail in the next
paragraphs. All variables are given in the observer's frame. 

A general question is the timing of photon and neutrino emission. Depending on
the environment of the astrophysical shock, neutrinos are not exactly emitted
at the same time as the photons. This is important in the case of temporal
selection of data in high-energy neutrino experiments. It is typically still
assumed that flaring sources emit neutrinos during the same time interval,
but the time window is often chosen to be slightly larger than the photon
emission time to account for this effect. Transient astrophysical shocks in
the context of temporal evolution are described
in~\cite{rachen_meszaros98}. 
%===============================================
\subsection{Precursor neutrino emission}
%===============================================
Since the observation of GRB980425 and GRB030329, it is known that at least a
fraction of long GRBs originates in heavy supernova explosions, see
e.g.~\cite{grb030329}. The emission
is beamed, so that not every supernova explosion can be observed as a GRB as
well. The Gamma Ray Burst can only be detected when the jet is directed
towards Earth. The pulsar, remaining in the center of the SNR, is believed to
form prior to the GRB. The pre-GRB is propagating outwards, and a shock is
formed when it meets the magneto-hydrodynamic pulsar
wind. In this shock, protons can be accelerated. The fast protons in turn
interact
\begin{enumerate}
\item with the
supernova remnant's photon field,
\item with the photon field of the MHD wind itself, and
\item with cold protons from the slow plasma.
\end{enumerate}
In the these interactions, neutrinos are
produced~\cite{precursor_nus}. The total spectrum is calculated by assuming
that a fraction of the fast proton spectra, $dN_{p}/d\ep$, are converted into neutrinos,
\be
\frac{dN_{\nu}}{d\en}=\frac{1}{4\,\pi\,d_{L}^{2}}\cdot
\left\{\begin{array}{lll}
\int f_{pp}\cdot M_{\nu}(\ep)\frac{dN_{p}}{d\ep}&&\mbox{ for }\ep<{\ep}^{th}\\
\frac{f_{\pi}}{4}\frac{dN_p}{d\ep}&&\mbox{ for }\ep>{\ep}^{th}\,.
\end{array}
\right.
\ee
Here, $d_L$ is the luminosity distance and ${\ep}^{th}\approx 10^{6}$~GeV is the threshold energy of the protons for the production of
the $\Delta$-resonance. Below this threshold, $p\,p$ interactions dominate,
where $f_{pp}$ is the proton-proton efficiency. Above the threshold, the total spectrum is given by neutrinos generated in $p\,\gamma$
interactions. For the case of photo-hadronic interactions, the proton spectrum
needs to be weighted by the fraction of protons converted into pions,
$f_{\pi}\approx 1$, and it needs to be considered that the flux distributes equally
to the four decay
products $(e^{+},\,\nu_{e},\nu_{\mu},\overline{\nu}_{\mu})$. For the case of
proton-proton interactions, it needs to be considered that the neutrino
multiplicity for an incident proton, $M_{\nu}$, is energy dependent as
\begin{eqnarray}
M_{\nu}(E_{p})&=&\frac{1}{4}\,N_{\pi/K}\cdot \left\{(E/\mbox{GeV})\cdot
  \left[1/2\cdot \ln\left(10^{11}\mbox{ GeV}/\ep\right)\right]  \right\}^{-1}\cdot\nonumber\\
&\cdot&\Theta\left(1/4\cdot \frac{m_{\pi/K}}{\mbox{GeV}}\cdot \gamma_{cm}<1/4\cdot \frac{\ep}{\mbox{GeV}}\right)\,.
\end{eqnarray}
Here, $\Theta$ is a step-function, which is $\Theta=1$ for the given range and
$\Theta=0$ elsewhere. The product $m_{\pi/K}\cdot \gamma_{cm}$ is the lower energy
threshold for pion/kaon production, with  $m_{\pi/K}$ as the mass of the
particle and $\gamma_{cm}$ as the Lorentz boost factor at the center of mass
in the lab-frame. The normalization factors $N_{\pi}$ or $N_{K}$ are
 given by integrating the distribution of pions,
resp.~kaons. The total numbers are given as $N_{\pi}\approx 7$ and
$N_{K}\approx 0.6$. The proton-proton efficiency is assumed to be $f_{pp}=1$, see~\cite{precursor_nus}.

The proton spectra $dN_{p}/d\ep$ are calculated in~\cite{precursor_nus},
assuming an ${\ep}^{-2}-$\linebreak shaped spectrum. The resulting flux is shown in the
lower left graph of Fig.~\ref{grb_spectra:fig}. The precursor is not observable
in photons, since the environment is optically thick. Thus, the detection of
neutrinos from precursor GRBs in the future provides the possibility to alert
photon experiments before the actual GRB
occurs. In~\cite{marek_opticalfollowup2007}, it is proposed to use neutrino
events in {\sc
  IceCube} as a possible trigger for optical telescopes. If the neutrinos
arrive in multiples within a short time window, they might originate from
a core-collapse supernova. Such a trigger would allow for the observation of
SNe in the early phase of the collapse. 
%===============================================
\subsection{Prompt GRB spectra \& neutrinos \label{grb_spectra}}
%===============================================
The prompt GRB photon spectrum, $dN_{\gamma}/d\eg(\eg)$, is usually given by a Band
fit~\cite{band},
\begin{footnotesize}
\begin{equation}
\frac{dN_{\gamma}}{d\eg}=\left\{ 
\begin{array}{ll}
A\,\left(\frac{\eg}{100\mbox{\footnotesize\,  keV}}\right)^{\alpha_{\gamma}}\,e^{\left(-\eg/E_{0}\right)}&\eg\leq(\alpha_{\gamma}-\beta_{\gamma})\,E_0\\
A\,\left(\frac{\eg}{100\mbox{\footnotesize\,
    keV}}\right)^{\beta_{\gamma}}\,\left[\frac{(\alpha_{\gamma}-\beta_{\gamma})E_0}{100\,\mbox{\footnotesize\,  keV}}\right]^{\alpha_{\gamma}-\beta_{\gamma}} e^{\left(-(\alpha_{\gamma}-\beta_{\gamma})\right)}&\eg>(\alpha_{\gamma}-\beta_{\gamma})\,E_0\,.
\end{array}
\right.
\end{equation}
\end{footnotesize}
Here, $\eg$ is the photon energy, $E_0$ is the reference energy with
$(\alpha_{\gamma}-\beta_{\gamma})\cdot E_0=\egb$ as the break energy in the
photon spectrum.
The function was designed empirically in order to match the GRB spectra. It is
the best mathematical description for broadband GRB spectra.
The energy range of detection may not cover
both parts of the spectrum and spectral fits with simpler
powerlaw approximations are often applied. 
The break energy for the photon spectrum is given as $\egbM=(\alpha_{\gamma}-\beta_{\gamma})\cdot E_0/$MeV.

The spectrum is presumably produced by synchrotron radiation of electrons in
the internal shock fronts of the jet, see e.g.\ \cite{hh_02} and references therein. There are two approaches to
explain the break in the 
spectrum at a break energy of typically $\egb\sim 250$~keV. The most common
explanation is the steepening of the spectrum by one power due to the cooling of electrons at high energies, see for example~\cite{zhang04,piran05} as a
review. The break can also be explained by assuming an Inverse
Compton scattering scenario, see e.g.~\cite{dar_rujula} and references
therein. Throughout this review, all energies concerning GRB spectra are given in the observer's frame at Earth unless declared otherwise. 
For a regular GRB, the spectral indices are usually distributed around average
values of $\alpha_{\gamma}\sim -1$ and $\beta_{\gamma}\sim -2$. These values
scatter over a wide range. Short GRBs tend to have harder
spectra with $\alpha_{\gamma}\sim 0$ and $\beta_{\gamma}\sim -1$.
Assuming hadronic acceleration in the jet, a prompt neutrino flux that is correlated
to the photon spectrum results from photo-hadronic interactions
in the source. 

The neutrino spectrum, $dN_{\nu}/d\en$, can be derived assuming
that the proton spectrum follows the electron spectrum of the
source. Furthermore, it needs to be assumed that electron losses can be
neglected. The neutrino flux in turn follows the proton spectrum in a first order
approximation, since it can be connected to the observed synchrotron spectrum
of the sources. An
improvement of the calculations is generally possible by calculating the
electron losses~\cite{deyoung06}. Here, this effect is neglected and the
result is taken as a lower limit for the neutrino flux.

Protons accelerated in astrophysical shocks can interact with the source's
photon field and produce a $\Delta-$resonance which in turn leads to neutrino production.
Proton and neutrino energy are directly proportional, $\en=\ep/20$. The
product of proton and photon energy remains constant, since the energy of the $\Delta$ mass
has to be produced in the center of mass system, $\ep\cdot \eg
=const$. Hence, photon and neutrino energy are inversely proportional,
$\en\propto {\eg}^{-1}$. 
The neutrino flux is then given as
\begin{equation}
\frac{dN_{\nu}}{d\en}\,\en^{2}=A_{\nu}\cdot \left\{ \begin{array}{lll}
(\en/\enb)^{-\alpha_{\nu}}&&\mbox{ for } \en<\enb\\
(\en/\enb)^{-\beta_{\nu}}&&\mbox{ for } \enb<\en\leq \ens\,.
\end{array}
\right.
\end{equation}
The photon spectral indices can be used to describe $\alpha_{\nu}=\beta_{\gamma}+1$ and
$\beta_{\nu}=\alpha_{\gamma}+1$. It should be noted that the usage of a broken
powerlaw instead of the original Band function can lead to increased
uncertainties in the calculation as pointed out in~\cite{ignacio2007}. The use
of the original Band function is more accurate, since it is able to fit the
GRB spectrum very precisely over the whole energy band, $(20,\,2000)$~keV. A
complete Band fit, however,
cannot be extracted for many bursts. In the case of {\sc BATSE}, it was still
possible to reconstruct the entire Band function, while with {\sc Swift}, only
single powerlaw fits are possible in most of the cases. Also, in combination
with the uncertainties in the measurements, the difference between using a
broken powerlaw compared to a Band function may be negligible in many
cases. Therefore, for simplicity, a broken powerlaw is used in the following
paragraphs for cases where the entire bandwidth is detected. 

A second break at $\en=\ens$ appears in the neutrino spectrum which is caused by synchrotron radiation
of the neutrino-producing pions. Pions with sufficiently high energies suffer
synchrotron losses before decaying and do not produce neutrinos. Thus, the
high-energy tail of the neutrino spectrum is steepened by one power,
\begin{equation}
\frac{dN_{\nu}}{d\en}\,\en^2\propto (\en/\enb)^{-\beta_{\nu}}\,(\en/\ens)^{-1}
\mbox{ for }\en\geq \ens\,.
\end{equation}

The spectrum is normalized to the $\gamma$-ray fluence $F_{\gamma}$ which is
assumed to be proportional to the neutrino luminosity,
\begin{equation}
x\cdot
F_{\gamma}=\int_{E_{\min}}^{E_{\max}}\frac{dN_{\nu}}{dE_{\nu}}\,dE_{\nu}\approx
\ln(10)\cdot A_{\nu}\,.
\end{equation}
All parameters occurring in the following calculations are listed in table \ref{parameters}.
The factor $x$ is given by the product of the energy
transferred to the pions, $f_{\pi}$, a factor $1/8$ since half of the
photo-hadronic interactions result in four neutrinos and a factor
$1/f_e$ to account for the fraction of total energy in electrons compared to
protons in the jet~\cite{guetta}. 
The normalization constant $A_{\nu}$ is therefore given as
\begin{equation}
A_{\nu}=\frac{1}{8}\frac{1}{f_e}\frac{F_{\gamma}}{\ln(10)}f_{\pi}\,.
\label{anu:equ}
\end{equation}
In the following, the normalization of a single burst will be modified to a
quasi-diffuse normalization by multiplying $A_{\nu}$ with the number of bursts
per year ($1000$ bursts per year, of which $\eta=2/3$ are long and $\eta=1/3$
are short) and dividing the
result by $4\,\pi$ sr,
\begin{equation}
A_{\nu}'=\eta\cdot \frac{1000}{\mbox{yr }4\,\pi\mbox{ sr}}A_{\nu}\,.
\end{equation}
This number is comparable to the normalization which is achieved when using
the flux of UHECRs to normalize the spectrum, since the power in UHECRs is of
the same order as the electromagnetic emission of GRBs.
The total number of 1000 bursts per year is chosen to be consistent with
calculations given by Waxman and Bahcall in~\cite{wb97,wb99}. A estimate of $\sim 700$
bursts per year, however, is a more reasonable number when looking at the {\sc BATSE}
catalog~\cite{batse_cat}. In that case, all results could simply be
weighted by another factor of $2/3$.

The first break energy in the spectrum, $\enb$, is related to the break energy in the photon spectrum by
\begin{equation}
\enb= \frac{(m_{\Delta}^{2}-m_{p}^{2})\cdot \Gamma^{2}}{4\cdot (1+z)^{2}\cdot \egb}\,.
\end{equation}
It is determined through the minimal energy necessary to produce a
$\Delta$-resonance in the shock fronts of the bursts. 
The numerical values given in~\cite{ppb} for the proton mass, $m_{p}=0.94$~GeV,
and the $\Delta$ mass, $m_{\Delta}=1.23$~GeV, lead to 
\begin{equation}
\enb=7\cdot10^{5}\cdot (1+z)^{-2}\,\frac{\g25^{2}}{\egbM}\mbox{ GeV}
\label{enb_equ}
\end{equation}
in the observer's frame.

The second break energy is connected to the pion's synchrotron loss time. It depends on the neutrino flavor and for muon neutrinos, it is given as
\begin{equation}
\ens=\sqrt{\frac{3\,\pi\,\epsilon_e}{4\,\tau_{\pi}^{0}\,\sigma_{T}\,\epsilon_B\,L_{\gamma}}}\cdot
\frac{c^4\,t_v}{(1+z)\cdot m_{e}}\,\Gamma^{4}\,.
\label{ens_1}
\end{equation}
with the Thomson cross section $\sigma_T=0.665\cdot 10^{-24}$~cm$^{2}$ and
$\tau_{\pi}^{0}=2.6\cdot 10^{-8}$~s as the pion life time at rest.
For electron and anti-muon neutrinos, the break energy $\ens$ is about an order of
magnitude lower, since these neutrinos originate from muon decays. The muon
lifetime is about a factor of 100 higher than the pion lifetime, which gives a
lower energy threshold for synchrotron losses.
Here,
$\epsilon_b$ and $\epsilon_e$ are the fractions of the burst's internal energy going into the
magnetic field, respectively into electrons. 
The equipartition fractions have been set to $\epsilon_e=0.1$ and
$\epsilon_b=0.1$. There is no good way of determining the equipartition
fractions theoretically, yet. However, afterglow observations indicate values on
the order of $0.1$ for prompt emission~\cite{wax_equi}. The remaining parameters in
Equ.~(\ref{ens_1}) are listed in table~\ref{parameters}, with the values as used in the
following calculations. Inserting all numerical values gives
\begin{equation}
\ens=\frac{10^{8}}{1+z} \epsilon_{e}^{1/2} \, \epsilon_{b}^{-1/2}\, \g25^{4}
\,t_{v,-2}/\sqrt{\lumi} \mbox{ GeV}\,.
\label{ens_equ}
\end{equation}
A detailed derivation of the neutrino spectrum as presented above is given in
\cite{wax_equi}. 
%===========================================
\subsubsection{The average spectrum}
%===========================================
The average spectrum of prompt neutrinos from regular GRBs was calculated
in~\cite{wb99}, hereafter referred to as {\sl WB}. The following parameters were used,
\begin{eqnarray}
A_{\nu}&=&3\cdot 10^{-9}\,\diffunits\,,\\
\alpha_{\nu}&=&-1\,,\\
\beta_{\nu}&=&0\,,\\
\enb&=&10^{5}\mbox{ GeV}\,,\\
\ens&=&10^{7}\mbox{ GeV}\,.
\end{eqnarray}
Though the sensitivity of current neutrino detectors is not sufficient yet to detect
this flux (see also Section~\ref{nu_detection_methods}), they come close to the
predicted flux. With {\sc IceCube}, the model can therefore be tested in
detail.

A numerical calculation of the prompt neutrino spectrum from GRBs is given
in~\cite{murase_nagataki2006}. Multi-pion production is included in the
calculations by using the GEANT4 generator, and proton cooling was
calculated. This more precise approach results in a spectrum similar to the
one calculated in~\cite{wb99}, but with smooth transitions at the break
energies and a slightly different normalization.
%===========================================
\subsubsection{The parameters \label{params_section}}
%===========================================
The parameters in the neutrino flux calculations have been adjusted according
to the different classes of GRBs. A
summary of the mean values of the used parameters is given in
table~\ref{parameters}, for Gamma Ray Bursts, X-Ray Flashes (XRFs), Short Hard Bursts
(SHBs) and Short High-Energy (SHE) bursts. The different source classes are
discussed in section~\ref{classes}. 
\begin{table}[h!]
\centering{ 
\begin{footnotesize}
\begin{tabular}{|l|lllll|}
\hline 
Parameter&Symbol&GRB&XRF&SHB&SHE\\ \hline \hline
$\gamma$ fluence&$F_{\gamma}$ [erg/cm$^2$]&$10^{-5}$&$10^{-5}$&$10^{-5}$&$10^{-5}$\\ \hline
Redshift&$z$& $2$&$2$&$0.1$&$0.1$ \\ \hline
Lumi.~dist.&$d_l$ [Gpc]&$15$&$15$&$0.45$&$0.45$\\ \hline
$\gamma$ energy&$\eg$ [keV]&$(10,300)$&$(0.1,100)$&$(10^2,10^3)$&$(10^3,10^4)$\\ \hline
$\nu$ energy&$\en$ [GeV]&$(10^3,10^7)$&$(10^3,10^7)$&$(10^2,10^7)$&$(10^2,10^7)$\\\hline
Equipartition&$\epsilon_b$&$0.1$&$0.1$&$0.1$&$0.1$\\
fractions&$\epsilon_e$&$0.1$&$0.1$&$0.1$&$0.1$\\ \hline
$e^{-}-p$ energy ratio &$f_e$&$0.1$&$0.5$&$0.1$&$0.1$\\\hline 
Energy transferred to $\pi$ &$f_{\pi}$&0.2&$1$&0.2&0.2\\\hline 
Burst luminosity&$L_{\gamma}$ [erg/s]&$10^{51}$&$<10^{50}$&$10^{51}$&$10^{51}$\\ 
&$\lumi:=\frac{L_{\gamma}}{10^{52}\mbox{ erg/s}}$&0.1&$<0.01$&$0.1$&0.1\\ \hline
Boost factor&$\Gamma$&$300$&30&300&3000\\ \hline
&$\Gamma_{2.5}$&$1$&0.1&1&10\\ \hline
Spectral indices&$\alpha_{\nu}$&$-1$&$-1$&$0$&$0$\\
&$\beta_{\nu}$&$0$&$0$&$1$&$1$\\ \hline
$\gamma$ break energy&$\egb$ [MeV]&$0.1-1$&$0.1-0.001$&$0.1-1$&$>1$\\
&$\egbM:=$&$1$&$0.1$&$1$&$>1$\\ 
&$\egb/$MeV&&&&\\ \hline
$\nu$ break energies&$\enb$ [GeV]&$10^{5}$&$10^{6}$&$10^{5}$&$10^{5}$\\ 
&$\enbG:=\frac{\enb}{\mbox{GeV}}$&$10^{5}$&$10^{6}$&$10^{5}$&$10^{5}$\\\hline
&$\ens$ [GeV]&$10^{7}$&$10^{7}$&$10^{11}$&$10^{11}$\\
&$\ensG:=\frac{\ens}{\mbox{GeV}}$&$10^{7}$&$10^{7}$&$10^{11}$&$10^{11}$\\\hline
Time variability&$t_v$ [ms]&10&$10^{5}$&1&1\\ 
&$t_{v,-2}:=\frac{t_{v}}{10^{-2}\mbox{ s}}$&1&$10^{4}$&0.1&0.1\\ \hline 
Burst duration&$t_{90}$ [s]&40&40&0.2&0.2\\ \hline
\end{tabular}
\end{footnotesize}
\caption[Parameters for the prompt photon and neutrino emission]{Parameters
  for the prompt photon and neutrino emission in GRBs. The numbers quoted as
  typical values for GRBs, XRFs, Short Hard Bursts (SHBs) and Short High-Energy (SHE) bursts are to be taken as rough bench marks, since all of these
  parameters fluctuate strongly as emphasized in the text. The calculation of
  the luminosity distance at a redshift between $z=1-2$ for long bursts and
  $z=0.1$ for short events is done using
  cosmological parameters of $\Omega_m=0.3$, $\Omega_{\Lambda}=0.7$ and $h=0.72$.}
\label{parameters}
}
\end{table}

The boost factor $\Gamma$ is constrained to $100<\Gamma<1000$ for regular GRBs, since
for boost factors less than $100$, the medium would be optically thick to photons and
for $\Gamma>1000$, protons lose most of their energy to synchrotron radiation~\cite{hh_02}. The possibility of fluctuating
$\Gamma$ using the photon break energy is given as demonstrated in \cite{guetta}, but there are several arguments for
using a constant value: bursts can be misaligned which would lead to a
misinterpretation of the boost factor. Also, varying the break energy for each
single burst might implicitly already include boost factor fluctuations. Therefore,
a constant boost factor of $\Gamma=300$ is used in following calculations.

In previous publications, e.g.~\cite{guetta}, a variation of the energy
going into pions $f_{\pi}$ was discussed. Such a variation would further increase
the width of the distribution of the neutrino spectrum normalization. The burst
  luminosity, the boost factor, the photon break energy and the variability
  time influence $f_{\pi}$, as they are correlated as
\begin{equation}
f_{\pi}\sim 0.2\cdot \frac{\lumi}{\g25^{4}\,t_{v,-2}\,\egbM}\,.
\label{fpi}
\end{equation}
In the following calculations, this fraction will be kept at a constant value of
$f_{\pi}\sim0.2$ for classical GRBs for the following reasons: 
\begin{enumerate}
\item $f_{\pi}$ strongly depends on the
boost factor $\Gamma$ which will be used as a constant as discussed
before. The dependence on the other three parameters is only linear. The
variation of $\Gamma$ is more striking, since it is amplified by the
$\Gamma^4$ behavior. 
\item The main uncertainties in the current calculations result from the lack
  of knowledge of the parameters $\epsilon_e$ and $\epsilon_B$ as well as from
  uncertainties in the redshift relation which leads to uncertainties in
  $L_{\gamma}$. These three parameters are all important in the determination
  of the spectral normalization and thus a constant value is favorable.
\end{enumerate}
In the case of SHBs, the variability time is smaller, $t_{v,-2}\sim 0.1$ and
also the luminosity is about an order of magnitude lower,
$\lumi\sim0.1$. Thus, the pion energy constant stays the same and is assumed to be
$f_{\pi}=0.2$ for SHBs as well.

X-Ray Flashes (XRFs) will produce very high pion efficiencies due to their low
boost factors and break energies. $f_{\pi}\gtrsim 1$ indicates an optically
thick source in comparison to a classical GRB and yields a good target for
neutrino production.

The prediction of the diffuse neutrino flux derived from the model above is given in
\cite{wb97,wb99}, where the authors use average parameters to determine the shape
of the spectrum. Cosmological evolution of the sources was considered in
that model by taking into account the redshift evolution of GRBs. It is
assumed that GRBs follow the star formation rate, since they appear to be
connected to supernova-Ic explosions. Thus, the diffuse Waxman-Bahcall flux has to be
weighted by a factor $\eta=2/3$ or $\eta=1/3$ for long and short burst
samples respectively.\vspace{-0.5cm} 
%===============================================
\subsubsection{Conventional versus peculiar bursts \label{classes}}
%===============================================
As discussed in Section~\ref{multimessenger}, GRBs can be divided into different
sub-classes, such as long and short events and a further division of long
bursts into XRR bursts, XRFs and regular GRBs. While short and long bursts
differ in duration and in the hardness of the spectrum, XRR/XRFs and regular
GRBs differ in the energy range of prompt emission. A further class of short
bursts is predicted with a high-energy component, which has not be observed
yet, but which would be observable with {\sc GLAST}. 

\begin{figure}[h!]
\centering{
\includegraphics[width=\linewidth]{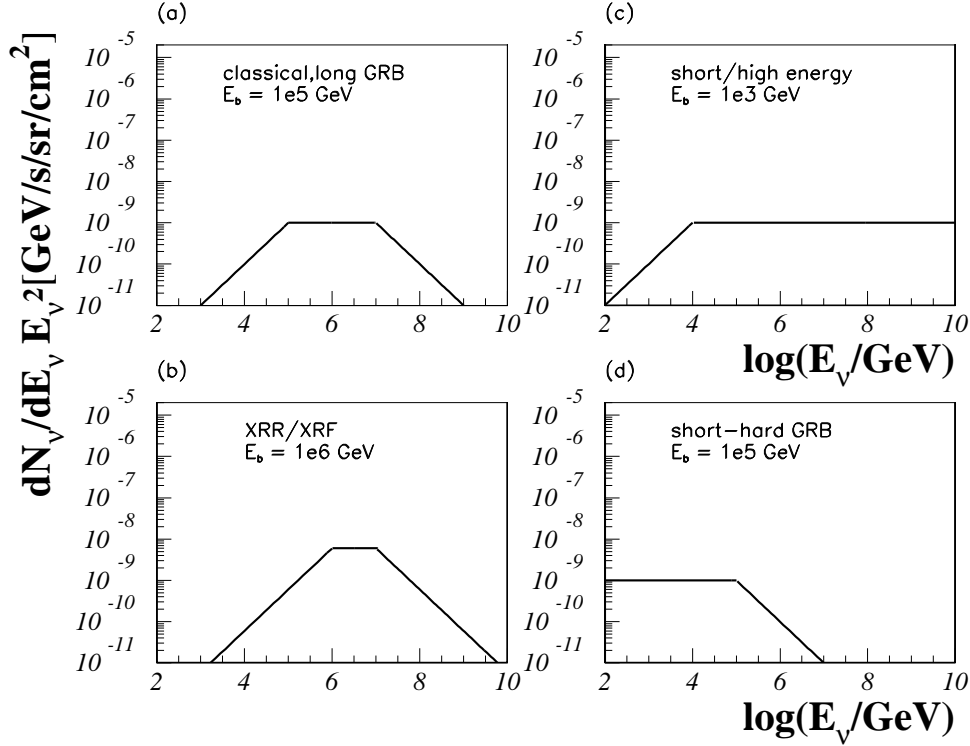}
\caption[Prompt neutrino emission from GRBs]{Prompt emission neutrino spectra from (a) a classical, long GRB as
  opposed to an XRR/XRF event (b) and from (c) short, high-energy bursts which
  are not observed yet compared to short, regular bursts as observed by {\sc Konus},
  {\sc Swift}, etc (d).}
\label{grb_prompt_nus}
}
\end{figure}

The photon spectral properties are reflected in the neutrino spectra. A
schematic representation of the neutrino spectra of the four different classes
is shown in Fig.~\ref{grb_prompt_nus}. Due to a low boost factor of
$\Gamma\sim 30$, XRR and XRFs have a lower break energy in the photon spectrum,
 which leads to higher neutrino break energy, $\enb\sim 10^{6}$~GeV
for XRR/XRFs (b) compared to $10^{5}$~GeV for regular GRBs (a). The second
break is at approximately the same energy, $10^{7}$~GeV due to the combination
of a lower boost factor and a higher variability time, $t_{v,-2}\approx
10^{4}$ due to large emission regions. XRRs and XRFs can actually be very
efficient neutrino emitters, given that for the low boost factor, the pion
parameter $f_{\pi}$ can become as large as $f_{\pi}=10^{3}$. Such a high value
of $f_{\pi}$ would, however, violate recent {\sc AMANDA} limits, see
Section~\ref{multimessenger}. Therefore, the flux is normalized to the current
limit in
Fig.~\ref{grb_prompt_nus}, which implies that the pion parameter must be
$f_{\pi}<4$. Other possibilities to get around the limit would be to use
smaller scales for the emission region and therefore smaller variability time
scales. This reduces the synchrotron break energy significantly. Also, there
could be much less XRR and XRF events than regular GRBs, which would affect
the normalization. The variation of the total luminosity plays an additional
role, affecting both the normalization and the second break energy.

The photon spectra of short bursts are typically harder than long ones which leads to a much softer neutrino spectrum of $\sim E_{\nu}^{-2}$ below the
break energy and $\sim E_{\nu}^{-3}$ above the break energy (see
Fig.~\ref{grb_prompt_nus} (c)). The class of high-energy short bursts would
produce a neutrino spectrum as displayed in
Fig.~\ref{grb_prompt_nus}~(d). There is, however, no experimental evidence yet
that such a class exists. While the photon spectrum would be observed at
$>10$~MeV energies, the corresponding neutrino spectrum is supposedly visible
at lower energies as indicated in the figure.

The reflections above show that the assumption of a (2:1) ratio of long to short bursts
and a (1:1:1) ratio of (GRB:XRR:XRF) would result in a modified spectrum which
can look significantly different from the spectrum which has so far been
classified as the typical neutrino GRB spectrum. 

In addition to the classification scheme as presented above, there are two
types of bursts which are of low photon luminosity, but can possibly be
sources of a significant flux of high-energy neutrinos: low-luminosity GRBs
like GRB060218 and choked bursts. The neutrino flux from low-luminosity GRBs
is estimated in~\cite{murase_lowlumi2006} to be at a level of
$dN_{\nu}/d\en\,{\en}^{2}\sim 10^{-9}\,\diffunits$ at $10^{6}$~GeV, which
would be observable by {\sc IceCube} after a few years of
observation. Neutrino emission from choked GRBs is predicted
in~\cite{meszaros_waxman2001}: While the
electromagnetic emission is not observed due to the optical thickness of the
event, e.g.~a supernova explosion into a dense SN-wind, neutrinos from proton
proton or proton photon interactions can still escape. Further acceleration of
secondary charged particles like muons, pions and kaons can lead to an
enhanced neutrino signal~\cite{koers_wijers2007}. Proton proton interaction
for instance would only produce high-energy neutrinos up to $\sim
0.1-1$~TeV. Assuming further acceleration, the maximum neutrino energy lies
above $10^{3}$~TeV and a flux of $dN_{\nu}/d\en\sim 5\cdot
10^{-8}\,\diffunits$ is expected for an ${\en}^{-2}-$type neutrino spectrum.

Apart from spectral fluctuations between different burst classes, there are
relatively large variations within the classes themselves. 

%===============================================
\subsubsection{Average versus single source variation}
%===============================================
A different approach to predicting the neutrino flux from GRBs is to look at each burst individually and add up the individual
spectra to make a prediction of the total flux from these sources. The study
of different catalogs are presented in~\cite{guetta,bshr2006,julia_phd} and
are summarized here. 

For each burst $i$ in a sample of $n$ GRBs, a prediction of the prompt
neutrino flux, $\Phi_i$, from
this source can be made as described above. With a
large volume neutrino telescope like {\sc AMANDA}, a large number of bursts can be analyzed simultaneously in order to
increase the detection significance. By stacking many bursts, the signal is increased more than the background. This makes it interesting to look at a
coincident spectrum of burst samples, i.e.~a quasi-diffuse flux. The total
flux $\Phi$ is given as
\begin{equation}
\Phi=\frac{\sum_{i}^{n}\Phi_i}{n}\,.
\end{equation}
Three different burst samples are used here for a systematic investigation:
\begin{enumerate}
\item[\ding{172}] {\sc BATSE} sample: the energy spectrum of 568 bursts from the {\sc BATSE} catalog are given
  in~\cite{guetta}. The neutrino spectra for the individual bursts was
  derived in~\cite{guetta} and in~\cite{bshr2006} using different parameter
  settings. Calculations here are following the latter
  publication~\cite{bshr2006}. Redshifts are determined using the variability
  method, see~\cite{fenimore_ramirez} for a description.
\item[\ding{173}] {\sc Konus} sample of short bursts: a catalog of 130 short bursts has
  been presented in~\cite{konus_short_cat}, of which 99 have fitted energy spectra
  and 34 are localized. Redshifts are set to $z=0.1$, if
  unknown\footnote{Short bursts are believed to occur when neutron-star/black hole
    binaries merge. These events typically happen in old regions, and the mean
  redshift is therefore expected to be as low as $z=0.1$.}.
\item[\ding{174}] {\sc Swift} sample of long bursts: measurements from {\sc Swift}-BAT
  have been published on the {\sc Swift} web-page~\cite{swift_web}. In this
  sample used here, those 188 long bursts with published spectral information
  until March 12, 2007 are included.
For the {\sc Swift} bursts without measured redshifts, $z=2$ is assumed.
\end{enumerate}
\begin{figure}[h!]
\centering{
\includegraphics[width=\linewidth]{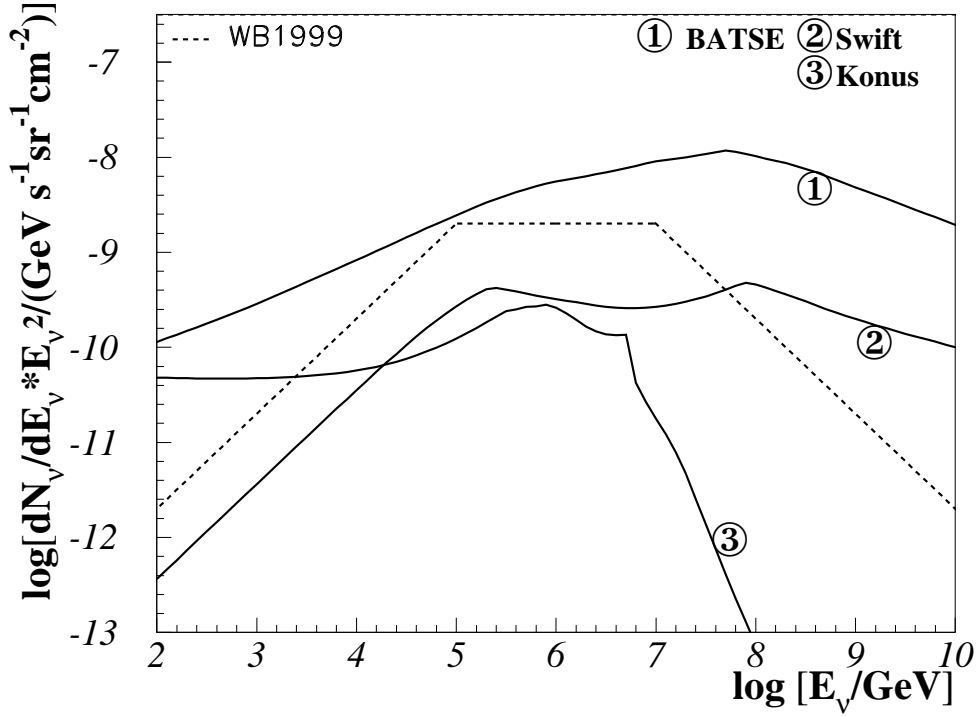}
\caption[Summary of the coincidence $\nu$ spectra of three GRB
samples]{Summary of the coincidence spectra of three GRB samples. Bursts in
  the {\sc BATSE} sample \ding{172} are typically quite strong. {\sc Swift}-detected bursts (sample \ding{173}) have a flat spectrum and a high
  second break energy, while the short {\sc Konus} bursts (sample \ding{174}) are generally very weak and produce a steep
  spectrum. All calculated mean spectra differ significantly from the
  average diffuse prediction made by Waxman and Bahcall (dashed line).}
\label{static_sum}
}
\end{figure}
To compare the different samples to each other, the coincidence fluxes of the
samples are shown in Fig.~\ref{static_sum}. Three major
conclusions can be drawn from the comparison of coincidence spectra.
\begin{enumerate}
\item While the spectral shape for the three samples of long bursts are {\em
  relatively} comparable, even small differences in the spectral indices
  result in order of magnitude deviations in certain parts of the
  spectrum. The main, systematic difference between the samples is the low
  normalization of the {\sc Swift} sample. Here, the reason lies in the
  properties of the {\sc BATSE} and {\sc Swift} experiments. While {\sc BATSE}
  had a wide field of view and could observe about $2\,\pi$~sr, its
  sensitivity was not as good as the one of {\sc Swift}. This fact leads to the
  observation of a large number of strong GRBs for {\sc BATSE} ($\sim 1$ per
  day) compared with a lower number ($\sim 100$~per year) of bursts with lower
  fluence for {\sc Swift}. Consequently, the neutrino flux which is
  proportional to the photon fluence is also lower. The differences in the
  field of view are corrected by assuming a GRB rate of 1000/yr for all of the
  samples. The coincidence
  spectra of the two {\sc BATSE}-samples are more than two orders of magnitude
  more intense than the {\sc Swift} sample only due to the fact that the
  increased number of bursts in the {\sc Swift} sample is not due to a large field
  of view but due to a good sensitivity. There are fewer strong bursts but a
  large number of low-intensity ones which reduces the total contribution of
  the signal. Such a bias from experimental data will be present in all of the
  experiments, since the results in each case will be dependent on
  properties such as energy range, sensitivity, field of view etc. 
\item The energy spectrum for short bursts is about one power steeper than the general
  flux for long bursts. This general conclusion is observable in the
  coincidence spectra. The neutrino spectrum from {\sc Konus} bursts is much
  steeper than the ones from long duration bursts. The reason lies in the hardness of the photon spectra. Since
  the short keV photon spectra are typically much harder than the long burst
  spectra, neutrino spectra for short bursts tend to be softer ($\gamma$ and
  $\nu$ energies are inversely proportional). This underlines that long and
  short bursts have to be treated separately. 
\item For all samples both the average and the coincidence spectrum differ from the
spectrum that is considered as the standard diffuse spectrum. 
\end{enumerate}

Due to the variation in the single source spectra, the detection rate for
single bursts in a neutrino detector like {\sc AMANDA} or {\sc IceCube} scatters over almost six orders of magnitude. There are a few
bursts with close to one event per km$^2$, while the mean of the distribution
rather lies around $R\sim 0.01$~yr$^{-1}$. The values of the mean for each sample
are given in table~\ref{average_rate_params}.

\begin{table}[ht]
\centering{\begin{footnotesize}
\begin{tabular}{|l||ll|}\hline
&mean $\nu_{\mu}$ rate/GRB [km$^{-2}$]&total number of $\nu_{\mu}$ [km$^{-2}$]\\
    \hline\hline
{\sc BATSE}&$10^{-2.05\pm0.67}$&14\\
WB [{\sc BATSE}]&$10^{-1.86\pm0.23}$&9.1\\ \hline
{\sc Konus}&$10^{-2.76\pm0.69}$&0.28\\
WB [{\sc Konus}]&$10^{-2.36\pm0.17}$&0.17\\ \hline
{\sc Swift}&$10^{-2.72\pm1.35}$& 3.4\\
WB [{\sc Swift}]&$10^{-1.95\pm0.17}$& 2.0\\ \hline
\end{tabular}
\end{footnotesize}
\caption[Mean neutrino spectra parameters]{Mean neutrino spectra parameters
  for the {\sc BATSE}, {\sc Konus} and {\sc Swift} samples. The
  standard deviation to the mean values was calculated as an error
  estimate.}
\label{average_rate_params}
}
\end{table}

Only 34 of the 99 sources with spectral information in the {\sc
  Konus} satellite are well-localized by the {\sc IPN3}. Consequently, the detection
  rates are only given for these 34 events. The single source parameterization
  causes a much wider scattering than the WB rates. Also, the mean detection
  rate is about half an order of
  magnitude lower in the case of single source parameterization (see table~\ref{average_rate_params}). The main
  reason lies in the steepness of the actual GRB spectra. These are typically
  about one power steeper than assumed in the WB spectrum and have a
  relatively low flux at the main detection energy range in neutrino Cherenkov
  telescopes ($\en>10^{4}-10^{5}$~GeV). This demonstrates again that at least
  systematic differences in the spectra need to be taken into account.
For the 165 long bursts with given localization, the results from previous samples
can be confirmed. Here, the systematic shift between WB rate and single source
parameterization comes from the lower normalization of {\sc Swift} bursts. 

While the detection of a single burst will be
difficult even with instrumented arrays of one cubic kilometer like the ones planned for
{\sc IceCube} or {\sc KM3NeT} (the most intense bursts in the detector have a
detection probability of $<1$~neutrino event per km$^2$ effective area), the stacking of photon-detected GRBs
increases the significance immensely, so that the model can be tested in the
very near future.
GRB analyses usually make use of the temporal and spatial
information from the GRB satellites. GRB030329 was analyzed and it turns
out that the single source sensitivity in this case is several orders
of magnitude above the predicted flux~\cite{mike_icrc05}. The burst was
analyzed adjusting the spectrum to the observed properties of the GRB, similar
to how it is described above. An alternative approach like the one done by {\sc AMANDA},
is the stacking of all bursts in a sample using the WB spectrum for all of the
bursts.
By stacking data from the
sample of neutrino-induced muons~\cite{kyler_paper} or
cascades~\cite{grbs_cascade07} according to the arrival direction and
their arrival time, the sensitivity to a diffuse flux from GRBs according to
WB can be determined. An analysis looking for the temporal clustering of
events within small time bins (100~s) is done for cascade-like events~\cite{grbs_cascade07}. Such a
search is independent of the photon-detection of GRBs and therefore also
includes choked bursts and bursts which are missed by GRB satellites.
%===============================================
\subsection{Afterglow neutrinos}
%===============================================
For the afterglow of a Gamma Ray Burst, it is typically assumed that the
emission arises when the burst hits the interstellar medium. The reverse
shocks accelerate charged particles and electrons emit synchrotron radiation at X-ray
energies. 
The spectrum is observed to decay as
\be
\frac{dN_{X}}{dE_{X}}\propto {E_{X}}^{-\alpha_{X}}\,.
\ee
The characteristic synchrotron frequency at
which the synchrotron cooling time becomes comparable to the ejecta expansion
time, $\exb$ lies below keV-energies,
\be
\exb\approx
0.3\,\epsilon_{B,-2}^{-3/2}\,n_{0}^{-1}\,E_{53}^{-1/2}\,t_{90}^{-1/2}\mbox{ keV}\,.
\ee
Here, $E=:E_{53}\cdot 10^{53}$~erg is the isotropic, kinetic energy of the
fireball and $\epsilon_{B,-2}:=\epsilon_B\cdot 10^{-2}$.
At energies $E_{X}>\exb$, the spectrum steepens
\be
\frac{dN_{X}}{dE_{X}}\propto {E_{X}}^{-\beta_{X}}
\ee 
and $\beta_{X}=\alpha_{X}+1/2$.
On average, the spectral indices scatter around $\alpha_X\sim 3/2$ and
$\beta_X\sim 2$. Just as for the
prompt emission, individual bursts can have spectral indices deviating
significantly from the average value as it is investigated for the {\sc Swift}
sample later this section.

Provided that proton and pion synchrotron losses can be
neglected\footnote{Reverse shocks happen at radii of $\sim 10^{17}$~cm, which
  is large enough to avoid synchrotron losses~\cite{afterglow_nus}.}, the
afterglow neutrino spectrum is then given as
\be
{\en}^{2}\cdot \frac{dN_{\nu}^{a}}{dE_{\nu}}=A_{\nu}^{a}\cdot\left\{
  \begin{array}{lll}
{\en}^{-\alpha_{\nu}^{a}}&&\mbox{for} \en<\enb\\
{\en}^{-\beta_{\nu}^{a}}&&\mbox{for} \en>\enb\,.
\end{array}
\right. 
\ee
The normalization $A_{\nu}$ can be determined by assuming that UHECRs are
produced by GRBs, implying that the power in
UHECRs is comparable to the electromagnetic output from Gamma Ray Bursts, 
\be
\int E_{CR}\,\frac{dN_{CR}}{dE_{CR}}\,dE_{CR}\approx 10^{44}\mbox{ erg\,Mpc}^{-3}\mbox{yr}^{-1}\,.
\ee
The neutrino break energy is
inversely correlated to the photon critical energy $\exb$ as
described in Equ.~(\ref{enb_equ}) with $\exb\approx 0.1-1$~keV. Therefore, 
\be
\enb= 0.7-7\cdot 10^{9} (1+z)^{-2}\mbox{ GeV}
\ee
for a boost factor of $\Gamma\sim 300$.
The spectral indices are given as $\alpha_{\nu,a}=-\beta_{X}+1$ and
$\beta_{\nu,a}=-\beta_{X}+1.5$. With an average value of $\beta_{X}\sim 2$,
the neutrino spectrum behaves as
\be
{\en}^{2}\cdot \frac{dN_{\nu}^{a}}{dE_{\nu}}=A_{\nu}^{a}\cdot\left\{
  \begin{array}{lll}
{\en}^{1}&&\mbox{for} \en<\enb\\
{\en}^{1/2}&&\mbox{for} \en>\enb\,.
\end{array}
\right. 
\ee
The break energy is fixed in~\cite{afterglow_nus} to $\enb=10^{8}$~GeV, and
the normalization factor is calculated to be
\be
A_{\nu}=10^{-10}\,\diffunits\,.
\ee
The maximum energy is again given by the maximum proton energy with $\ep^{\max}\sim
10^{21}$~eV and $\en^{\max}=\ep^{\max}/20\approx 5\cdot 10^{9}$~GeV.
\subsection{Single source spectra}
\begin{figure}[h!]
\centering{
\includegraphics[width=\linewidth]{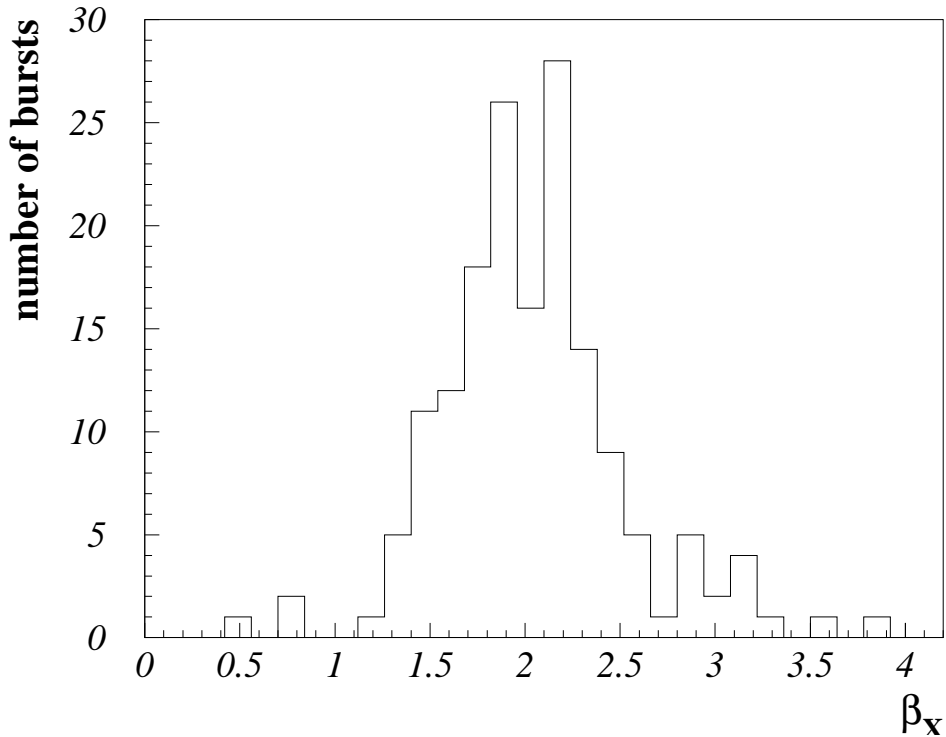}
\caption[Histogram of the afterglow spectral index $\alpha_X$ for long {\sc
  Swift} bursts.]{Histogram of the afterglow spectral index $\alpha_X$ for long {\sc
  Swift} bursts.
\label{alphax}
}
}
\end{figure}

\begin{figure}[h!]
\centering{
\includegraphics[width=\linewidth]{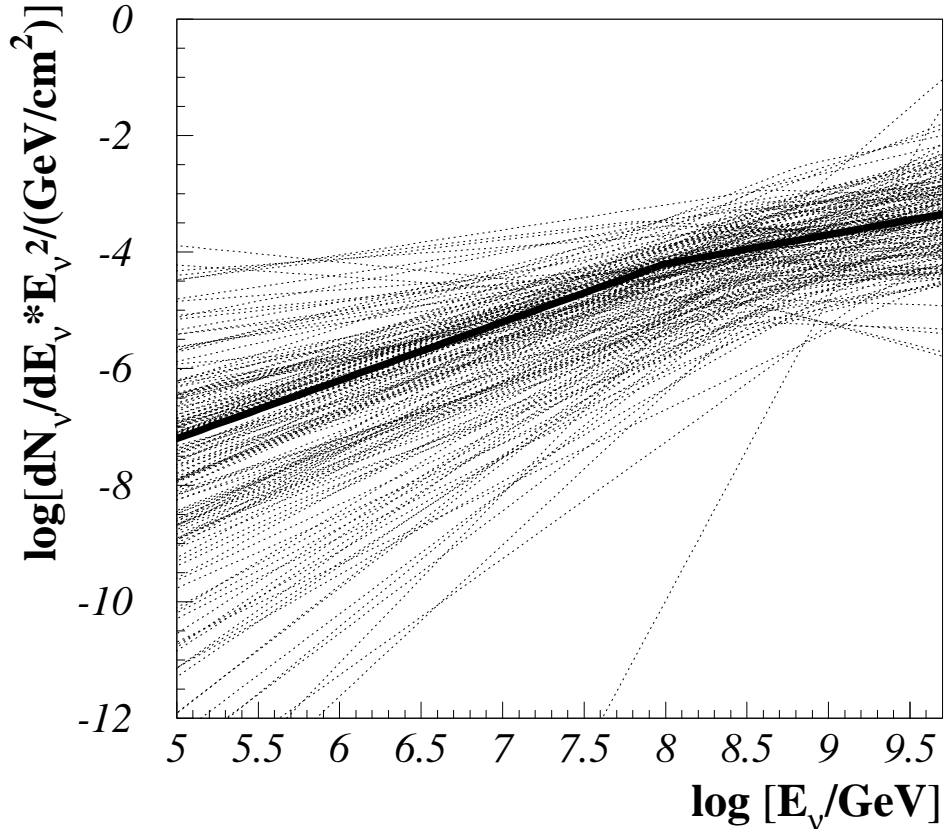}
\caption[Single source spectra from {\sc Swift}-detected GRB
  afterglows.]{Single source spectra from {\sc Swift}-detected GRB
  afterglows (dotted lines). The solid, thick line indicates the mean spectrum as calculated in~\cite{afterglow_nus}.
\label{single_afterglow}
}
}
\end{figure}

The result as presented above relies on the averaged parameters as observed for
GRBs. Single source spectra can in some cases deviate significantly from the
mean spectrum. {\sc Swift}-XRT-detected bursts are given on the {\sc Swift}
web-page~\cite{swift_web} (as of August 2007). Since XRT is sensitive in the energy range of
$(0.2,\,10)$~keV. Single powerlaw fits are presented and for the break
energy, the lower energy threshold $\exb\approx 0.2$~keV can be used,
resulting in a neutrino break energy of
\be
{\enb}^{Swift}\approx3.5\cdot 10^{9}\cdot (1+z)^{-2}\mbox{ GeV}\,.
\ee
For those bursts without given redshift, $z=2$ is used. The distribution of
spectral indices is shown in Fig.~\ref{alphax}. The mean value lies near
$\beta_X\approx 2$.
Figure~\ref{single_afterglow} shows the spectrum for each individual GRB in
the sample.
The normalization calculates as it was done for the prompt emission,
assuming that a single burst produces a fraction of the total flux of charged
Cosmic Rays, resulting in the neutrino normalization factor of
\be
A_{\nu}^{a}=\frac{1}{8}\frac{1}{f_e}\frac{F_{\gamma}}{\ln(10)\cdot t_{90}}f_{\pi}\,.
\ee
Again, $f_{\pi}\approx 0.2$, but this should be considered
as an upper limit, since the efficiency for pion production in afterglow
emission can be reduced significantly~\cite{afterglow_nus,rice_grbs07}. The
maximum energy for neutrino emission is ${\en}^{\max}\approx 10^{9.5}$~GeV,
given that protons are accelerated up to $10^{12}$~GeV, and neutrinos carry
$1/20$th of the proton energy.
Figure~\ref{single_afterglow} shows that the single source spectra (dotted lines) 
deviate from the mean spectrum (solid, thick line) by up to two orders of
magnitude. Furthermore, the spectral index varies from burst to burst by more
than one power.
\subsection{Neutrinos and early afterglow}
As discussed in Section~\ref{multimessenger}, the rapid slew of {\sc XRT} after GRB
alerts from {\sc BAT}, both on board of the {\sc Swift} satellite, made early
afterglow detection possible. It is discussed
in~\cite{murase_early_afterglow2007} that in the case of late internal shocks,
neutrino production via proton-photon interactions can be very efficient,
while reverse shocks cannot produce high-energy neutrinos.
In particular, short X-ray flares as observed by {\sc Swift} can be connected to neutrino emission in case of hadronic particle
acceleration within the frame of the late internal shock model~\cite{murase_xrayflares2006}. For neutrino telescopes like {\sc IceCube}, a
diffuse flux from X-ray flares can be tested, the flux level is near the {\sc
  IceCube} sensitivity level for one year of detection ($dN_{\nu}/d\en\cdot
{\en}^{2}\sim 10^{-8.5}\,\diffunits$ at $10^{6}$~GeV).
%%%%%%%%%%%%%%%%%%%%%%%%%%%%%%%%%%%%%%%%%%%%
\subsection{Neutrinos from GRBs: presence and future}
%%%%%%%%%%%%%%%%%%%%%%%%%%%%%%%%%%%%%%%%%%%%
Neutrinos can originate in the three phases of a GRB. Neutrino emission is
expected prior to the GRB, when the pre-GRB interacts with the central pulsar
or the SNR shell~\cite{precursor_nus}. The prompt emission is accompanied by a
neutrino flux if the observed spectrum of charged Cosmic Rays comes from
GRBs~\cite{wb97,wb99}. The afterglow emission can for the same reason be
accompanied by a neutrino signal~\cite{afterglow_nus}.

Searches for neutrinos from GRBs have been performed by the {\sc AMANDA}
experiment. In the case of muon neutrinos, directional information provided by
 {\sc IPN3} was used in order to reduce the atmospheric
background. No significant GRB signal was detected. The neutrino flux limit
from {\sc AMANDA} data as presented in~\cite{kyler_paper} comes
close to the flux as predicted in~\cite{wb99},
\be
\dl=6\cdot 10^{-9}\,\diffunits\,,
\ee
for a WB-like spectrum.
With {\sc IceCube}, it will
be possible to constrain the model. In the case of cascade-like events, two
analyses have been performed. The first one is also based on the selection of
events according to IPN-3-detected bursts. The second one does not use
photon-based measurements, but simply looks for unexpected temporal clustering
of events. The cascade-limits are still far above the predictions and no
general conclusions can be drawn yet.
The {\sc RICE} experiment has investigated five bursts with respect to a
possible flux connected to the afterglow emission~\cite{rice_grbs07}, with
limits a few orders of magnitude above the prediction, derived for each burst
individually, following~\cite{afterglow_nus}. With the stacking of more
bursts, it can be possible to improve such limits. Larger detection arrays
like {\sc AURA} or {\sc SalSA} will have increased effective areas and will
hopefully be possible to detect neutrinos from GRB afterglows.

The investigation of individual burst fluctuations has revealed that the
detection rate fluctuates within each class between $\sim
10^{-5}-1$~events/km$^{-2}$/burst, see~\cite{guetta,bshr2006,julia_phd}. The systematic
variations between the flux predictions for different burst samples come from
the bias connected to each of the satellites. The energy band pass
and the field of view of the detector
are important features which lead to a selective sampling of events. For
instance, when comparing {\sc BATSE} and {\sc Swift} bursts, the latter are on
average less luminous, since {\sc Swift} has a smaller field of view and a
better sensitivity than {\sc BATSE}. Short bursts detected by {\sc Swift}
typically do not belong to the class of short-hard bursts. This is a new
sub-GRB-class detected by {\sc Swift}, which has not been observed
before~\cite{sakamoto}. {\sc Swift} can detect these events due to its low energy band pass of
$(15,\,150)$~keV.

In the future, {\sc GLAST} will open the possibility of observing the
high-energy component of GRBs, probably revealing new classes of GRBs. TeV
photon detectors like {\sc MILAGRO} and {\sc MAGIC} can observe the component
of very high-energy photons in coincidence with {\sc GLAST} in order to
improve the understanding of high-energy emission from GRBs, see e.g.~\cite{milagro_gamma_santafe2007,magic_gamma_santafe2007,glast_gamma_santafe2007}. It is
likely that the differences between these classes are intrinsic as they are
for long and short bursts. The detection of neutrinos in the three different
phases can contribute to distinguishing the different sub-classes and it can also to
explain the underlying physics. With the completion of {\sc IceCube} in 2011,
first results can be expected, constraining current neutrino flux models.

%%%%%%%%%%%%%%%%%%%%%%%%%%%%%%%%%%%%%%%%%%%%

%starburst Neutrinos
%%==============================================================
\section{Neutrinos from starburst galaxies\label{starbursts}}
%%==============================================================
The emission of neutrinos from starburst galaxies was suggested
by~Loeb \& Waxman~\cite{lw06} (LW). The observation of synchrotron radiation
at radio wavelengths implies the existence of relativistic electrons, which
have a characteristic energy of $\sim 1$~GeV, assuming that magnetic fields in
starburst galaxies exceed the galactic magnetic field by two orders of
magnitude. If protons are co-accelerated along with the electrons, neutrinos
can be produced in proton-proton interactions. For high densities of
$n>100$~cm$^{-3}$, starburst galaxies serve as a ''proton calorimeter'', in
which all protons are absorbed in $p\,p$ interactions to produce
neutrinos. There are three crucial parameters for the calculation of the neutrino spectrum
from starburst galaxies:
\begin{enumerate}
\item {\it The spectral behavior} -- It is assumed by LW that the neutrino
  spectrum follows the induced proton spectrum, leading to a spectral index of
  $\alpha_{\nu}\approx\alpha_{p}\sim2.00-2.25$. The maximum energy for the neutrino spectrum is
  given by ${\en}^{\max}=0.1$~PeV, derived from the fact that starbursts serve
  as proton calorimeters up to a proton energy of $\ep\lesssim3$~PeV. Here, the
  ratio between proton and neutrino energy is $\ep/\en=20$.  
\item {\it The normalization of the total spectrum} -- In order to estimate
  the intensity of the neutrino flux, the synchrotron radiation from electrons
  is used. This can be done using the following connection between the different
  luminosities: The observed
  synchrotron emission is directly connected to the energy flux of the primary
  electrons,
\be
E_{e}^{2}\,\frac{dN_e}{dE_e}\approx 2\,\nu_{synch}\,L_{\nu,synch}\,.
\label{ne_lsynch}
\ee
Here, $dN_{e}/dE_{e}$ is the energy spectrum of the primary electrons with an
energy $E_e$, $\nu_{synch}$ is the characteristic frequency of the
synchrotron photons and $L_{\nu,synch}$ is the synchrotron luminosity per
frequency interval. The factor $2$ arises from the fact that
$\nu_{synch}\propto E_{e}^{2}$, so that the energy interval $\Delta E_{e}$
corresponds to twice the width of the synchrotron frequency interval, see
Section~\ref{radiation:general_crs}. Assuming a fixed ratio of protons to
electrons, $\eta_{p/e}=6$ for starbursts, the proton luminosity is connected
to the electron luminosity as
\be
\ep\,\frac{dL_p}{d\ep}=\eta_{p/e}\cdot E_{e}^{2}\,\frac{dN_e}{dE_e}\,.
\label{lp_ne}
\ee
If all protons are absorbed in $p\,p-$interactions, the proton luminosity
corresponds to the neutrino luminosity, with a reduction factor $1/3$,
accounting for the fact that $2/3$ of the $\Delta-$resonance processes goes
into charged pions and $1/2$ of the charged pion's energy goes into neutrinos,
\be
\en\,\frac{dL_n}{d\en}=\frac{1}{3}\cdot f\cdot \ep\,\frac{dL_p}{d\ep}\,.
\label{lp_ln}
\ee
Here, $f$ accounts for the fraction of starbursts serving as a
calorimeter. It is assumed by LW that all starbursts are calorimeters,
resulting in $f=1$.
Combining Equations~(\ref{ne_lsynch}), (\ref{lp_ne}) and (\ref{lp_ln}) gives a
connection between synchrotron radiation and neutrino emission as
\be
\en\,\frac{dL_n}{d\en}=\frac{1}{3}\cdot f \cdot \eta_{p/e}\cdot
2\,\nu_{synch}\,L_{\nu,synch}=4\cdot \nu_{synch}\,L_{\nu,synch}\,.
\ee
The normalization is performed by using the observed synchrotron spectrum at
$\nu_{synch}=1.4$~GHz, which corresponds to an electron energy of
$E_e=1$~GeV. The value is calculated by using the correlation between radio
and FIR emission. Here, it is assumed that the entire FIR background arises
from the emission from starburst galaxies. 
\item {\it The source evolution function with redshift} -- The distribution of
  starburst galaxies in the sky is taken into account, using recent models of
  star formation rate as discussed in~\cite{lw06} and \cite{tqwl06}.
\end{enumerate}
The resulting neutrino flux is given as
\be
{\en}^{2}\,\frac{dN_{\nu}}{d\en}(1\mbox{GeV})=\frac{c}{4\,\pi}\cdot \zeta\cdot
t_H\cdot \left[4\,\nu\,L_{\nu}\right]_{\nu_{synch}=1.4\mbox{GHz}}
\label{sb_flux:equ}
\ee
assuming that the characteristic electron energy corresponds to the neutrino energy,
$E_e\approx \en\approx 1$~GeV. The factor $\zeta\approx 0.5$ takes into
account the source evolution which follows the star formation rate and $t_H$
is the age of the Universe. This flux gives the total muon-neutrino
signal for a flavor ratio at the source. If, on the one hand, a mixing of neutrinos from
$(\nu_e,\,\nu_{\mu},\,\nu_{\tau})=(1:2:0)$ at the source to
$(\nu_e,\,\nu_{\mu},\,\nu_{\tau})=(1:1:1)$ is assumed, the muon-neutrino flux
must be divided by a factor of two. On the other hand, if muon-neutrinos and anti-muon neutrinos are
considered, the flux is enhanced by a factor of two. Thus, the flux which can
be measured with large volume neutrino detectors in the muon-neutrino channel
corresponds to the result of Equ.~(\ref{sb_flux:equ}). Numerically, the
neutrino flux is given as
\be
{\en}^{2}\,\frac{dN_{\nu}}{d\en}= 10^{-7}\,\left(\frac{\en}{1\mbox{ GeV}}\right)^{2-\alpha_p}\,\diffunits\,.
\ee
An ${\en}^{-2}-$signal is already close to recent {\sc AMANDA}
limits~\cite{jess_diffuse}. If the induced proton flux is steeper,
i.e.~$\alpha_{p}\approx 2.15$ as suggested by Cosmic Ray data, the flux is
more difficult to observe with {\sc AMANDA}, and next generation experiments
like {\sc IceCube} and {\sc KM3NeT} are needed to test the model.

The question on the fraction of the diffuse far infrared (FIR) flux
coming from starbursts is still unclear. While LW assume that $100\%$
of the detected signal comes from starbursts, Stecker~\cite{stecker_sbg06} presents that
the fraction is only 23\% on average. On the whole, the diffuse flux from
starbursts is probably lower than predicted: LW assume that starbursts are loss
dominated, which means that most primaries interact and do not escape the
source. This enhances the neutrino flux, since basically all protons lose
their energy in proton-proton interactions and produce
neutrinos. This fact is reflected in the variable $f$ which, for the
loss-dominated case of a proton calorimeter is $f=1$. Observations of the spectral radio index of the sources ($S\sim
\nu^{-0.8}$) indicate, however, that starbursts are in the diffusion limit, indicating
that a negligible fraction of protons interact and only few neutrinos are
produced. With {\sc IceCube} soon reaching the sensitivity level of the
prediction of LW, it will be possible to solve this question. 

There is another possibility to expect enhanced neutrino emission
from starbursts. In the past few years, it could be shown that long GRBs are
typically connected to the explosion of Wolf-Rayet stars into a supernova
Ic. These occur preferably in star forming regions. Thus, a diffuse flux of
GRBs similar to the prediction of~\cite{wb97,wb99} should originate from the direction of these galaxies. There
are two different ways to normalize the diffuse GRB spectrum. One method is to
assume that the observed keV-photon flux is proportional to the neutrino
flux. In that case, the normalization is dependent on the number of observed GRBs
per year. This number is strongly dependent on the instrument and
the number is not very exact. Under the assumption that GRBs accelerate
protons up to the highest energies, $E_p\sim 10^{21}$~eV, the neutrino
spectrum can also be normalized to the flux of ultra high-energy cosmic rays
(UHECRs). In this case, the normalization is independent of GRB
observations. It should be kept in mind that the spectral index of the
spectrum still varies from burst to burst - in the model of Waxman\&Bahcall,
an average spectral index was used. 

It is possible to look for a neutrino signal from GRBs by stacking starburst
galaxies. This method has one advantage over a triggered-GRB search: it is
a systematic search, since independent of GRB
data. A disadvantage is that only nearby events can be included, since the
sample of starbursts only reaches out to redshifts of $z=0.07$. It should,
however, be possible to use {\sc IRAS}
data to identify starburst galaxies at higher redshift. The search for a GRB
signal from starbursts should be considered as a systematic search for choked and
undetected GRBs. The
sources can be selected according to their FIR-flux, since this is a measure
of the SN rate in a starbursts. A higher FIR flux indicates a high star formation rate,
thus more SNe and therefore, also more GRBs.

%%%%%%%%%%%%%%%%%%%%%%%%%%%%%%%%%%%%%%%%%%%%

%%%%%%%%%%%%%%%%%%%%%%%%%%%%%%%%%%%%%%%%%%%%
%GZK Neutrinos

%==============================================================
\section{Cosmogenic Neutrinos\label{nus_gzk}}
%==============================================================
The prediction of the GZK cutoff at energies in the
spectrum of UHECRs due to proton interactions with the CMB implies the existence of a flux of ultra high-energy
neutrinos. The production scenario and the expected event rates of such a flux
of cosmogenic neutrinos are reviewed here.
%-------------------------------------------
\subsection{Production of cosmogenic neutrinos\label{gzk_nus:general}}
\begin{figure}
\centering{
\epsfig{file=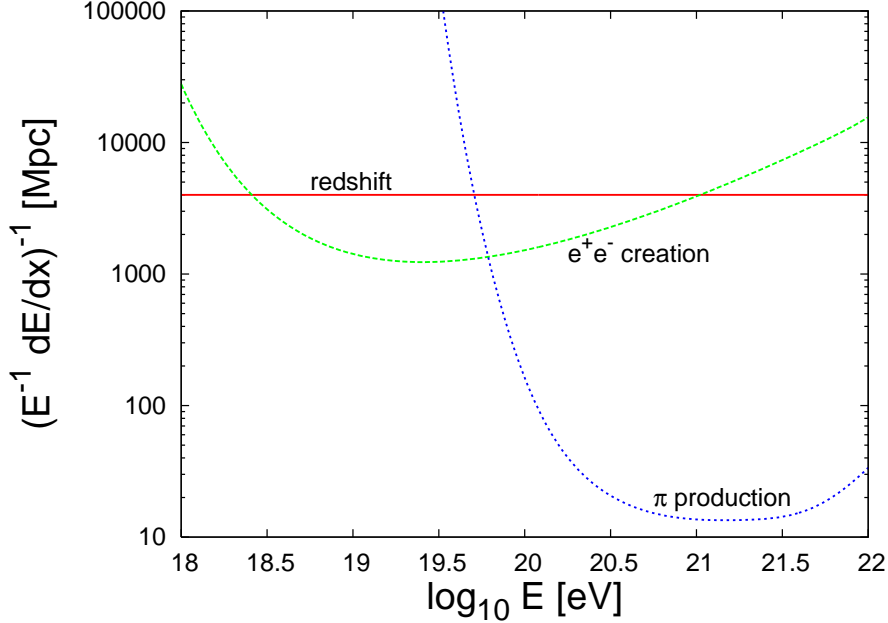,angle=270,width=12cm}
\caption{Attenuation length as a function of energy. The straight line
  represents adiabatic energy losses due to the expansion of the Universe. The
  upper (green) line shows energy losses due to Bethe-Heitler pair production,
  while the lower curve is the attenuation length resulting from
  pion-production processes. Reprinted from~\cite{hooper_taylor2007}, with permission from Elsevier.\label{energy_losses_gzk}}
}
\end{figure}

%-------------------------------------------
High-energy protons interact with the CMB on
their way to Earth via Bethe-Heitler pair production,
\begin{equation}
p\,\gamma_{CMB}\rightarrow p\,e^{+}\,e^{-}
\end{equation}
and via the $\Delta-$resonance,
\be
p\,\gamma_{CMB}\rightarrow \Delta\,.
\ee
The energy threshold for Bethe-Heitler pair production is $E_{th}^{BH}=5\cdot 10^{18}$~eV,
while it is higher for the $\Delta-$resonance, $E_{th}^{\Delta}\approx 5\cdot 10^{19}$~eV. The
attenuation length $\lambda=\ep/(-d\ep/dx)$ determines the propagation radius of
the protons from a given source, with $E_{p}$ as the energy of the proton and
$dE_{p}/dx$ as the mean energy loss rate. The latter depends on the photohadronic
cross sections as well as on the CMB photon field density, and is determined
by Monte Carlo
methods. Figure~\ref{energy_losses_gzk} shows the simulated attenuation length
as presented in~\cite{hooper_taylor2005}. Below $10^{19}$~eV, Bethe-Heitler pair production is
the dominant process, before the production of the $\Delta-$resonance becomes
dominant above its threshold energy, $\ep>5\cdot 10^{19}$~eV. Pair production
results in an attenuation length of $\lambda \sim 1000$~Mpc above $10^{19}$~eV, while the
$\Delta-$ resonance allows for the undisturbed propagation over a range
of only $\sim 10$~Mpc. The latter is therefore the dominant process leading to
the prediction of the so-called GZK cutoff at the highest energies,
i.e.~$\ep>5\cdot 10^{19}$~eV.

Neutrinos are produced in the decay of the $\Delta-$resonance via the channel of charged pion production, see
Section~\ref{multimessenger}. The resulting diffuse neutrino flux
is usually referred to
as the flux of {\it cosmogenic neutrinos} or {\it GZK neutrinos}. The neutrino flux ranges
from approximately $\en\sim 10^{16}$~eV up to $\en\sim 10^{21}$~eV. The shape
and strength of the neutrino flux depends on various parameters which are
described in the next paragraph.
\subsubsection{The parameter space}
The strength and shape of the cosmogenic neutrino flux depends on  seven factors:
\begin{itemize}
\item[(i)] the energy spectrum of primary protons - spectral shape, normalization and energy cutoff,
\item[(ii)] the proton-photon cross section $\sigma_{p\,\gamma}$,
\item[(iii)] the CMB photon density $n_{\gamma}^{CMB}$,
\item[(iv)] the evolution of proton-emitting sources with redshift,
\item[(v)] the composition of UHECRs,
\item[(vi)] magnetic field deflections,
\item[(vii)] neutrino oscillations\,.
\end{itemize}
\paragraph{Primary spectrum}
The spectral shape and normalization of the sources determines the cosmogenic
neutrino flux. Typically, it is assumed that all sources follow a powerlaw
with an exponential cutoff at a critical energy $E_{c}$,
\be
\frac{dN_{p}}{d\ep}=A_{p}\cdot \ep^{-\alpha_p}\cdot \exp\left(-\frac{\ep}{E_{c}}\right)\,.
\ee
Such a spectrum can be produced in shock fronts of AGN or GRBs, see
Section~\ref{multimessenger}. Monte Carlo simulations indicate a spectral
index close to $\alpha_p\approx 2$. A generic spectrum with an exact value of $\alpha=2$ is
therefore used for all sources in calculations of the cosmogenic neutrino flux. However, it
must be noted that recent results from test particle shock acceleration
indicate that this value can deviate from $\alpha_p=2$ depending on the exact
shock configuration, see e.g.~\cite{baring04,mbq2007,stecker_baring2007}.
The spectral powerlaw behavior is in accordance with the powerlaw structure of
the UHECRs, which are used for the normalization of the total proton spectrum
at Earth. This way, no concrete normalization for the single sources needs to
be given.
\paragraph{Cross section}
Cross section measurements from accelerator data are used in order to
estimate the fraction of energy going into neutrinos and the number of neutrinos produced in one interaction. This is done for
the complete energy spectrum at the given redshift $z$, and the spectra are
propagated from the source to Earth, using a large number of discrete steps of
$\Delta z$. The value of $\Delta z$ is chosen to be much smaller than the
average mean free path. 
\paragraph{CMB photon density}
The photon density of the cosmic microwave background, $n_{\gamma}^{CMB}$, is well known since its
observation with the {\sc COBE}\footnote{{\bf CO}smic {\bf B}ackground {\bf E}xplorer} satellite~\cite{cobe}.
The spectrum can be fitted with a blackbody spectrum of $T\sim2.7$~K,
\be
n_{\gamma}^{CMB}(\nu)\propto \nu^{3}\cdot\left(\exp\left[\frac{h\nu}{k\cdot T}\right]-1\right)^{-1}\,.
\ee 
\paragraph{Evolution of source population with redshift}
Typically, it is assumed that ultra high-energy protons are
produced in AGN or GRBs, see Section~\ref{multimessenger}. These objects are
believed to follow the evolution of star formation and the redshift behavior
is often assumed as $\rho \sim (1+z)^{m}$ up to a redshift $z=z_{\max}$, with
$m>0$, e.g.~\cite{schmidt72}. At high redshifts, $z>3$~to~$4$, the source number counts become small
and the exact behavior of for example the distance distribution of AGN is difficult to
determine. 
\paragraph{Composition of UHECRs}
In most neutrino flux predictions, it is assumed that a dominant fraction to
the spectrum of UHECRs are protons. While protons are very efficient neutrino
generators via $p\,\gamma-$interactions, heavier nuclei lose energy via
photo-disintegration~\cite{stecker_photodisint1969}. For instance, iron nuclei
interacting with CMB photons induce the following reaction:
\begin{eqnarray}
^{56}\mbox{Fe}\,\gamma_{CMB}\rightarrow &^{55}\mbox{Mn}&\,p\\
&^{55}\mbox{Mn}&\,\gamma_{CMB}\rightarrow ^{54}\mbox{Mn}\,n\\
n\rightarrow p\,e^{-}\,\overline{\nu}_{e}\,.
\end{eqnarray}
In these processes, only very few neutrinos are produced, i.e.~via neutron
decays. Most
models neglect the fraction of heavy nuclei in the spectrum of
UHECRs. However, if heavy nuclei dominate the Cosmic Ray spectrum at the
highest energies, the neutrino flux can be small as pointed out in~\cite{hooper_taylor2005}.
\paragraph{Magnetic field deflections}
Protons are deflected by intergalactic magnetic fields and by the magnetic
field of the Milky Way on their way to Earth. While this effect could play a role, it
is usually neglected in the calculation of cosmogenic neutrino flux
predictions. The modelling of extragalactic fields is very difficult, since
they are not well-determined. 
\paragraph{Neutrino oscillations}
Neutrino oscillations are usually not considered in the calculation of the GZK
neutrino flux. For a precise determination the changes in the spectrum,
oscillation effects would need to be applied after each
single propagation effect. However, a good approximation is add up the
$\nu_{\mu}$ and $\nu_{e}$ spectra at Earth and assume that each of the three
flavors are present in equal amounts, as it is the case if the neutrinos are
produced far away from Earth.
\subsubsection{On the evolution of the GZK neutrino flux model}
The existence of the cosmogenic neutrino flux was proposed by Berezinsky \&
Zatsepin~\cite{bz1969} shortly after the prediction of the GZK cutoff
was made~\cite{greisen,kuzmin_zatsepin68}. The first prediction presented
in~\cite{bz1969} was revised during the years by applying the gained
knowledge on the different parameters described
above. Table~\ref{gzk_models:tab} summarizes the development of the
model. Each row includes at least one new development in physics or astronomy which was
implemented by one or more authors: 

\begin{itemize}
\item[1969]
When the first prediction was made in 1969, the
mass of the $W^{\pm}-$bosons, $M_{W^{\pm}}$, was not yet determined experimentally. 
\item[1972] The
implementation of the $W-$boson propagator with the correct mass term in the
calculation of the cosmogenic neutrino flux lead to the improvement of the
prediction, see
e.g.~\cite{wdowczyk1972,stecker1973,berezinsky_smirnov1975}. 
\item[1985] Hill \& Schramm
normalized the spectrum according to recent measurements of UHECRs. At the
same time, the cognition
that sources are not homogeneously distributed over the the sky, but evolve
with redshift was considered for the first time~\cite{hill_schramm1985,hill_schramm1986}. 
\item[1991] The evolution scenario, cross
sections and normalization factors were further improved using recent results
from astrophysical and particle physics data, see
e.g.~\cite{stecker1991,yoshida_teshima,protheroe_johnson1996}. 
\item[2001] The observation
of UHECRs above $10^{20}$~eV lead to the prediction that UHECRs are
accelerated up to at least $E_{\max}\approx 10^{21}$~eV. The change in the
maximum energy of the primary spectrum also changes the cosmogenic neutrino
flux which was pointed out in~\cite{ess01}.
\item[2005] It was shown
in~\cite{hooper_taylor2005} that the cosmogenic neutrino flux is reduced
significantly if a large fraction of heavy nuclei is present in the observed
flux of UHECRs.
\item[2007] It is not clear yet which are the sources of UHECRs. The two
  primary candidates are AGN and GRBs. These source classes follow different
  evolution scenarios, which is pointed out in~\cite{yuksel07}. GRBs evolve
  stronger than AGN which leads to an enhanced signal of cosmogenic neutrinos.
\end{itemize} 

\begin{table}
\centering{
\begin{tabular}{ll|llllll}
year&ref&$\sigma_{p\,\gamma}$&$A_p$&$E_{\max}$&evolution&B-&photo-\\
&&&&&&dfl.&dis.\\\hline\hline
1969&\cite{bz1969}&unknown $M_{W^{\pm}}$&estimate&$10^{20}$~eV&no&no&no\\
'72-'75&e.g.~\cite{wdowczyk1972},&{\bf incl.~$M_{W^{\pm}}$}&estimate&$10^{20}$~eV&no&no&no\\
&\cite{stecker1973,berezinsky_smirnov1975}&&&&&&\\
'85-'86&\cite{hill_schramm1985},&incl.~$M_{W^{\pm}}$&{\bf  CRs}&$10^{20}$~eV&{\bf
  yes}&no&no\\
&\cite{hill_schramm1986}&&&&&&\\
'92-'96&e.g.~\cite{stecker1991},&yes&{\bf impr.}&$10^{20}$~eV&{\bf impr.}&no&no\\
&\cite{yoshida_teshima,rachen93},&&&&&&\\
&\cite{protheroe_johnson1996}&&&&&&\\
2001&e.g.~\cite{ess01}&yes&CRs&{\bf $10^{21}$~eV}&yes&no&no\\
2005&\cite{hooper_taylor2005}&yes&CRs&{\bf $10^{21-22.5}$~eV}&yes&no&{\bf
  yes}\\
2007&\cite{yuksel07}&yes&CRs&$10^{21}$~eV&{\bf impr.}&no&no\\\hline
\end{tabular}
\caption{Evolution of GZK models. Boldface: improvement compared to previous
  models. Impr.=Improved; dfl.=deflections; dis.=disintegration.\label{gzk_models:tab}}
}
\end{table}

\begin{figure}[ht]
\centering{
\epsfig{file=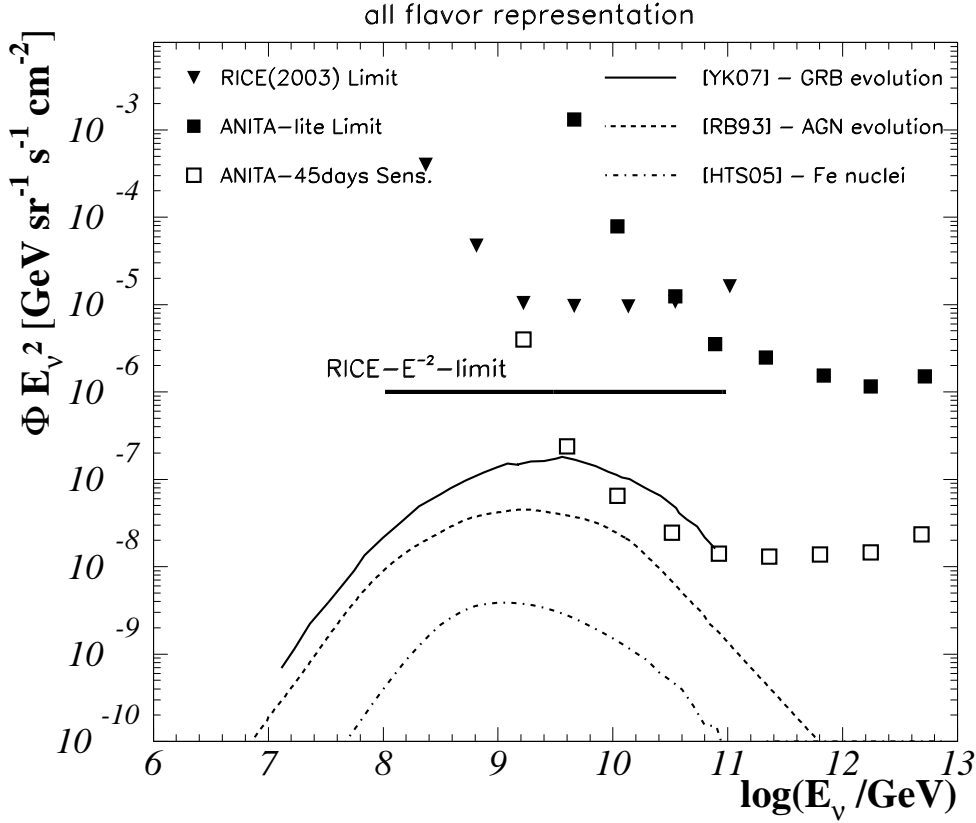,width=\linewidth}
\caption{Comparison of different GZK neutrino flux models, for all neutrino
  flavors, neutrinos and anti-neutrinos. Uppermost, solid line:
  prediction using an evolution function matching the one from
  GRBs~\cite{yuksel07}. Middle, dashed line: model presented
  in~\cite{rachen93}, using an evolution function matching those of AGN. Lower, dot-dashed line: pure Fe-based
  spectrum~\cite{hooper_taylor2005,andrew_priv}. The model-independent limits from~\cite{rice2003}
  and {\sc ANITA-lite} are shown together with the sensitivity for
  a 45~day long flight with the full {\sc ANITA} experiment~\cite{anita_limits06}. \label{gzk_models:fig}}
}
\end{figure}

In Fig.~\ref{gzk_models:fig}, the energy spectrum of cosmogenic neutrinos is
shown for three different models. The
uppermost, solid line shows the resulting spectrum if GRBs are responsible for the flux of
UHECRs. GRBs are observed to evolve even stronger than different AGN classes,
which leads to the enhancement of the flux as shown in~\cite{yuksel07}. The
current {\sc Auger} limit comes close to the strong evolution model and with data
from the completed {\sc Auger} array, it will be possible to test this model
in detail. The middle, dashed line shows the spectrum for an AGN-type
evolution of sources as calculated in~\cite{rachen93}. The
lower, dot-dashed line represents the hypothetical spectrum using that UHECRs consist of
iron only as calculated in~\cite{hooper_taylor2005,andrew_priv} to show that the
cosmogenic neutrino flux is decreased significantly with increasing
contributions from heavy nuclei in the Cosmic Ray spectrum. It is shown
in~\cite{hooper_taylor2005} that, the higher the mass number of the nucleus,
the less neutrinos are produced during propagation. A maximum energy of
$\ep=10^{12.4}$~GeV is used for the primary iron nuclei, taking into account
the increase of the maximum energy with the charge $Z$, see Equ.~(\ref{emax:equ})
in Section~\ref{multimessenger}. For lower maximum energies, the flux is
further diminished. 
%=================================
\subsection{Comparison of event rates in underground detectors}
%=================================
In this section, the expected neutrino rates is calculated for underground
detectors in order to get an estimate of the necessary array size for
the detection of ultra high-energy neutrinos. Four detector
properties have to be considered:
\begin{enumerate}
\item The \emph{threshold energy} $E_{\min}$ of the detector determines the
  total integral neutrino rate per year and km$^2$. The threshold energy is varied between
$E_{\min}=10^{7}$~GeV and $E_{\min}=10^{10}$~GeV in the following calculations.
\item The \emph{depth} of the detector below the Earth's surface $d$ determines the
fraction of neutrinos absorbed before reaching the detector.
\item  The
\emph{effective range} of neutrino-induced cascade, $r_{eff}$. It is determined by the
size of the detector, since the cascade has to be induced within the
detection array in order to be observed. All following calculations are performed
using $r_{eff}=1$~km. For a constant effective range, the results scale
linearly with the range a simple factor can be used to account for different
detection array sizes.
\item The \emph{density} of the
detector material, $\rho'=\rho/\rho_{H_2 O}$ has to be considered as
well. The density will be given in terms of $\rho'=1$, assuming water or ice
as the detector medium. For detection arrays in salt, the result can be
multiplied by a factor $2.3$ in order to receive the corresponding event rates.
\end{enumerate}
The angle integration was performed over the upper hemisphere ($2\,\pi$) only, since at these energies, no contribution from the lower hemisphere due to
absorption. The result is given in units of yr$^{-1}$ km$^{-3}$
GeV$^{-1}$. Three models are compared here,
\begin{itemize}
\item[{\bf [YK07]}] GRB evolution scenario~\cite{yuksel07}. With a source evolution matching the
  one of GRBs, a high neutrino flux is expected. This model is used as the
  maximum flux possible, since it is close to, but still below, most recent
  neutrino flux limits. 
\item[{\bf [RB93]}] AGN evolution scenario~\cite{rachen93}. The less strong evolution of
  AGN leads to a reduced neutrino flux compared to the GRB evolution
  scenario. 
\item[{\bf [HTS05]}] Fe iron nuclei as primaries~\cite{hooper_taylor2005,andrew_priv}.  It is assumed that the dominant fraction of Cosmic Rays
  consists of heavy nuclei. A pure iron spectrum is an absolute lower limit -
  the neutrino flux increases with the decrease of the mass of the primary
  nucleus, and it is likely that the spectrum of UHECRs is a composition of
  light and heavy nuclei. The question of the fraction of heavy nuclei in
  UHECRs is not answered yet. Therefore, this model is used
  as a lower limit here.
\end{itemize}

\begin{figure}[ht]
\centering{
\includegraphics[width=11.5cm]{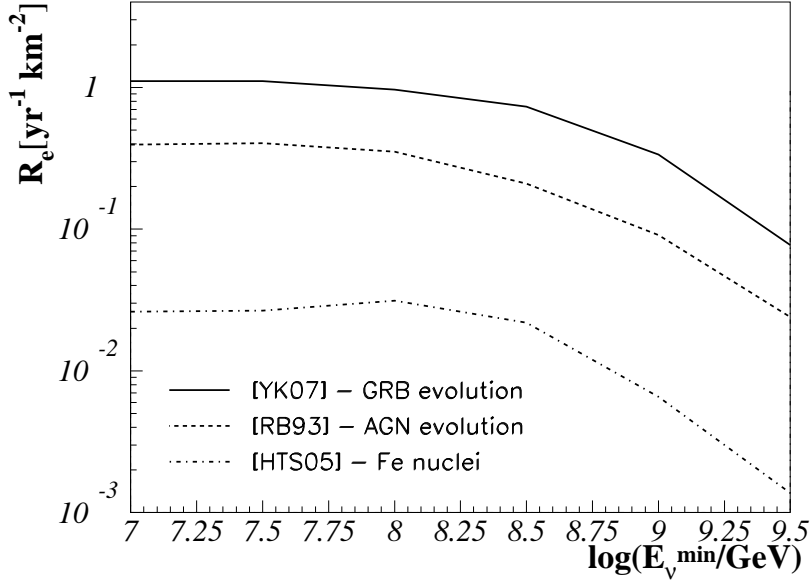}
\caption[Energy-integrated neutrino detection rate ($\nu_e$)]{Integrated neutrino rate, representing the number of detected
  electron neutrino induced cascades above a certain energy $E_{\min}$. Three
  different models are shown. The solid line represents the model
  from~\cite{yuksel07}, using GRBs as the sources of UHECRs, the dashed line shows
an AGN-based model from~\cite{rachen93}. The dot-dashed line represents a pure
iron spectrum as calculated in~\cite{hooper_taylor2005,andrew_priv}.}
\label{nue_rate_emin}
}
\end{figure}

Since
the cross sections for nucleon neutrino interactions do not change
significantly with neutrino flavor, the results are basically the same
for both flavors. The calculations presented here are performed for electron
neutrinos. For muon neutrinos, the results are comparable. In the case of
tau-neutrinos, the regeneration of tau leptons needs to be considered as well.
Figure~\ref{nue_rate_emin} shows the
integrated electron neutrino detection rates depending on the threshold energy of the detector for
the four different GZK neutrino flux predictions. A depth of $d=0.5$~km
is assumed in these calculations. With a detection threshold below $E_{\min}\sim
10^{8}$~GeV, a rate of $R\sim 1$~events/yr/km$^{3}$ can be expected for the
maximum model [YK07]. The AGN evolution model [RB93] yields a factor of $\sim
5$ less neutrinos and the minimal model using Fe nuclei as primaries results
in $\sim 0.03$~events/yr/km$^{3}$. This shows that the instrumentation of
volumes significantly larger than $1$~km$^3$ is eligible in order to achieve a
significant rate of GZK neutrinos per year. This is possible with radio and
acoustic detection arrays, since the cascades have a large extension, so that
a sparsely instrumented area suffices. Simulations for the extension of
{\sc IceCube} to {\sc IceCube Hybrid} have been performed, assuming a
radio/acoustic array of $10\times10$~km$^2$ in the ice around the {\sc IceCube}
detector, together with additional {\sc IceCube} strings around
the original array~\cite{vandenbroucke_arena}. The spacing between the acoustic/radio strings is assumed
to be $1$~km. While the size of the cascades in the radio and in the acoustic
channel is an advantage of these new techniques, the high energy threshold
 at $E_{\min}=10^{8.5}$~GeV for radio measurements, and at $E_{\min}=10^{9.5}$ for
acoustic detection of neutrino-induced showers diminishes the rate by up to an
order of magnitude as it can be
seen from Fig.~\ref{nue_rate_emin}. The total event rate per year is still a
few up to 10 events per year for an array of $>100$~km$^{3}$ and an AGN
evolution scenario.

Table~\ref{depth_table_nue} shows how the event rate for electron neutrinos changes with the depth of the
detector below the Earth's surface. Results for five different depths,
$d=0.4,\,0.6,\,1.0,\,1.5,\,2.5$~km, are shown for a threshold energy of
$E_{\min}=10^{7}$~GeV. The rate decreases with the depth. It is reduced by a factor of two when the detector is built at $1.5$~km depth
instead of $0.5$~km. For threshold energies above $E_{\min}\sim 10^{8}$~GeV as they
are expected for radio and acoustic detection, the background of atmospheric
muons and neutrinos is negligible. In the case of optical detection, energy
cuts can be applied in order to remove lower-energy atmospheric events. Consequently, it is not necessary to deploy the
detector deep into the Earth to reduce any background and a detector has the
best detection potential at shallow depths of $d\sim 0.5$~km.

\begin{table}[ht]
\centering{
\begin{tabular}{l|lllll}
\hline
&$R_0$&[km$^{-3}\,$yr$^{-1}$]&\\ 
$d$[km]$=$&$0.4$&$0.6$&$1.0$&$1.6$&$2.5$\\ \hline \hline
YK07&1.2&0.97& 0.69&0.45&0.27\\ 
RB93&0.44&0.34&0.25&0.17&0.10\\
HTS05&0.029& 0.022&0.016& 0.0010&0.0055\\ \hline
\end{tabular}
}
\caption[$\nu_{e}$-induced events - ice/water -
$E_{\min}=10^{7}$~GeV]{Electron neutrino event rates for different detector depths, $d=0.4,0.6,1.0,1.6,2.5$~km. The detector threshold energy is $E_{\min}=10^{7}$~GeV. The rate decreases with the depth, since neutrinos are absorbed in the Earth.}
\label{depth_table_nue}
\end{table}

%=================================
\subsection{Conclusions and outlook}
%=================================
Shortly after the prediction of the GZK cutoff, the first estimates of the
cosmogenic neutrino flux were made~\cite{bz1969}. During the past 40 years,
advances in particle physics, e.g.~the determination of the $W$ mass, as well
as in
astro- and astroparticle physics, e.g.~improving evolution functions, have
lead to the improvement of predicted GZK neutrino flux. If UHECRs are protons
originating from AGN, an detection rate of about $1-2$~events can be expected in
{\sc IceCube} per year. This event rate is relatively low, since the GZK
neutrino flux becomes important at energies above $E_{\min}=10^{8}$~GeV, while the
sensitivity of {\sc IceCube} peaks at lower energies around $\en\sim
$~PeV. Still, the flux should become visible within only a few years of
operation due to the extremely low background at these high energies. The
event rate for the GZK neutrino flux is expected to be significantly higher
with future detection
methods, like the acoustic and radio detection. Here, much larger arrays of
the order of $>100$~km$^3$ are
planned to be instrumented. For acoustic detection, the lower energy threshold
lies at around $\sim 10^{9.5}$~GeV, for radio detection, it is close to $\sim
10^{8.5}$~eV. Apart from underground arrays, neutrino-induced
electromagnetic and hadronic showers can be detected with balloon experiments,
e.g.~{\sc ANITA}, airshower experiments like {\sc Auger} and Cherenkov
telescopes like {\sc MAGIC}. All these methods have relatively high detection
thresholds. Therefore, they are not suited for the observation of point source
fluxes from e.g.~AGN, but are well-suited for the detection of the diffuse GZK
neutrino flux.

Within the next years, it will be possible to further improve the model of the
GZK neutrino flux either by the detection of a signal or by setting
limits. The sensitivity of a 45~days-flight by {\sc ANITA} is indicated in
Fig.~\ref{gzk_models:fig}. Here, it is already possible to constrict strong
evolution models, like those assuming that UHECRs come from GRBs. Together
with the detailed examination of the GZK cutoff observed with charged Cosmic
Rays by the {\sc Auger} observatory and other airshower experiments, the
search for the GZK neutrino flux will
render possible the constraint or the determination of the evolution function
of the sources of UHECRs. 

%%%%%%%%%%%%%%%%%%%%%%%%%%%%%%%%%%%%%%%%%%%%

%%%%%%%%%%%%%%%%%%%%%%%%%%%%%%%%%%%%%%%%%%%%
\section{Summary and conclusions \label{summary}}
%%%%%%%%%%%%%%%%%%%%%%%%%%%%%%%%%%%%%%%%%%%%
The concept of multimessenger physics becomes more and more important these
days in order to get a complete picture of the sources of investigation. In
particular, the non-thermal Universe can only be explained by combining pieces of
information from all three messengers, i.e.~photons, charged Cosmic Rays and
neutrinos. The connection between the observation of charged Cosmic Rays and
high-energy photons can only be established by means of high-energy neutrino
astrophysics. Hence, the deployment of $1$~km$^{3}$ neutrino detectors is
crucial in order to further increase the sensitivity to the flux from
extraterrestrial neutrinos. 

The modeling of neutrino emission from different sources
is essential for the selection of data in large volume neutrino detectors.
The different source classes reviewed here,
i.e.~galactic sources, AGN, GRBs, starbursts and GZK neutrinos, are all
analyzed with different methods. These include the search for a diffuse
signal at the highest energies, single point source searches, transient source
searches and stacking approaches. The different methods help to reduce the
background towards the signal expectation. 

The fact that neutrino
flux limits can be used for physics constraints can be formulated as the {\it
  Olbers paradox for neutrinos}. Olbers pointed out in the 19th century that the
fact of a dark night sky contradicted the prevailing theory of the
Universe: a homogeneous distribution of stars in an spatially and temporally
infinite Universe in thermal equilibrium implied an infinitely bright sky
during day and night time. The simple limit of the dark night sky was a
first indication that the world view at the time was incomplete. Today, it is
known that the Universe is not in thermal equilibrium, which, among other
things, leads to the dark night sky. In analogy to this optical limit, the so
far dark neutrino sky provides first indications of how to improve current
neutrino flux models. The current neutrino flux limits
can already be used to restrict different models. In particular, a correlation
between X-rays from AGN and neutrinos can be excluded, since models based on
this assumption violate recent neutrino flux limits.

Next generation of neutrino telescopes like {\sc IceCube} and {\sc KM3NeT} enables a deeper view into the neutrino sky by more than an
order of magnitude. With new detection techniques, like
the observation of neutrino-induced cascades at radio wavelengths, or the
detection of an acoustic cascade in ice or water, enables the investigation of
neutrino emission at the highest energies. The combination of different
observational techniques renders possible the detailed investigation of a
neutrino signal and thereby further advancement in the understanding of multimessenger astrophysics.

%During the past decade, it has been shown that natural large water or ice
%reservoirs can be used for the instrumentation of large volume neutrino
%detection arrays. The atmospheric neutrino spectrum has been measured up to
%energies of $\en\approx 100$~TeV, which implies an improved sensitivity by
%{\bf ????} orders of magnitude. The most restrictive limit to the detection of
%extraterrestrial neutrinos is given by the
%{\sc AMANDA} experiment, using the muon-neutrino detection channel,
%\be
%\dl=2.6\cdot 10^{-8}\,\diffunits\,.
%\ee
%The limit is valid in the energy range of
%$300$~TeV$<\en<1$~PeV~\cite{kirsten_icrc07}. The achieved limits can be used
%to restrict different scenarios of neutrino production, but still leave room
%for alternative scenarios. For instance, the postulated connection between X-rays and
%neutrinos from AGN could be rejected, which is in agreement with recent
%observations showing that a large fraction of the emitted X-rays from AGN is
%actually thermal. 
%{\bf plot main models HE neutrinos; 1 AGN, 1 GRB, 1 GZK; Kirsten spectrum \&
%  limit; HE limit (Auger or HiRes or both}\\
%{\bf Olber's paradox for neutrinos! Aufh\"anger fuer Schlu{\ss}}
%%% Local Variables: 
%%% mode: latex
%%% TeX-master: t
%%% End: 

\subsection*{Acknowledgments}
The outcome of this review article has profited very much from the interaction
 with different colleagues. In particular, I would
like to thank Francis Halzen for intense discussions throughout the
years. Many thanks also to Wolfgang Rhode, Peter Biermann, Per-Olof Hulth,
Kirsten M\"unich, Athina Meli, Andrew Taylor, Eli Waxman, Kohta Murase, Lutz K\"opke,
Andreas Haungs, Justin Vandenbroucke, Christian Spiering, Spencer Klein, Lisa
Gerhard, Yolanda Sestayo de la Cerra, Ralf Wischnewski, Jenni Adams, Karl Mannheim, Reinhard Schlickeiser and many more for valuable remarks. Additionally,
I would like to thank the {\sc IceCube} and the {\sc MAGIC} collaborations for 
fruitful interactions. Thank you also to the entire astroparticle physics group in
Dortmund, for inspiring discussions. Finally, I would like to thank the
anonymous referee for very constructive comments and ideas, which helped to improve this
review significantly.
 This work is supported by the Chaudoire award, granted by the Chaudoire
 foundation, connected to the University of Dortmund. 
%\bibliography{review_lib1,review_lib2,review_lib3,review_lib4,review_lib5,review_lib6}
\bibliographystyle{alpha}

\newcommand{\etalchar}[1]{$^{#1}$}

\end{document}